\documentclass[12pt]{article}
\usepackage[dvips]{epsfig}
\usepackage{cite}
\newcommand{\ep}{\varepsilon}
\newcommand{\Gl}[2]{{\mbox{Gl}}_{#1}\left(#2\right)}
\newcommand{\Li}[2]{{\mbox{Li}}_{#1}\left(#2\right)}
\newcommand{\Cl}[2]{{\mbox{Cl}}_{#1}\left(#2\right)}
\newcommand{\Ls}[2]{{\mbox{Ls}}_{#1}\left(#2\right)}
\newcommand{\LS}[3]{{\mbox{Ls}}_{#1}^{(#2)}\left(#3\right)}                    
\newcommand{\Lsc}[2]{{\mbox{Lsc}}_{#1\!}\left(#2\right)}
\newcommand{\Ti}[2]{{\mbox{Ti}}_{#1}\left(#2\right)}
\newcommand{\tfrac}[2]{{\textstyle{\frac{#1}{#2}}}}
 
\newcommand{\SN}[3]{{\mbox{S}}_{#1,#2}\left(#3\right)}

\newcommand{\Snp}[2]{{\mbox{S}}_{#1\!}\left(#2\right)}

\def\H{\hbox{H}}
\def\d{\hbox{d}}
\def\f{\hbox{f}}

\begin{document}
\renewcommand{\thefootnote}{\fnsymbol{footnote}}

\begin{flushright}
 {DESY/02-214} \\[3mm]
 {hep-th/0303162} \\[3mm]
% {Oktober 2003}
% {Februar 2004}
\end{flushright}
 \vspace*{2.0cm}
 \begin{center}
 {\Large \bf
 Massive Feynman diagrams and inverse binomial sums
 }
 \end{center}
 
\vspace{3mm}
 
 \begin{center}
 A.~I.~Davydychev$^{a,}$\footnote{
 Current address: Schlumberger, SPC, 110 Schlumberger Dr., MD-5,
 TX~77478, USA.
 Email address:
 davyd@thep.physik.uni-mainz.de}
 \quad  and \quad
 M.~Yu.~Kalmykov$^{b,c}$\footnote{
Supported by DFG under Contract SFB/TR~9-03
and in part by RFBR grant \# 04-02-17192.
Email address: kalmykov@thsun1.jinr.ru}\\
 
\vspace{3mm}

$^{a}${\em
Institute for Nuclear Physics, 
Moscow State University, \\
 119992, Moscow, Russia}
\\
\vspace{.3cm}
$^{b}${\em
 DESY--Zeuthen,
Theory Group, Platanenallee 6, \\ D-15738 Zeuthen, Germany}
\\
\vspace{.3cm}
$^{c}${\em
BLTP, JINR, 141980 Dubna, Russia.}
\\
\end{center}
  
 \hspace{3in}
 \begin{abstract}
 When calculating higher terms of the $\ep$-expansion
 of massive Feynman diagrams, one needs to evaluate
 particular cases of multiple inverse binomial sums.
 These sums are related to the derivatives of certain
 hypergeometric functions with respect to their parameters.
 We explore this connection and
 analytically calculate a number of such infinite sums,
 for an arbitrary value of the argument which corresponds to
 an arbitrary value of the off-shell external momentum.
 In such a way, we find a number of new results for
 physically important two-loop two- and three-point
 Feynman diagrams. The results are presented in terms
 of generalized polylogarithmic functions.
\end{abstract}

\vspace{3mm}

%\noindent
%PACS: 11.15.Bt, 02.30.Gp, 02.30.Lt, 12.15.Lk
%%%% extra: 12.38.Bx 
%\\
%Keywords: Feynman diagrams, 
%          Hypergeometric functions, 
%          Multiple inverse binomial sums, 
%          Higher order $\ep$-expansion,
          
\thispagestyle{empty} 
\newpage
%=====================================================================
\renewcommand{\thefootnote}{\arabic{footnote}}
\setcounter{footnote}{0}

%%%%%%%%%%%%%%%%%%%%%%%%%%%%%%%%%%%%%%%%%%%%%%%%%%%%%%%%%%%%%%%%%%%%%%%%%
\section{Introduction}
\setcounter{equation}{0}
%%%%%%%%%%%%%%%%%%%%%%%%%%%%%%%%%%%%%%%%%%%%%%%%%%%%%%%%%%%%%%%%%%%%%%%

In many cases,
the results of analytical calculation of Feynman diagrams can be 
represented as combinations of hypergeometric functions. 
However, the problem of constructing the $\ep$-expansion of
hypergeometric functions within dimensional regularization~\cite{dimreg} 
(where $n=4-2\ep$ is the space-time dimension) is not completely solved.
Rather often,  
results for the terms of the $\ep$-expansion can be expressed in terms of 
polylogarithms~\cite{Lewin} as well as 
Nielsen polylogarithms~\cite{Nielsen}. 
Recently it was demonstrated that new types of functions, 
harmonic polylogarithms~\cite{RV00} and 
multiple polylogarithms~\cite{GR2001}, appear in 
multiloop calculations. Furthermore, the nested sums~\cite{nested}
were proposed as a generalization of multiple polylogarithms, and 
the $\ep$-expansion for a large class of hypergeometric function 
was constructed \cite{expansionII}. 

In this paper we study the {\it multiple inverse binomial sums}
defined as\footnote{Some particular results for 
the {\it multiple binomial sums} are presented in \cite{poleII}. 
Those sums are defined similarly to Eq.~(\ref{binsum}),
but with $\left( 2j \atop j\right)$ in the numerator, rather
than in the denominator.}
\begin{equation}
\Sigma_{a_1,\ldots,a_p;
\; b_1,\ldots,b_q;c}^{\; i_1,\ldots,i_p; \;j_1,\ldots,j_q}(u)
\equiv
\sum_{j=1}^\infty \frac{1}{\left( 2j \atop j\right) } \frac{u^j}{j^c}
[S_{a_1}(j\!-\!1)]^{i_1}\ldots [S_{a_p}(j\!-\!1)]^{i_p}\;
[S_{b_1}(2j\!-\!1)]^{j_1}\ldots [S_{b_q}(2j\!-\!1)]^{j_q},
\label{binsum}
\end{equation}
where $S_a(j)\equiv\sum_{k=1}^j k^{-a}$ is the harmonic sum\footnote{
Through the rest of this paper, the notations $S_a$ and $\bar{S}_a$
will always mean $S_a(j-1)$ and $S_a(2j-1)$, respectively, 
even we do not mention this explicitly.
When there are no sums of the type $S_{a}$ or $\bar{S}_{b}$
on the r.h.s.\ of Eq.~(\ref{binsum}),
we shall put a ``$-$'' sign instead of the indices $(a,i)$ or $(b,j)$
of $\Sigma$, respectively.} 
and $u$ is an arbitrary argument.
In what follows, we will also use
$z\equiv \tfrac{1}{4}u$
as the argument of the occurring hypergeometric functions.
For sums of the type~(\ref{binsum}), the
{\em weight} ${\bf J}$
can be defined as 
${\bf J} = c + \sum_{k=1}^{p} a_k i_k + \sum_{k=1}^{q} b_k j_k$, 
whereas the {\em depth} can be associated with the sum 
$\sum_{k=1}^{p} i_k + \sum_{k=1}^{q} j_k$.

All multiple binomial sums~(\ref{binsum}) can be presented in
terms of function $\psi(z)=\frac{\rm d}{{\rm d}z}\ln\Gamma(z)$ 
and its derivatives by means of the following relation
\[
\psi^{(k-1)}(j) = (-1)^k (k-1)! \left[\zeta_k - S_k(j-1) \right],
  \qquad  k>1,
\]
where $\psi^{(k)}(z)$ is the $k$-th derivative of the $\psi$-function.
In particular, for $k=1$ we have $\psi(j) = S_1(j-1)-\gamma_E$,
where $\gamma_E$ is Euler's constant.

The sums~(\ref{binsum})
appear in the calculation of massive Feynman diagrams 
within several different approaches: for instance, as solutions of 
differential equations for Feynman amplitudes~\cite{dem},
through a na\"ive $\ep$-expansion of hypergeometric functions 
within Mellin--Barnes technique~\cite{BD-TMF}, or 
in the framework of recently proposed algebraic approach~\cite{Isaev}.
Physical applications include the one-, two- and three-loop 
massive Feynman diagrams with two massive cuts 
\cite{FKV98,DD,FKV99,DK-bastei,DK01,oneloop}. 
The case $u=1$ corresponds to the single-scale propagator-type 
diagrams~\cite{single}. 
Although there are many publications 
concerning harmonic series~\cite{harmonic},
only a limited number of results are available for the inverse binomial sums. 
Some particular results 
for $u=1,2,3$ can be extracted from~\cite{KV00,DK01}.
The sums with $u=1,3$ are expressible in terms 
of an ``odd'' basis~\cite{odd}, whereas the case $u=2$
corresponds to an ``even'' basis~\cite{DK01,euler}\footnote{The connection 
between ``sixth root of unity''~\cite{B99}
and ``odd''/``even'' bases was discussed in~\cite{DK01,review}.}. 
Several new results for a generalization of the 
sums (\ref{binsum}) are presented in \cite{newI}.
However, only for special type of sums the analytical 
results are available~\cite{KV00,sums,BBK},
\begin{equation}
\label{KV}
\sum_{j=1}^\infty \frac{1}{\left( 2j \atop j\right) } \frac{u^j}{j^c}   = 
 - \sum_{i=0}^{c-2} \frac{(-2)^i}{i! (c-2-i)!} 
( \ln u )^{c-2-i} \LS{i+2}{1}{\theta} \;,
\end{equation}
where
$c \geq 2$,
\begin{equation}
\label{def_theta}
\theta \equiv 2\arcsin\left(\tfrac{1}{2}\sqrt{u}\right) 
= 2\arcsin\sqrt{z}\; , 
\end{equation}
\begin{equation}
\label{def_Ls}
\LS{j}{k}{\theta} =   - \int\limits_0^\theta {\rm d}\phi \;
   \phi^k \ln^{j-k-1} \left| 2\sin\frac{\phi}{2}\right| \, ,
\quad 
\Ls{j}{\theta} = \LS{j}{0}{\theta} \; 
\end{equation}
is the generalized log-sine function~\cite{Lewin}.
Some examples when such functions occur in the $\ep$-expansion
of Feynman diagrams can be found in Refs.~\cite{DT2,Ls_ex,odd,D-ep}.
Usually, the occurring angles~(\ref{def_theta}),
possess certain geometrical meaning~\cite{DD}. 

The main aim of the present publication is the analytical 
calculation of inverse binomial sums~(\ref{binsum}),
including some relevant examples of physically important 
Feynman diagrams.

The paper is organized as follows. 
In Section~2, employing the connection with the $\ep$-expansion
of hypergeometric functions,
we obtain analytical results for sums of the type~(\ref{binsum}),
valid for $u\leq 4$ ($z\leq 1$). Mainly the sums of
the {\em weights}~${\bf 3}$ and ${\bf 4}$ are considered.  
In Section~3 the analytical continuation 
to other values of $u$ is constructed. 
Section~4 contains some applications of our results related to 
the $\ep$-expansion of Feynman diagrams, 
mainly two-loop master integrals. In Appendix~A we briefly summarize 
the relevant properties of the harmonic polylogarithms 
of complex arguments and related functions.
In Appendix~B we show how certain identities between hypergeometric
functions can be used to establish relations between the corresponding sums.
In Appendix~C we have collected explicit 
results for inverse binomial sums of lower weights. 
In Appendix~D relations between 
binomial, harmonic and inverse binomial sums are explored, and analytical results 
for multiple binomial sums up to weight {\bf 3} are presented. 
Appendix~E contains a realistic example of a physical application
of the considered integrals. {\tt (Remember to change when appendices 
are ready!)}

%%%%%%%%%%%%%%%%%%%%%%%%%%%%%%%%%%%%%%%%%%%%%%%%%%%%%%%%%%%%%%%%%%%%%%%%%
\section{Inverse binomial sums and hypergeometric functions}
\setcounter{equation}{0}

\subsection{The $\ep$-expansion of the $_2F_1$ function}

Here, our analysis is based on comparing
two representations of hypergeometric function whose parameters
depend on $\varepsilon$. 
One of them is the series representation in terms of the harmonic sums,
whereas the second one is the exact result in terms of the
functions related to the polylogarithms. 

In particular, we consider the $_{2}F_1$ 
hypergeometric function of a special type,
\begin{equation}
\label{2F1_def}
_{2}F_1\left(\begin{array}{c|}
1+a_1\ep, 1+a_2 \ep\\
\tfrac{3}{2} + b \ep  \end{array} ~z \right)
= \sum_{j=0}^{\infty} \frac{u^j}{j!}\;
\frac{(1+a_1\ep)_j (1+a_2\ep)_j (1+b\ep)_j}  
{(2+2b\ep)_{2j}} \; ,
\end{equation}
where $u=4z$, $(\alpha)_j\equiv\Gamma(\alpha+j)/\Gamma(\alpha)$ is
the Pochhammer symbol,
and we have used the duplication formula
$(2\beta)_{2j}=4^j(\beta)_j(\beta+\tfrac{1}{2})_j$.
To perform the $\ep$-expansion we use the well-known 
representation
\begin{equation}
(1+a\ep)_j = j! \; \exp\left[ -\sum_{k=1}^{\infty} \frac{(-a\ep)^k}{k}
S_k(j) \right] \; ,
\label{pochhamer}
\end{equation}
which yields
\begin{eqnarray}
&& \hspace*{-15mm}
_{2}F_1\left(\begin{array}{c|}
1+a_1\ep, 1+a_2 \ep\\
\tfrac{3}{2} + b \ep  \end{array} ~z \right)
= \frac{2(1 + 2 b\ep)}{u} 
\sum_{j=1}^\infty \frac{1}{\left( 2j \atop j\right) }  \frac{u^j}{j}  
\Biggl\{
1 
+ \ep \left[ 
   (A_1+b)S_1 - 2 b \bar{S}_1
      \right] 
\nonumber \\ && 
+  \ep^2 \Bigl[ 
2 b^2 \left( \bar{S}_2 + \bar{S}_1^2 \right) 
-  2 b (A_1+b) S_1 \bar{S}_1 
- \tfrac{1}{2} (A_2+b^2) S_2
+ \tfrac{1}{2} (A_1+b)^2 S_1^2
\Bigr]
\nonumber \\ && 
+ \ep^3 \Bigl[ 
\tfrac{1}{6} (A_1+b) ^3 S_1^3 
- b (A_1+b)^2 S_1^2 \bar{S}_1 
- \tfrac{1}{2} (A_1+b)(A_2+b^2)S_1 S_2
\nonumber \\ && \hspace{10mm}
+ 2 b^2 (A_1+b) S_1 \left( \bar{S}_2 + \bar{S}_1^2 \right)
+ b (A_2+b^2) S_2 \bar{S}_1
+ \tfrac{1}{6} (2 b^3 - A_1^3 + 3 A_1 A_2) S_3
\nonumber \\ && \hspace{10mm}
- \tfrac{4}{3} b^3 
\left( 2 \bar{S}_3 + 3 \bar{S}_2 \bar{S}_1 + \bar{S}_1^3 \right)
\Bigr]
+ {\cal O} (\ep^4)
\Biggr\} \; ,
\label{hyper:f21}
\end{eqnarray}
where $A_k \equiv a_1^k+a_2^k.$
Note that the coefficient of the $\ep^3$ term can be represented as
\begin{eqnarray}
\label{combinations}
&& \hspace*{-7mm}
\tfrac{1}{6}(A_1^3+b^3) {\cal C}_0 
+2 b^2 (A_1+b) {\cal C}_1
+4 b^3 {\cal C}_2
+\tfrac{1}{2} b (A_1-b) (A_1-2b) {\cal C}_3
\nonumber \\ &&
+\tfrac{1}{2} (A_1^2-A_2) (A_1+b) \left(S_1 S_2-S_3\right)
+\tfrac{1}{2} b (A_1^2-A_2) \left(S_3-2 S_2 \bar{S}_1\right) \; ,
\end{eqnarray}
where we have introduced the following combinations of 
harmonic sums:
\begin{eqnarray}
\label{combinations2}
{\cal C}_0&=& S_1^3-3S_1S_2+2S_3\; ,
\nonumber \\
{\cal C}_1&=& 
S_1^3-S_1 S_2+S_1 (\bar{S}_1^2+\bar{S}_2)
-\tfrac{5}{2} S_1^2 \bar{S}_1+\tfrac{3}{2} S_2 \bar{S}_1 \; ,
\nonumber \\
{\cal C}_2&=&  \tfrac{3}{4}S_1 S_2 - \tfrac{3}{4}S_1^3
+\tfrac{3}{2}S_1^2\bar{S}_1 - S_2\bar{S}_1 
-\tfrac{1}{3} \left( 2\bar{S}_3 + 3 \bar{S}_1\bar{S}_2
+\bar{S}_1^3\right) \; ,
\nonumber \\
{\cal C}_3&=& S_1^3-2 S_1^2 \bar{S}_1 - S_1 S_2 + 2 S_2 \bar{S}_1 \; .
\end{eqnarray}
To completely define the $\ep^3$ order of the expansion
of $_2F_1$, we need results for six combinations
of sums, according to the number of independent
combinations of the parameters $A_i$ and $b$.

Let us first consider the case $0\leq u \leq 4 $ ($0\leq z\leq 1$).
In this region  the following parametrization 
can be used:  
$u = 4 \sin^2\tfrac{\theta}{2}$ ($z= \sin^2\tfrac{\theta}{2}$),
where $0 \leq \theta \leq \pi.$
In the rest of this paper, we will use the short-hand notation 
\begin{equation}
L_\theta \equiv
\ln \left( 2 \cos \tfrac{\theta}{2} \right) \; , \qquad  
l_\theta \equiv
\ln \left( 2 \sin \tfrac{\theta}{2} \right) \; .
\end{equation}

For a few special cases, the $\ep$-expansion of the 
$_2F_1$ functions is known. 
First of all, the following relation holds~\cite{PBM3}:
\begin{equation}
{}_{2}F_1 \left(\begin{array}{c|} \!\! 1\!+\!\ep, 1 \!-\! \ep\! \\
\frac{3}{2} \end{array} ~\sin^2 \tfrac{\theta}{2} \right)\! =
\frac{\sin \left( \ep \theta \right)}{\ep \sin(\theta)} \; .
\label{basis_I}
\end{equation}
Its expansion contains only the even powers of~$\ep$.
Using it, it is easy to get 
\begin{eqnarray}
\sum_{j=1}^\infty \frac{1}{\left( 2j\atop j\right)} \frac{u^j}{j} S_2 
&=&  \tfrac{1}{6} \theta^3 
\tan\tfrac{\theta}{2} \; ,
\label{S2}
\\
\sum_{j=1}^\infty \frac{1}{\left( 2j\atop j\right)} 
\frac{u^j}{j} \left( S_2^2 - S_4 \right)
&=&  \tfrac{1}{60} \theta^5 
\tan\tfrac{\theta}{2} \; ,
\label{S4}
\end{eqnarray}
etc. Then, there are some cases~\cite{D-ep,DK01} when 
an arbitrary term of the $\ep$-expansion can be calculated
in terms of log-sine functions, 
%%%%%%%%%%%%%%%%%%%%%%%%%%%%%%%%%%%%%%%%%%%%%%%%%%%%%%%%%%%%
\begin{eqnarray}
_2F_1 \left(\begin{array}{c|} 1, 1+\ep  \\
\tfrac{3}{2} \end{array} ~\sin^2 \tfrac{\theta}{2} \right)
&=& \frac{\left( 2 \cos \tfrac{\theta}{2} \right)^{-2\ep}}{\sin \theta} 
\sum_{j=0}^{\infty} \frac{(2\ep)^j}{j!}
\left[ \Ls{j+1}{\pi-\theta} - \Ls{j+1}{\pi} \right] \; ,
\label{basis_II}
%%%%%%%%%%%%%%%%%%%%%%%%%%%%%%%%%%%%%%%%%%%%%%%%%%%%%%%%%%%%
\\
_2F_1 \left(\begin{array}{c|} 1, 1+\ep  \\
\tfrac{3}{2}+\ep \end{array} ~\sin^2 \tfrac{\theta}{2} \right)
 &=& -\frac{(1+2\ep)}{\sin\theta (2 \sin\tfrac{\theta}{2})^{2\ep}}
 \sum_{j=0}^{\infty} \frac{(2\ep)^j}{j!} \Ls{j+1}{\theta} \; ,
\\
_2F_1 \left(\begin{array}{c|} 1, 1+\ep  \\
\tfrac{3}{2}+\tfrac{1}{2}\ep \end{array} ~\sin^2 \tfrac{\theta}{2} \right)
 &=& -\frac{1+\ep}{(2\sin\theta)^{1+\ep}}
 \sum_{j=0}^{\infty} \frac{\ep^j}{j!} \Ls{j+1}{2\theta} \; .
\end{eqnarray}
Moreover, for a more general case an integral representation 
can be obtained~\cite{PBM3,DK01},
\begin{equation}
{}_{2}F_1 \left(\begin{array}{c|} 1, 1 \!+\! a \ep \\
\frac{3}{2}+b \ep \end{array} ~\sin^2 \tfrac{\theta}{2} \right) =
\frac{(1\!+\! 2 b \ep)\left( 2 \cos \tfrac{\theta}{2} \right)^{2b\ep-2a\ep}}
{ \sin \theta \left( 2 \sin \tfrac{\theta}{2} \right)^{2b\ep}}
\int\limits_0^\theta {\rm d}\phi
\left(2 \sin \tfrac{\phi}{2} \right)^{2b\ep} 
\left(2 \cos \tfrac{\phi}{2} \right)^{2a\ep-2b\ep} . 
\label{basis_III}
\end{equation}
Any order of the $\ep$-expansion of Eq.~(\ref{basis_III})
can be expressed in terms of 
the generalized ``log-sine-cosine'' function $\Lsc{i,j}{\theta}$
whose properties are listed in Appendix~B of Ref.~\cite{DK01},
\begin{equation}
\label{Lsc}
\Lsc{i,j}{\theta}=-\int\limits_0^{\theta}
{\rm d}\phi\; \ln^{i-1}\Bigl|2\sin\tfrac{\phi}{2}\Bigr| \;
\ln^{j-1}\Bigl|2\cos\tfrac{\phi}{2}\Bigr| \; .
\end{equation}
Up to the level $i+j=5$, only one independent function
appears, $\Lsc{2,3}{\theta}$, that cannot be expressed
in terms of ordinary log-sine integrals. 
Using the definition~(\ref{Lsc}), one can obtain
the following representation:
\begin{equation}
\Lsc{2,3}{\theta} = \tfrac{1}{12}\Ls{4}{2\theta}
- \tfrac{1}{3}\Ls{4}{\theta}
- \tfrac{1}{6} \int\limits_0^{\theta} {\rm d}\phi\; 
\ln^3\left( \tan\tfrac{\phi}{2} \right) \; .
\label{Lsc23_via_int}
\end{equation}
The integral occurring in Eq.~(\ref{Lsc23_via_int}) can be expressed 
in terms of the inverse tangent integrals (see in Ref.~\cite{Lewin}),
\begin{equation}
\label{Ti_N}
\Ti{N}{z} 
= {\rm Im}\left[ \Li{N}{{\rm i}z}\right]
= \frac{1}{2  {\rm i} } 
\Bigl[\Li{N}{ {\rm i}z} - \Li{N}{- {\rm i}z}\Bigr] \;,
\qquad
\Ti{N}{z} = \int\limits_0^z \frac{{\rm d}x}{x}\; \Ti{N-1}{x} \; .
\end{equation}
For $\Ti{N}{z}$, the following integral representation
(see, e.g., in Ref.~\cite{DT2}) is useful:
\begin{equation}
\Ti{N}{z} 
= \frac{(-1)^{N-1}z}{(N-1)!}
\int\limits_0^1 {\rm d}\xi \; \frac{\ln^{N-1}\xi}{1+z^2 \xi^2} \; .
\end{equation}
In this way, we arrive at the following representation 
of $\Lsc{2,3}{\theta}$ in terms of the inverse tangent integrals
(\ref{Ti_N}):
\begin{eqnarray}
\label{Lsc-Ti}
\Lsc{2,3}{\theta} &=& \tfrac{1}{12}\Ls{4}{2\theta}
- \tfrac{1}{3}\Ls{4}{\theta}
+ 2 \Ti{4}{\tan\tfrac{\theta}{2}}
- 2 \ln\left( \tan\tfrac{\theta}{2} \right)\;
 \Ti{3}{\tan\tfrac{\theta}{2}}
\nonumber \\ &&
+ \ln^2\left( \tan\tfrac{\theta}{2} \right)\;
 \Ti{2}{\tan\tfrac{\theta}{2}}
- \tfrac{1}{6} \theta \ln^3\left( \tan\tfrac{\theta}{2} \right) \; .
\end{eqnarray}
Note that $\Lsc{2,3}{\pi-\theta}$ can be reduced 
to $\Lsc{2,3}{\theta}$ by using 
Eq.~(A.27) of Ref.~\cite{DK01}.

Using Eqs.~(\ref{basis_II})--(\ref{basis_III}), we obtain 
the following results for the weight-{\bf 2} and weight-{\bf 3}
sums, in addition to Eq.~(\ref{S2}): 
%%%%%%%%%%%%%%%%%%%%%%%%%%%%%%%%%%%%%%%%%%%%%%%%%%%%%%%%%%%%%%%%%%%%%%%%%%%%%%%%%%%%%%%%%%%%%%%%%%%%%%%%%%%
\begin{eqnarray}
\sum_{j=1}^\infty \frac{1}{\left( 2j\atop j\right)} \frac{u^j}{j} S_1 
&=&  2 \tan\tfrac{\theta}{2}
\Bigl[\Ls{2}{\pi- \theta}  - \theta L_\theta \Bigr] \;,
\label{S1a}
\\  
\sum_{j=1}^\infty \frac{1}{\left( 2j\atop j\right)} \frac{u^j}{j} \bar{S}_1
&=&  
\tan\tfrac{\theta}{2}
\Bigl[
2 \Ls{2}{\pi- \theta}
+  \Ls{2}{\theta}
+  \theta l_\theta
- 2 \theta L_\theta
\Bigr] \; ,
\label{S1bar}
\\  
\sum_{j=1}^\infty \frac{1}{\left( 2j\atop j\right)}  \frac{u^j}{j} S_1^2 
&=&  4 \tan\tfrac{\theta}{2}
\Bigl[
\Ls{3}{\pi\!-\!\theta} - \Ls{3}{\pi}
- 2 \Ls{2}{\pi\!-\!\theta} L_\theta
+ \theta L^2_\theta
+ \tfrac{1}{24} \theta^3 
\Bigr] , 
\hspace*{7mm}
\label{S1S1}
\\ 
\sum_{j=1}^\infty \frac{1}{\left( 2j\atop j\right)}
\frac{u^j}{j} S_1 \bar{S}_1
&=&
\tan\tfrac{\theta}{2}  
\Bigl\{
  5  \left[ \Ls{3}{\pi\!-\!\theta} - \Ls{3}{\pi} \right]
-    \Ls{3}{\theta}
+ \tfrac{1}{2} \Ls{3}{2\theta}
- 2 \Ls{2}{\theta} L_\theta
\nonumber \\ && 
+ 2  \Ls{2}{\pi\!-\!\theta} l_\theta
- 8  \Ls{2}{\pi\!-\!\theta} L_\theta
- 2 \theta l_\theta  L_\theta 
+ 4 \theta  L_\theta^2 
+ \tfrac{1}{12} \theta^3
\Bigr\} , \hspace*{7mm}
\label{S1S1bar}
\\  
\!\!
\sum_{j=1}^\infty \frac{1}{\left( 2j\atop j\right)}
\frac{u^j}{j} 
\left( \bar{S}_2 + \bar{S}_1^2 \right) \!\!
&=&
\tan\tfrac{\theta}{2}  
\Bigl\{
  6  \left[ \Ls{3}{\pi-\theta} - \Ls{3}{\pi} \right]
- 3  \Ls{3}{\theta}
+ 4  \Ls{2}{\pi-\theta} l_\theta
\nonumber \\ &&
- 8  \Ls{2}{\pi-\theta} L_\theta
+ 2  \Ls{2}{\theta} l_\theta
+ \Ls{3}{2\theta}
- 4 \Ls{2}{\theta} L_\theta
\nonumber \\ && 
+ \theta l^2_\theta
- 4 \theta  L_\theta l_\theta
+ 4 \theta  L^2_\theta 
+ \tfrac{1}{12} \theta^3
\Bigr\} \; ,
\label{S2S1bar}
\end{eqnarray}
where $\Ls{3}{\pi}=-\tfrac{1}{2}\pi\zeta_2$.
In fact, all $\Ls{j}{\pi}$ are expressible 
in terms of the $\zeta$-function~\cite{Lewin}.

For the weight {\bf 4}, the results for some combinations
of sums involved in~(\ref{combinations}) can also be expressed
in terms of log-sine functions, using 
Eqs.~(\ref{basis_II})--(\ref{basis_III}),
\begin{eqnarray}
&& \hspace*{-10mm} 
\sum_{j=1}^\infty \frac{1}{\left( 2j\atop j\right)} \frac{u^j}{j} {\cal C}_0
= 
8 \tan\tfrac{\theta}{2}
\Bigl\{
\Ls{4}{\pi-\theta} \!-\! \Ls{4}{\pi} 
\nonumber \\ && \hspace{15mm} 
\!-\! 3 \Bigl[ \Ls{3}{\pi \!-\! \theta} \!-\! \Ls{3}{\pi} \Bigr] L_\theta
\!+\! 3 \Ls{2}{\pi \!-\! \theta} L^2_\theta
\!-\! \theta L^3_\theta
\Bigr\} \; ,
\label{combination:1}
\\ && \hspace*{-10mm}
%%%%%%%%%%%%%%%%%%%%%%%%%%%%%%%%%%%%%%%%%%%%%%%%%%%%%%%%%%%%%%%%%%%%%%%%
\sum_{j=1}^\infty \frac{1}{\left( 2j\atop j\right)} \frac{u^j}{j} {\cal C}_1 
= 
\tan\tfrac{\theta}{2}  
\Bigl\{
\tfrac{2}{3} \Ls{4}{\theta}
- \tfrac{1}{3} \Ls{4}{ 2\theta}
- \tfrac{14}{3} \left[ \Ls{4}{ \pi - \theta} - \Ls{4}{ \pi} \right]
\nonumber \\ && 
+ 14 \left[ \Ls{3}{ \pi - \theta} - \Ls{3}{\pi} \right] L_\theta
- 2 \Ls{3}{\theta} l_\theta
+ \Ls{3}{2 \theta} \left[ L_\theta + l_\theta \right]
- 14 \Ls{2}{\pi-\theta} L_\theta^2
\nonumber \\ && 
+ 2 \Ls{2}{\pi-\theta} l_\theta^2
- 2 \Ls{2}{2 \theta} L_\theta l_\theta   
- \Ls{2}{2 \theta} L_\theta^2
- 2 \theta L_\theta l_\theta^2
- 2 \theta L_\theta^2 l_\theta
+ 4 \theta L_\theta^3 
\Bigr\} \;, 
\label{combination:C1}
\\ && \hspace*{-10mm}
%%%%%%%%%%%%%%%%%%%%%%%%%%%%%%%%%%%%%%%%%%%%%%%%%%%%%%%%%%%%%%%%%%%%%%%%%%%%%%%
\sum_{j=1}^\infty \frac{1}{\left( 2j\atop j\right)} \frac{u^j}{j} {\cal C}_2 
= 
\tan\tfrac{\theta}{2}  
\Bigl\{
\tfrac{1}{3} \Ls{4}{2 \theta}
- \Ls{4}{\theta}
+ 4 \left[ \Ls{4}{\pi - \theta} - \Ls{4}{\pi}  \right] 
- \Ls{3}{2 \theta}\left( l_\theta + L_\theta  \right) 
\nonumber \\ &&
- 12 \left[\Ls{3}{\pi-\theta} - \Ls{3}{\pi} \right] L_\theta
+ 3 \Ls{3}{\theta} l_\theta 
+ 12 \Ls{2}{\pi-\theta} L_\theta^2  
- 3 \Ls{2}{\pi-\theta} l_\theta^2
\nonumber \\ &&
- \tfrac{1}{2} \Ls{2}{2\theta} l_\theta^2 
+ \Ls{2}{2\theta} L_\theta^2 
+ 2 \Ls{2}{2\theta} l_\theta L_\theta 
+ 2 \theta  l_\theta^2 L_\theta
+ 2 \theta  l_\theta L_\theta^2
- \tfrac{1}{3} \theta  l_\theta^3
- \tfrac{10}{3} \theta L_\theta^3 
\Bigr\} \; ,
\label{combination:C2}
\end{eqnarray}
%%%%%%%%%%%%%%%%%%%%%%%%%%%%%%%%%%%%%%%%%%%%%%%%%%%%%%%%%%%%%%%%%%%%%%%%%%%%%%%
where $\Ls{4}{\pi}=\tfrac{3}{2}\pi\zeta_3$
and the combinations of the harmonic sums $C_j$ are defined in (\ref{combinations2}).
Finally, using (\ref{basis_III}) the result for the sum involving $C_3$ 
can be expressed in terms of the ${\mbox{Lsc}}$-function (\ref{Lsc-Ti}),
\begin{eqnarray}
\label{last4}
&& \hspace*{-10mm}
\sum_{j=1}^\infty \frac{1}{\left( 2j\atop j\right)} \frac{u^j}{j} {\cal C}_3
= 
- 4\tan\tfrac{\theta}{2}  
\Bigl\{ 
  2 \Lsc{2,3}{\theta}
+ 2  \left[ \Ls{4}{\pi-\theta} - \Ls{4}{\pi} \right]
\nonumber \\ && 
+ 2 \left[ \Ls{3}{\pi-\theta} - \Ls{3}{\pi} \right] l_\theta
- 8 \left[ \Ls{3}{\pi-\theta} - \Ls{3}{\pi} \right] L_\theta
- \Ls{3}{2 \theta} L_\theta 
+ 2 \Ls{3}{\theta} L_\theta 
\nonumber \\ && 
+ 8 \Ls{2}{\pi-\theta} L_\theta^2 
- 4 \Ls{2}{\pi-\theta} L_\theta  l_\theta   
+ \Ls{2}{2 \theta} L_\theta^2
- 2 \theta  L_\theta^3
+ 2 \theta   L_\theta^2 l_\theta
\Bigr\}  \; .
\end{eqnarray}

Some of the results (\ref{S1a})--(\ref{last4}) can be written 
in a slightly different form by means of relations~\cite{Lewin}
$$
\Ls{2}{\theta} = \Cl{2}{\theta} \; , \quad 
\Cl{2n}{\pi+\theta} = - \Cl{2n}{\pi-\theta} \; , \quad
\Cl{2n+1}{\pi+\theta} = \Cl{2n+1}{\pi-\theta} \; . 
$$

However, in this way one cannot obtain
results for the two remaining combinations 
in~(\ref{combinations}),
involving $A_1^2-A_2=2a_1a_2$,
since in Eq.~(\ref{basis_III})
we always have $a_1a_2=0$. An interesting relation
between these two functions is obtained in Appendix~B. 
It should also be noted that for higher hypergeometric
functions other combinations of sums may arise
at this level (see below).

%%%%%%%%%%%%%%%%%%%%%%%%%%%%%%%%%%%%%%%%%%%%%%%%%%%%%%%%%%%%%%%%%%%%%%%%%
\subsection{Expansion of higher functions}

Let us consider 
the hypergeometric function of the following type:
\begin{equation}
\hspace*{-3mm}
_{P+1}F_P\left(\begin{array}{c|}
\tfrac{3}{2} + b_1 \ep,\ldots, \tfrac{3}{2} + b_{J-1} \ep, \; 
1+a_1\ep, \ldots, 1+a_K \ep, \; 2+d_1 \ep \ldots, 2+d_L \ep \\
\tfrac{3}{2} + f_1 \ep, \ldots, \tfrac{3}{2} + f_J \ep, \; 
1+e_1\ep, \ldots, 1+e_{R}\ep, \; 
2+c_1\ep, \ldots, 2+c_{K+L-R-2} \ep
\end{array} ~z \right),
\label{PFQ}
\end{equation}
where $P=K+L+J-2$.
Using the representation (\ref{pochhamer})
its $\ep$-expansion can be reduced to inverse binomial sums~(\ref{binsum}).
The original hypergeometric function (\ref{PFQ}) can be written as
(see details in Appendix~B of Ref.~\cite{DK01})
\begin{eqnarray}
&& \hspace*{-10mm}
_{P+1}F_P\left(\begin{array}{c|}
\{ \tfrac{3}{2} +b_i\ep\}^{J-1}, \; 
\{ 1+a_i\ep\}^K, \; \{ 2+d_i\ep\}^L  \\
\{ \tfrac{3}{2} + f_i\ep\}^J, \; 
\{ 1+e_i\ep \}^R,
\{ 2+c_i\ep \}^{K+L-R-2} 
\end{array} ~z \right)
\nonumber \\ && \hspace*{20mm}
= \frac{2}{u}
\frac{
\Pi_{s=1}^{K+L-R-2} (1+c_s\ep)  
\Pi_{k=1}^{J} (1+ 2 f_k\ep)}
{\Pi_{i=1}^{L} (1+d_i\ep)
\Pi_{r=1}^{J-1} (1 + 2 b_r\ep)}
\sum_{j=1}^\infty \frac{1}{\left( 2j \atop j \right) }  
\frac{u^j}{j^{K-R-1}}  
\Delta \;, 
\label{PFQ2}
\end{eqnarray}
where $u=4z$,  
\begin{eqnarray}
\Delta &=& 
\exp \left[ \sum_{k=1}^{\infty} \frac{(-\ep)^k}{k}  
\left( S_k T_k + 2^k U_k \bar{S}_k + W_k j^{-k} \right) 
\right]
\nonumber \\
&=&  
1 
- \ep \left( 
W_1 j^{-1} 
+  T_1 S_1 
+ 2 U_1 \bar{S}_1
\right)
+  \ep^2 \Bigl[ 
\tfrac{1}{2} j^{-2}\left( W_2 + W_1^2 \right) 
+ W_1 j^{-1} 
 \left( T_1 S_1 
+ 2 U_1 \bar{S}_1  \right)
\nonumber \\ &&
+  2 T_1 U_1 S_1 \bar{S}_1   
+ \tfrac{1}{2} T_2  S_2 
+ \tfrac{1}{2} T_1^2 S_1^2
+ 2 \left( U_2 \bar{S}_2 
+ U_1^2 \bar{S}_1^2 \right)
\Bigr]
\nonumber \\ && 
-  \ep^3 \Bigl\{ 
\tfrac{1}{6} j^{-3} \left(2 W_3 + 3 W_1 W_2 + W_1^3 \right)
+ \tfrac{1}{2} j^{-2} \left( W_2 + W_1^2\right)
\left( T_1 S_1 + 2 U_1 \bar{S}_1  \right) 
\nonumber \\ && 
+ \tfrac{1}{2} W_1 j^{-1} \left[   
T_2 S_2   
+ T_1^2 S_1^2   
+ 4 T_1 U_1 S_1 \bar{S}_1  
+ 4 \left( U_2 \bar{S}_2 + U_1^2 \bar{S}_1^2 \right)
                \right]
\nonumber \\ && 
+ \tfrac{1}{6} T_1^3 S_1^3 
+ T_1^2 U_1 S_1^2 \bar{S}_1  
+ \tfrac{1}{2} T_1 T_2 S_1 S_2 
+ T_2 U_1 S_2 \bar{S}_1 
+ \tfrac{1}{3} T_3 S_3  
\nonumber \\ && 
+ 2 T_1 S_1 \left( U_2 \bar{S}_2 + U_1^2 \bar{S}_1^2 \right)
+ \tfrac{4}{3}
\left(
  U_1^3 \bar{S}_1^3  
+ 3 U_1 U_2 \bar{S}_1 \bar{S}_2 
+ 2 U_3 \bar{S}_3 
\right)
\Bigr\}
+ {\cal O}(\ep^4)\; , 
\label{expansion}
\end{eqnarray}
and we introduced the constants 
\begin{eqnarray*}
&& \hspace*{-8mm}
% A_k \equiv \sum_{i=1}^{K} a_i^k, \quad 
% B_k \equiv \sum_{i=1}^{J-1} b_i^k, \quad 
% C_k \equiv \sum_{i=1}^{K+L-R-2} c_i^k, \quad 
% D_k \equiv \sum_{i=1}^{L} d_i^k, \quad 
% E_k \equiv \sum_{i=1}^{R} e_i^k, \quad
% F_k \equiv \sum_{i=1}^{J} f_i^k, 
A_k \equiv \sum a_i^k, \quad 
B_k \equiv \sum b_i^k, \quad 
C_k \equiv \sum c_i^k, \quad 
D_k \equiv \sum d_i^k, \quad 
E_k \equiv \sum e_i^k, \quad
F_k \equiv \sum f_i^k, 
\nonumber \\
&& \hspace*{-8mm}
T_k \equiv B_k + C_k + E_k - A_k - D_k - F_k, \quad
U_k \equiv F_k - B_k, \quad
W_k \equiv C_k - D_k \; , 
\end{eqnarray*}
where the summations extend over all possible values of the parameters
in Eq.~(\ref{PFQ}).
We can see that for the general values of the parameters 
of the $_{P+1}F_P$ function (\ref{PFQ}) we need all the
occurring sums separately. However,
for many applications, it is sufficient to consider 
the case $J=1$ only. In this case, $B_k=0$ and $U_k=F_k=f_1^k$.
Substituting this into Eq.~(\ref{expansion}), we see 
that $\bar{S}_2$ and $\bar{S}_3$ would only appear 
in combinations $\left( \bar{S}_2 + \bar{S}_1^2 \right)$
and $\left( \bar{S}_1^3 + 3 \bar{S}_1 \bar{S}_2 + 2 \bar{S}_3 \right)$.
Furthermore, using the notations~(\ref{combinations2}),
Eq.~(\ref{expansion}) in the case $J=1$ can be presented as
\begin{eqnarray}
\Delta\Bigr|_{J=1} &=& 
1 
- \ep \left( 
W_1 j^{-1} 
+  T_1 S_1 
+ 2 f_1 \bar{S}_1
\right)
+  \ep^2 \Bigl[ 
\tfrac{1}{2} j^{-2}\left( W_2 + W_1^2 \right) 
+ W_1 j^{-1} 
 \left( T_1 S_1 
+ 2 f_1 \bar{S}_1  \right)
\nonumber \\ &&
+  2 f_1 T_1 S_1 \bar{S}_1   
+ \tfrac{1}{2} T_2  S_2 
+ \tfrac{1}{2} T_1^2 S_1^2
+ 2 f_1^2 \left( \bar{S}_2 + \bar{S}_1^2 \right)
\Bigr]
\nonumber \\ && 
-  \ep^3 \Bigl\{ 
\tfrac{1}{6} j^{-3} \left(2 W_3 + 3 W_1 W_2 + W_1^3 \right)
+ \tfrac{1}{2} j^{-2} \left( W_2 + W_1^2\right)
\left( T_1 S_1 + 2 f_1 \bar{S}_1  \right) 
\nonumber \\ && 
+ \tfrac{1}{2} W_1 j^{-1} \left[   
T_2 S_2   
+ T_1^2 S_1^2   
+ 4 f_1 T_1 S_1 \bar{S}_1  
+ 4 f_1^2 \left( \bar{S}_2 + \bar{S}_1^2 \right)
                \right]
\nonumber \\ &&
+ \tfrac{1}{6} T_1 \left( T_1^2 + 3 f_1 T_1 + 3 f_1^2 \right) S_1^3
+ \tfrac{1}{2} T_1 \left( T_2 - f_1 T_1 -f_1^2 \right) S_1 S_2
+ \tfrac{1}{3} T_3 S_3
\nonumber \\ &&
+ f_1 \left( T_2 + T_1^2 +2 f_1 T_1 + 2 f_1^2 \right) S_2 \bar{S}_1
+ 2 f_1^2 T_1 {\cal C}_1 
- 4 f_1^3 {\cal C}_2 
\nonumber \\ &&
- \tfrac{1}{2} f_1 (T_1 + 2 f_1) (T_1 + 3 f_1) {\cal C}_3
\Bigr\}
+ {\cal O}(\ep^4)\; .
\label{expansion2}
\end{eqnarray}

Eq.~(\ref{expansion}) also shows that we may need 
the sums with higher values of $c$ (the power of $1/j$), 
which would come from
$1/j^{K-R-1}$ in Eq.~(\ref{PFQ2}) and extra powers of $1/j$
in Eq.~(\ref{expansion}). Note that in the case $K-R=2$,
when we have the same number of $c_i$ and $d_i$,
we have $1/j$ factor in the sum in Eq.~(\ref{PFQ2}). 
Moreover, when the parameters $\{2+c_i\}$ and $\{2+d_i\}$
are at all missing, we get no extra factors of $1/j$
in Eq.~(\ref{expansion}), for all $W_k$ would vanish.

%%%%%%%%%%%%%%%%%%%%%%%%%%%%%%%%%%%%%%%%%%%%%%%%%%%%%%%%%%%%%%%%%%%%%%%%%
For the sums of the type~(\ref{binsum}) with an arbitrary integer power 
$c\geq 2$  and  $0 \leq u \leq 4$, the following one-fold integral 
representation \cite{BBK} is useful:
\begin{eqnarray}
\sum_{j=1}^\infty  \frac{u^j}{j^c} \frac{f(j)}{\left( 2j\atop j\right)} =
\frac{1}{(c-2)!} \int\limits_0^{\theta}
{\rm d}\phi \frac{\cos \tfrac{\phi}{2}}{\sin\tfrac{\phi}{2}}
\Bigl[\ln{u} - 2\ln \left(2 \sin \tfrac{\phi}{2} \right)\Bigr]^{c-2} 
\sum_{j=1}^\infty \frac{\left( 4 \sin^2\tfrac{\phi}{2} \right)^{j}}{j} 
\frac{f(j)}{\left( 2j\atop j\right)}  \; ,
\label{higher_c}
\end{eqnarray}
where $f(j)$ stands for an arbitrary combination of the harmonic sums.
In this manner, the original sum of the type~(\ref{binsum}) 
with an arbitrary positive integer parameter $c\geq 2$ is reduced
to a one-fold integral representation containing
a sum with $c=1$ in the integrand. 
Using the representation~(\ref{higher_c}) we can generalize our 
results~(\ref{S2}), (\ref{S4}), (\ref{S1a})--(\ref{last4})
to the case of an arbitrary integer $c>1$.
However, only for special types of sums these results are 
expressible in terms of generalized log-sine functions~(\ref{def_Ls}):
\begin{eqnarray}
&& \hspace*{-7mm}
%%%%%%%%%%%%%%%%%%%%%%%%%%%%%%%%%%%%%%%%%%%%%%%%%%%%%%%%%%%%%%%%%%%%%%%%%%
\sum_{j=1}^\infty  \frac{u^j}{j^c} \frac{1}{\left( 2j\atop j\right)} S_2 =
- \tfrac{1}{6} \sum_{i=0}^{c-2} \frac{(-2)^i}{i!(c-2-i)!} 
\left(\ln u \right)^{c-2-i} \LS{i+4}{3}{\theta} \;, 
\label{S2G}
\\ && \hspace*{-7mm}
%%%%%%%%%%%%%%%%%%%%%%%%%%%%%%%%%%%%%%%%%%%%%%%%%%%%%%%%%%%%%%%%%%%%%%%%%%%%
\sum_{j=1}^\infty  \frac{u^j}{j^c} \frac{1}{\left( 2j\atop j\right)}
\left( S_2^2-S_4 \right) =
- \tfrac{1}{60} \sum_{i=0}^{c-2} \frac{(-2)^i}{i!(c-2-i)!} 
\left(\ln u \right)^{c-2-i} \LS{i+6}{5}{\theta} \;, 
\label{S4G}
%%%%%%%%%%%%%%%%%%%%%%%%%%%%%%%%%%%%%%%%%%%%%%%%%%%%%%%%%%%%%%%%%%%%%%%%%%%%
\end{eqnarray}
where $c \geq 2$, $0 \leq u \leq 4$, and the angle $\theta$ is
defined in Eq.~(\ref{def_theta}). 

Further results can be extracted 
from integral representation~(\ref{higher_c}) by using 
the integration rules for Clausen's function~\cite{Lewin} and 
the following relation:
%%%%%%%%%%%%%%%%%%%%%%%%%%%%%%%%%%%%%%%%%%%%%%%%%%%%%%%%%%%%
\[
k \int\limits_0^\theta {\rm d}\phi\;
{\phi}^{k-1} \LS{j}{i}{m \phi}
=  \theta^k \LS{j}{i}{m\theta} - \frac{1}{m^k} \LS{j+k}{i+k}{m \theta}
\; . 
\]
%%%%%%%%%%%%%%%%%%%%%%%%%%%%%%%%%%%%%%%%%%%%%%%%%%%%%%%%%%%%%%%%%%%%%%%%%%%%%%%%%%%%%%%%%
In this way, we obtain 
\begin{eqnarray}
\label{S1c2}
\sum_{j=1}^\infty \frac{1}{\left( 2j \atop j\right) } 
\frac{u^j}{j^2} S_1   &=&  
  4 \Cl{3}{\pi - \theta} - 2 \theta \Cl{2}{\pi - \theta} + 3 \zeta_3 , 
\\ 
\sum_{j=1}^\infty \frac{1}{\left( 2j \atop j \right) } 
\frac{u^j}{j^2} \bar{S}_1
 &=&  -2 \Cl{3}{\theta} + 4  \Cl{3}{\pi-\theta}
- 2 \theta \Cl{2}{\pi- \theta}
-   \theta \Cl{2}{\theta}
+ 5 \zeta_3 \; , \hspace*{10mm}
\label{S1barc2}
\\ 
%%%%%%%%%%%%%%%%%%%%%%%%%%%%%%%%%%%%%%%%%%%%%%%%%%%%%%%%%%%%%%%%%%%%%%%%%%%%%%%%%%%%%%%%%
\sum_{j=1}^\infty \frac{1}{\left( 2j\atop j\right)} 
\frac{u^j}{j^2} S_1^2 
&=&  
4 \theta \left[ \Ls{3}{\pi-\theta} - \Ls{3}{\pi} \right]
-  4\left[ \Ls{2}{\pi-\theta} \right]^2
+ \tfrac{1}{24} \theta^4 \; ,
\label{S1S1c2}
\\
%%%%%%%%%%%%%%%%%%%%%%%%%%%%%%%%%%%%%%%%%%%%%%%%%%%%%%%%%%%%%%%%%%%%%%%%%%%%%%%%%%%%%%%%%%%%%%%%%  
\sum_{j=1}^\infty \frac{1}{\left( 2j \atop j\right) } 
      \frac{u^j}{j^2} S_1 \bar{S}_1
&=& 
 5 \theta \left[\Ls{3}{\pi-\theta} - \Ls{3}{\pi} \right]
- \theta \Ls{3}{\theta} 
+ \tfrac{1}{2} \theta \Ls{3}{2 \theta}
\nonumber \\ && 
- 4 \left[\Ls{2}{\pi-\theta} \right]^2
+ \tfrac{1}{48} \theta^4 
- 2 \Ls{2}{\pi-\theta}  \Ls{2}{\theta} \; ,
\label{S1S1barc2} 
\\
%%%%%%%%%%%%%%%%%%%%%%%%%%%%%%%%%%%%%%%%%%%%%%%%%%%%%%%%%%%%%%%%%%%%%%%%%%%%%%%%%%%%%%%%%%%%%%%%%  
\sum_{j=1}^\infty \frac{1}{\left( 2j \atop j\right) } 
\frac{u^j}{j^2} 
\left( \bar{S}_1^2 + \bar{S}_2 \right)
&=& 
  6 \theta \left[\Ls{3}{\pi-\theta} - \Ls{3}{\pi} \right]
- 3 \theta \Ls{3}{\theta}
+  \theta \Ls{3}{2 \theta}
\nonumber \\ &&
- 4 \left[\Ls{2}{\pi-\theta} \right]^2
- \left[\Ls{2}{\theta} \right]^2
+ \tfrac{1}{48} \theta^4
- 4 \Ls{2}{\pi-\theta}  \Ls{2}{\theta} \; .
\label{S2bar_2}
\end{eqnarray}

However, other sums are not expressible in terms of 
the generalized log-sine functions.
One needs to introduce a new function\footnote{Alternatively,
instead of $\Phi(\theta)$ one may introduce the generalized
Glashier function $\Gl{4}{\theta;1}$, see Eq.~(\ref{Gl4(theta,1)}) 
in Appendix~A.},
\begin{equation}
\label{def_Phi}
\Phi(\theta) \equiv
\int\limits_0^\theta {\rm d}\phi\; 
\Ls{2}{\phi} \ln \left(2 \cos \tfrac{\phi}{2} \right) \; , 
\end{equation}
which obeys the following symmetry property:
\begin{equation}
\label{sym_Phi}
\Phi(\theta) + \Phi(\pi-\theta) 
= \Phi(\pi) + \Ls{2}{\pi-\theta} \Ls{2}{\theta} \;, 
\end{equation}
where 
\begin{eqnarray}
\hspace{-10mm}
\Phi(\pi) & = &  
\tfrac{1}{6} \ln^4 2 - \zeta_2 \ln^2 2 + \tfrac{7}{2} \zeta_3 \ln 2 
- \tfrac{53}{16} \zeta_4 + 4 \Li{4}{\tfrac{1}{2}} 
= 0.64909 \ldots 
\; . 
\end{eqnarray}

The following sums of the weight {\bf 4} are expressible 
in terms of the function $\Phi(\theta)$:
\begin{eqnarray}
&& \hspace*{-10mm}
\sum_{j=1}^\infty \frac{1}{\left( 2j \atop j\right) } 
\frac{u^j}{j^3} S_1   
=  
  2  \left[  \LS{4}{1}{\pi-\theta} - \LS{4}{1}{\pi} \right]  
- \tfrac{1}{2} \LS{4}{1}{2 \theta}
+ 2 \LS{4}{1}{\theta}
- 4 \theta \Ls{2}{\pi-\theta} l_\theta
\nonumber \\ && 
- 2 \pi \left[\Ls{3}{\pi-\theta} - \Ls{3}{\pi} \right]
+ 8 \left[ \Cl{3}{\pi-\theta} - \Cl{3}{\pi} \right] l_\theta
+ 4 \Phi(\pi-\theta) - 4 \Phi(\pi) \;, \hspace*{6mm}
\label{S1_3}
%%%%%%%%%%%%%%%%%%%%%%%%%%%%%%%%%%%%%%%%%%%%%%%%%%%%%%%%%%%%%%%%%%%%%%%%%%%%%%%%%%%%%%%%%%%%%%%%%
\\ && \hspace*{-10mm} 
\sum_{j=1}^\infty \frac{1}{\left( 2j \atop j\right) } 
\frac{u^j}{j^3} \bar{S}_1   
=  
  2 \left[ \LS{4}{1}{\pi-\theta} - \LS{4}{1}{\pi} \right]
- \tfrac{1}{2} \LS{4}{1}{2 \theta}
+ 4 \LS{4}{1}{\theta}
+ \left[ \Ls{2}{\theta} \right]^2
\nonumber \\ && 
- 2 \pi \left[\Ls{3}{\pi-\theta} - \Ls{3}{\pi} \right]
+ 8 \left[ \Cl{3}{\pi-\theta} - \Cl{3}{\pi} \right] l_\theta
- 4 \left[ \Cl{3}{\theta} - \zeta_3 \right] l_\theta
\nonumber \\ && 
- 2 \theta \Ls{2}{\theta} l_\theta
- 4 \theta \Ls{2}{\pi-\theta} l_\theta
+ 4 \Phi(\pi-\theta) - 4 \Phi(\pi) \;,
%%%%%%%%%%%%%%%%%%%%%%%%%%%%%%%%%%%%%%%%%%%%%%%%%%%%%%%%%%%%%%%%%%%%%%%%%%%%%%%%%%%%%%%%%%%%%%%%%
\label{S1bar_3}
\end{eqnarray}
where we have also used the following integral:
\begin{eqnarray}
\int\limits_0^\theta {\rm d}\phi\;
\phi \ln  \left(2 \cos \tfrac{\phi}{2} \right) 
     \ln  \left(2 \sin \tfrac{\phi}{2} \right)    
& =  & 
-\tfrac{1}{8} \LS{4}{1}{2\theta} + \tfrac{1}{2} \LS{4}{1}{\theta}
+ \tfrac{1}{2} \left[\LS{4}{1}{\pi - \theta} - \LS{4}{1}{\pi} \right]
\nonumber \\ && 
- \tfrac{1}{2}\pi \left[\Ls{3}{\pi - \theta} - \Ls{3}{\pi} \right] \; .
\nonumber 
\end{eqnarray}

Let us note that $\Phi(\theta)$ 
can be related to the real part of a certain
harmonic polylogarithm~\cite{RV00} of complex argument,
\begin{equation}
\label{PhiH}
\Phi(\theta) 
=  
\tfrac{1}{96} \theta^2 (2\pi-\theta)^2
- L_{\theta} \Cl{3}{\theta}
+ \zeta_3 \ln 2
-  \H_{-1,0,0,1}(1)   
+ \tfrac{1}{2}\Bigl[  
\H_{-1,0,0,1}( e^{{\rm i}\theta}) + \H_{-1,0,0,1}(e^{-{\rm i}\theta})   
\Bigr] ,
\end{equation}
where (for details, see Appendix~A) 
\begin{eqnarray}
\H_{-1,0,0,1}(y) &=& \int\limits_0^y {\rm d}x \; \frac{\Li{3}{x}}{1+x} ,
\label{h(-1001)}
\\
\H_{-1,0,0,1}(1) &=&  
- \tfrac{1}{12} \ln^4 2 + \tfrac{1}{2} \zeta_2 \ln^2 2
- \tfrac{3}{4} \zeta_3 \ln 2 + \tfrac{3}{2} \zeta_4 - 2 \Li{4}{\tfrac{1}{2}}
=  0.33955 \ldots \; .
\end{eqnarray}
The function $\H_{-1,0,0,1}(y)$ has a branch cut starting 
at the point $y=-1$ ($\theta=\pm\pi$),
which is subtracted by logarithmic terms in~(\ref{PhiH}), so that 
$\Phi(\theta)$ is a smooth function of variable $\theta$.

%%%%%%%%%%%%%%%%%%%%%%%%%%%%%%%%%%%%%%%%%%%%%%%%%%%%%%%%%%%%%%%%%%%%%%%%%
\subsection{Further results for the sums}

For further investigation of the sums of the type (\ref{binsum}),
a recursive approach appears to be useful. Some ideas applied below
are described in the book~\cite{generatingfunction}.
Let us rewrite Eq.~(\ref{binsum}) in the following form:
\begin{equation}
\Sigma^{i_1,\ldots,\;i_p;\; j_1,\ldots, j_q}_{a_1,\ldots,a_p;\; b_1,\ldots,b_q;\;c}(u)
\equiv \Sigma_{A;B;c}(u) = \sum_{j=1}^{\infty} u^j \eta_{A;B;c}(j) \; ,
\end{equation}
where $A\equiv \left( {}^{i_1,\ldots,i_p}_{a_1,\ldots,a_p}\right)$
and $B\equiv \left( {}^{j_1,\ldots,j_q}_{b_1,\ldots,b_q}\right)$
denote the collective sets of indices, whereas $\eta_{A;B;c}(j)$ is
the coefficient of $u^j$ in Eq.~(\ref{binsum}), i.e.\ the product
of binomial sums divided by $\left( 2j \atop j\right) j^c$.

The idea is to find a recurrence relation~\footnote{About application of the difference 
equations in the physical calculations we refer to \cite{difference}.}, with respect to~$j$, 
for the coefficients $\eta_{A;B;c}(j)$, and then
transform it into a differential equation
for the {\em generating} function $\Sigma_{A;B;c}(u)$.
In this way, the problem of summing the series would be reduced to 
solving a differential equation.
Using the explicit form of $\eta_{A;B;c}(j)$ given in Eq.~(\ref{binsum}),
the recurrence relation can be written in the following form:
\begin{equation}
2 (2j+1) (j+1)^{c-1}
\eta_{A;B;c}(j+1) - j^c \eta_{A;B;c}(j) = r_{A;B}(j) \; ,
\label{rec:relation}
\end{equation}
where the explicit form of the ``remainder'' $r_{A;B}(j)$ is given by
\begin{eqnarray}
{\textstyle \left( 2j \atop j\right)} \;
r_{A;B}(j) &=&
\prod\limits_{k=1}^p \left[ S_{a_k}(j-1) + j^{-a_k}\right]^{i_k}
\prod\limits_{l=1}^q \left[ S_{b_l}(2j-1) + (2j)^{-b_l}
 + (2j+1)^{-b_l} \right]^{j_l}
\nonumber \\ &&
- 
\prod\limits_{k=1}^p \left[ S_{a_k}(j-1) \right]^{i_k} 
\prod\limits_{l=1}^q \left[ S_{b_l}(2j-1) \right]^{j_l} \; .
\label{rAB}
\end{eqnarray}
In other words, it contains all contributions generated by
$j^{-a_k}$, $(2j)^{-b_l}$ and $(2j+1)^{-b_l}$ which appear
because of the shift of the index~$j$. 

Multiplying both sides of Eq.~(\ref{rec:relation}) by $u^j$,
summing from 1 to infinity, and using the fact that any extra power
of~$j$ corresponds to the derivative $u({\rm d}/{\rm d}u)$,
we arrive at the following differential equation for the
generating function $\Sigma_{A;B;c}(u)$: 
\begin{eqnarray}
\left(\frac{4}{u} -1 \right) \left( u \frac{{\rm d}}{{\rm d} u} \right)^c  
\Sigma_{A;B;c}(u)
- \frac{2}{u} \left( u \frac{{\rm d}}{{\rm d}u} \right)^{c-1} 
\Sigma_{A;B;c}(u)  =
2 \eta_{A;B;c}(1) + R_{A;B}(u) \; ,
\label{diff:I}
\end{eqnarray}
where $\eta_{A;B;c}(1)=\tfrac{1}{2}\delta_{p0}$ and 
$R_{A;B}(u)\equiv\sum_{j=1}^{\infty} u^j r_{A;B}(j)$.
Using Eq.~(\ref{rAB}) we obtain  
\begin{eqnarray}
R_{a_1;-}(u)  & = & 
\sum_{j=1}^\infty \frac{u^j}{\left( 2j \atop j\right) }  \frac{1}{j^{a_1}} \;, 
\label{R_a1}
\\ 
R_{-;b_1}(u)  & = & \sum_{j=1}^\infty   \frac{u^j}{\left( 2j \atop j\right) }  
\Biggl[
\frac{1}{(2j)^{b_1}} + \frac{1}{(2j+1)^{b_1}} 
\Biggr] \;, 
\label{R_b1}
\\ 
R_{a_1,a_2;-}(u) & = & \sum_{j=1}^\infty \frac{u^j}{\left( 2j \atop j\right) }  
\Biggl[
  \frac{S_{a_1}}{j^{a_2}} 
+ \frac{S_{a_2}}{j^{a_1}} 
+ \frac{1}{j^{a_1+a_2}} 
\Biggr] \;, 
\label{R_a1a2}
\\ 
R_{a_1;b_1}(u) & = & \sum_{j=1}^\infty \frac{u^j}{\left( 2j \atop j\right) }  
\Biggl[
\frac{S_{a_1}}{(2j)^{b_1}} 
\!+\! \frac{S_{a_1}}{(2j+1)^{b_1}} 
\!+\! \frac{\bar{S}_{b_1}}{j^{a_1}} 
\!+\! \frac{1}{j^{a_1}(2j)^{b_1}} 
\!+\! \frac{1}{j^{a_1}(2j+1)^{b_1}} 
\Biggr] \;, 
\label{R_a1b1}
\\ 
R_{-;b_1,b_2}(u) & = & \sum_{j=1}^\infty \frac{u^j}{\left( 2j \atop j\right) }  
\Biggl[
\frac{\bar{S}_{b_1}}{(2j)^{b_2}} 
\!+\! \frac{\bar{S}_{b_2}}{(2j)^{b_1}} 
\!+\! \frac{\bar{S}_{b_1}}{(2j+1)^{b_2}} 
\!+\! \frac{\bar{S}_{b_2}}{(2j+1)^{b_1}}
\!+\! \frac{1}{(2j)^{b_1+b_2}}  
\nonumber \\ && \hspace{20mm}
\!+\! \frac{1}{j^{b_1}(2j+1)^{b_2}} 
\!+\! \frac{1}{j^{b_2}(2j+1)^{b_1}} 
\!+\! \frac{1}{(2j+1)^{b_1+b_2}} 
\Biggr] \;, 
\label{R_b1b2}
\\ 
R_{a_1,a_2,a_3;-}(u) & = & \sum_{j=1}^\infty \frac{u^j}{\left( 2j \atop j\right) }  
\Biggl[
   \frac{S_{a_1}S_{a_2}}{j^{a_3}} 
\!+\!  \frac{S_{a_1}S_{a_3}}{j^{a_2}} 
\!+\!  \frac{S_{a_2}S_{a_3}}{j^{a_1}} 
\!+\!  \frac{S_{a_1}}{j^{a_2+a_3}} 
\!+\!  \frac{S_{a_2}}{j^{a_1+a_3}} 
\!+\!  \frac{S_{a_3}}{j^{a_1+a_2}} 
\!+\!  \frac{1}{j^{a_1+a_2+a_3}} 
\Biggr].
\nonumber 
\label{R_a1a2a3}
\\ 
\end{eqnarray}

In terms of the geometrical variable~(\ref{def_theta}),
\begin{equation}
\label{u_theta}
u \equiv u_{\theta} = 4 \sin^2 \tfrac{\theta}{2} \;,  \quad
\frac{4}{u} -1 = \cot^2\tfrac{\theta}{2} \; , \quad
u \frac{{\rm d}}{{\rm d}u} = \tan\tfrac{\theta}{2}
\frac{{\rm d}}{{\rm d} \theta} \; ,
\end{equation}
the differential equation (\ref{diff:I}) takes the following form: 
\begin{eqnarray}
\frac{1}{2 \sin^2 \tfrac{\theta}{2}}
\left( 2 \sin \tfrac{\theta}{2} \cos\tfrac{\theta}{2} 
\frac{{\rm d}}{{\rm d}\theta} - 1 \right)
\left( 
\tan\tfrac{\theta}{2}
\frac{{\rm d}}{{\rm d} \theta} \right)^{c-1}
 \Sigma_{A;B;c}(u_{\theta})  = 
\delta_{p0} + R_{A,B}(u_{\theta}) \; .
\label{diff:II}
\end{eqnarray}
Furthermore, Eq.~(\ref{diff:II}) can be represented as
\begin{equation}
\left( 
\tan\tfrac{\theta}{2}
\frac{{\rm d}}{{\rm d} \theta} \right)^{c-1}
\Sigma_{A;B;c}(u_{\theta}) = 
\tan\tfrac{\theta}{2}\;
\sigma_{A;B}(\theta) \; ,
\label{basisIA}
\end{equation}
where
\begin{equation}
\frac{{\rm d}}{{\rm d} \theta} \sigma_{A;B}(\theta) =  
\delta_{p0}
+ R_{A,B}(u_{\theta}) \; , \qquad
\sigma_{A;B}(\theta) = \delta_{p0}\theta 
+ \int\limits_0^{\theta} {\rm d}\phi\; R_{A;B}(u_{\phi}) \; ,
\label{basisII}
\end{equation}
with $u_{\phi}=4\sin^2\tfrac{\phi}{2}$.
In particular, for $c=1$ and $c=2$ we have, respectively,
\begin{eqnarray}
\Sigma_{A;B;1}(u_{\theta}) & = &
\tan\tfrac{\theta}{2} \;
\sigma_{A;B}(\theta) 
=
\tan\tfrac{\theta}{2}
\left[  \delta_{p0}\theta 
+ \int\limits_0^{\theta} {\rm d}\phi\; R_{A;B}(u_{\phi}) \right] \; , 
\label{uniq}
\\
\Sigma_{A;B;2}(u_{\theta}) & = & 
\int\limits_0^{\theta} {\rm d}\phi \sigma_{A;B}(\phi) 
= \theta  \sigma_{A;B}(\theta) - \tfrac{1}{2}\delta_{p0}\theta^2 
- \int\limits_0^{\theta}  \phi {\rm d}\phi\; R_{A,B}(u_{\phi})\; .
\end{eqnarray}

\noindent
Introducing $l_{\theta}\equiv \ln\left(2\sin\tfrac{\theta}{2}\right)
=\tfrac{1}{2}\ln{u_{\theta}}$,
Eq.~(\ref{basisIA}) can be rewritten as
\begin{equation}
\left( \frac{1}{2} \frac{{\rm d}}{{\rm d} l_{\theta}} \right)^{c-k}
\Sigma_{A;B;c}(u_{\theta})
= \Sigma_{A;B;k}(u_{\theta}) \; ,
\label{basisIB}
\end{equation}
which is nothing but the differential form of the relation~(\ref{higher_c}).
The iterative solution of Eq.~(\ref{basisIB}) is 
\begin{equation}
\Sigma_{A;B;c}(u_{\theta})
= -\sum_{i=1}^{k} 
\frac{(-2)^i}{i!}
l_{\theta}^i \; \Sigma_{A;B;c-i}(u_{\theta})
+ \frac{(-2)^k}{k!} \int\limits_0^{\theta} {\rm d}\phi \;
l_{\phi}^k \; \frac{ {\rm d} \Sigma_{A;B;c-k}(u_{\phi})}{{\rm d}\phi} \; ,
\label{solution:I}
\end{equation}
where $l_{\phi}\equiv \ln\left(2\sin\tfrac{\phi}{2}\right)
=\tfrac{1}{2}\ln{u_{\phi}}$.

In particular, this solution allows us to formulate and prove
the following important {\em statement}. 
If for some $k$ the derivative $\Sigma_{A;B;c-k}(u_{\theta})$ 
is expressible only in terms of the powers of $\theta$ and $l_{\theta}$,  
then the sum $\Sigma_{A;B;c}(u_{\theta})$ can be presented
in terms of the generalized log-sine functions. Moreover, 
according to a statement proven in Appendix~A.1 of  Ref.~\cite{DK01}, 
the analytic continuation of any generalized log-sine 
function $\LS{j}{k}{\theta}$ can be expressed in terms of Nielsen 
polylogarithms. 

%%%%%%% Misha, ne strirai poka etot comment, ya hotel podumat',
%%%%%%% ne nado li chto-nibud' v texte rasshirit'. A.
%
% Dokastelstvo prostoe:
% Esli
% d Sigma /d theta  \ sim theta^a \ln^b (2 sin( theta/2) )
% to integral itself is equal to Ls^(i)_j (theta)
% togda posleduyuhsee integrirovanie,
% Ls^(i)_j (theta) d ln (2 sin( theta/2) )
% = Ls^(i)_j (theta) ln (2 sin( theta/2) )  - Ls^(i)_{j+1}(theta)
% poslduyshie integrirovanie s  d ln (2 sin( theta/2) ) daet nam
% ( Ls^(i)_j (theta) ln (2 sin( theta/2) )  - Ls^(i)_{j+1}(theta) )
% d ln (2  sin( theta/2) )
% =  Ls^(i)_j (theta) 1/2*ln^2 (2 sin( theta/2) )  - 1/2 Ls^(i)_{j+2}(theta)
% -  Ls^(i)_{j+1}(theta)  ln (2 sin( theta/2) )  +  Ls^(i)_{j+2}(theta)
% i t.d.

Let us now explain how this general method works for specific sums 
of interest. Consider 
\[
\Sigma_{3;-;1}^{1;-}(u) =
\sum_{j=1}^\infty \frac{1}{\left( 2j \atop j\right) } 
\frac{u^j}{j} S_3 
= \tan\tfrac{\theta}{2}\; 
\int\limits_0^{\theta} {\rm d}\phi\; R_{3;-}(u_{\phi}) \; ,
\]  
where $R_{3;-}(u)$ is defined in Eq.~(\ref{R_a1}) and
the result can be extracted from Eq.~(\ref{KV}):
\begin{equation}
\sum_{j=1}^\infty \frac{1}{\left( 2j \atop j \right) } \frac{u^j}{j^3} = 
2 \left[\Cl{3}{\theta} + \theta \Cl{2}{\theta} - \zeta_3 \right]
+ \theta^2 l_{\theta} \; ,
\label{j^3}
\end{equation}
Integrating over $\phi$ we obtain
%%%%%%%%%%%%%%%%%%%%%%%%%%%%%%%%%%%%%%%%%%%%%%%%%%%%%%%%%%%%%%%%%%%%%%%%%%%%%%%%%%%5
\begin{equation}
\sum_{j=1}^\infty \frac{1}{\left( 2j \atop j\right) } 
\frac{u^j}{j} S_3   =  
\tan\tfrac{\theta}{2}  
\Bigl\{ 
  6 \Cl{4}{\theta}
- \theta^2 \Cl{2}{\theta}
- 4 \theta \Cl{3}{\theta}
- 2 \theta \zeta_3 
\Bigr\} \; .
\label{S3c1}
\end{equation}
For another sum, $\Sigma_{2,1;-;1}^{1,1;-}(u)$, we get
\[
\Sigma_{2,1;-;1}^{1,1;-}(u) =
\sum_{j=1}^\infty \frac{1}{\left( 2j \atop j\right) } 
\frac{u^j}{j} S_1 S_2 
= \tan\tfrac{\theta}{2}\; 
\int\limits_0^{\theta} {\rm d}\phi\; R_{2,1;-}(u_{\phi}) \; ,
\]
where $R_{2,1;-}(u)$ is defined in Eq.~(\ref{R_a1a2}).
Using Eqs.~(\ref{S1c2}), (\ref{S2}) and (\ref{j^3}) to calculate $R_{2,1;-}(u)$, 
and integrating over $\phi$, we get
%%%%%%%%%%%%%%%%%%%%%%%%%%%%%%%%%%%%%%%%%%%%%%%%%%%%%%%%%%%%%%%%%%%%%%%%%%%%%%%%%%%5
\begin{eqnarray}
\sum_{j=1}^\infty \frac{1}{\left( 2j \atop j\right) } 
\frac{u^j}{j} S_1 S_2   &=&  
\tan\tfrac{\theta}{2}  
\Bigl\{ 
  \theta^2 \Cl{2}{\pi-\theta}
- \theta^2 \Cl{2}{\theta}
- 4 \theta \Cl{3}{\pi-\theta}
- 4 \theta \Cl{3}{\theta}
\nonumber \\ && \hspace{10mm} 
- 8 \Cl{4}{\pi-\theta}
+ 6 \Cl{4}{\theta}
+ \theta \zeta_3 
- \tfrac{1}{3} \theta^3 L_{\theta}
\Bigr\} \; .
\label{S1S2c1}
\end{eqnarray}

For the sum $\Sigma_{3;-;1}^{1;-}(u)$, this approach is also 
applicable. In this case we have 
\[
\Sigma_{3;-;1}^{1;-}(u) =
\sum_{j=1}^\infty \frac{1}{\left( 2j \atop j\right) } 
\frac{u^j}{j} S_1^3
= \tan\tfrac{\theta}{2}\; 
\int\limits_0^{\theta} {\rm d}\phi\; R_{1,1,1;-}(u_{\phi}) \; ,
\]
with $R_{1,1,1;-}(u)$ defined in Eq.~(\ref{R_a1a2a3}).
Using Eqs.~(\ref{S1S1}), (\ref{S1c2}) and (\ref{j^3}) to calculate $R_{1,1,1;-}(u)$, 
and integrating over $\phi$, we get
%%%%%%%%%%%%%%%%%%%%%%%%%%%%%%%%%%%%%%%%%%%%%%%%%%%%%%%%%%%%%%%%%%%%%%%%%%%%%%%%%%%5
\begin{eqnarray}
\sum_{j=1}^\infty \frac{1}{\left( 2j \atop j\right) } 
\frac{u^j}{j} S_1^3   &=&  
\tan\tfrac{\theta}{2}  
\Bigl\{ 
  6 \Cl{4}{\theta}
- 24 \Cl{4}{\pi-\theta}
- 12 \theta \Cl{3}{\pi-\theta}
- 4 \theta \Cl{3}{\theta}
\nonumber \\ && 
- 24 L_\theta \left[ \Ls{3}{\pi-\theta} - \Ls{3}{\pi} \right]
+ 8 \left[ \Ls{4}{\pi-\theta} - \Ls{4}{\pi} \right]
- \theta^2 \Cl{2}{\theta}
\nonumber \\ && 
+ 3 \theta^2 \Cl{2}{\pi-\theta}
+ 24 \Cl{2}{\pi-\theta}  L^2_\theta
- \theta^3 L_\theta  
- 8 \theta  L^3_\theta
+ 7 \zeta_3 \theta 
\Bigr\} \; . 
\label{S1S1S1c1}
\end{eqnarray}
Combining Eqs.~(\ref{S3c1}), (\ref{S1S2c1}) and  (\ref{S1S1S1c1}), we  
successfully reproduce Eq~(\ref{combination:1}).

As a further example,
let us consider the sum $\Sigma_{2;1;1}^{1;1}(u)$. We obtain
\[
\Sigma_{2;1;1}^{1;1}(u) =
\sum_{j=1}^\infty \frac{1}{\left( 2j \atop j\right) } 
\frac{u^j}{j} S_2 \bar{S}_1
= \tan\tfrac{\theta}{2}\; 
\int\limits_0^{\theta} {\rm d}\phi\; R_{2;1}(u_{\phi}) \; ,
\]
where $R_{2;1}(u)$ corresponds to Eq.~(\ref{R_a1b1}),
with some of the contributing sums having $(2j+1)$ in the denominator.
Let us denote them as
\[
{\widetilde{\Sigma}}_{A;B;c;d}(u) \equiv
\sum_{j=1}^{\infty} \frac{u^j}{(2j+1)^d} \; \eta_{A;B;c}(j) \; .
\]
To calculate such sums, it is convenient to apply the differential
operator that lowers the value of $d$,
\begin{equation}
2 u^{1/2} \frac{{\rm d}}{{\rm d}u} 
\left[ u^{1/2}\; {\widetilde{\Sigma}}_{A;B;c;d}(u) \right]
\Leftrightarrow 
\frac{2}{\cos{\tfrac{\theta}{2}}} 
\frac{{\rm d}}{{\rm d}\theta} \left[ \sin{\tfrac{\theta}{2}} \;
 {\widetilde{\Sigma}}_{A;B;c;d}(u_{\theta}) \right]
= {\widetilde{\Sigma}}_{A;B;c;d-1}(u_{\theta}) \; .
\label{diff_2j}
\end{equation}
In particular, for $d=1$ the sum on the r.h.s.\ is the standard
sum~(\ref{binsum}) which is supposed to be known. Then 
the result for ${\widetilde{\Sigma}}_{A;B;c;1}(u)$ 
can be obtained by integration,
\[
{\widetilde{\Sigma}}_{A;B;c;1}(u_{\theta})
= \frac{1}{2 \sin{\tfrac{\theta}{2}}}
\int\limits_0^{\theta} {\rm d}\phi\; \cos{\tfrac{\phi}{2}}\;
\Sigma_{A;B;c}(u_{\phi}) \; . 
\]
Using the following decomposition 
\[
\frac{1}{j^a (2j+1)^b} = 
\sum_{i=0}^{a-1} (-2)^i \left( b-1+i \atop  b-1 \right) \frac{1}{j^{a-i}}
+ \sum_{i=0}^{b-1} (-2)^a \left( a-1+i \atop a-1 \right) \frac{1}{(2j+1)^{b-i}} \;, 
\]
we are able to rewrite 
${\widetilde{\Sigma}}_{A;B;c;d}(u)$ as a linear combination of the 
sums, with one of the last two indices equal to zero:
\[
{\widetilde{\Sigma}}_{A;B;c;d}(u) = 
\sum_{i=0}^{c-1}  (-2)^i \left( d-1+i \atop  d-1 \right) {\widetilde{\Sigma}}_{A;B;c-i;0}(u) 
+ \sum_{i=0}^{d-1} (-2)^c \left( c-1+i \atop c-1 \right) {\widetilde{\Sigma}}_{A;B;0;d-i}(u)  
\]
For the calculation of the ${\widetilde{\Sigma}}_{A;B;0;d}(u)$ (with $c=0$) 
we will apply d-times  the procedure (\ref{diff_2j})
\begin{equation}
\Biggl( \frac{2}{\cos{\tfrac{\theta}{2}}} 
\frac{{\rm d}}{{\rm d}\theta} \Biggr)^d \left[ \sin{\tfrac{\theta}{2}} \;
 {\widetilde{\Sigma}}_{A;B;0;d}(u_{\theta}) \right]
= \left( \tan\tfrac{\theta}{2}  \frac{{\rm d}}{{\rm d}\theta}\right) {\Sigma}_{A;B;1;0}(u_{\theta}) \; .
\end{equation}
This equation can be easily integrated for some particular cases
\begin{eqnarray} 
\sum_{j=1}^\infty \frac{1}{\left( 2j \atop j \right) } \frac{u^j}{2j+1} 
  &=&  \frac{\theta}{\sin\theta} - 1 
  = \frac{2}{\sin{\theta}} \; \Ti{1}{\tan{\tfrac{\theta}{2}}}- 1 \; , 
\\  
\sum_{j=1}^\infty \frac{1}{\left( 2j \atop j \right) } \frac{u^j}{(2j+1)^2} 
&=& 
\frac{2}{\sin{\tfrac{\theta}{2}}} \; \Ti{2}{\tan{\tfrac{\theta}{4}}}- 1 \; , 
\\  
\sum_{j=1}^{\infty}  \frac{1}{\left( 2j \atop j\right) } \frac{u^j}{(2j+1)} S_1 
&=&
\theta \cot{\tfrac{\theta}{2}} 
+ \frac{2}{\sin \theta} \left[ \Ls{2}{\pi-\theta} - \theta L_\theta \right] - 2 \; , 
\\  
\sum_{j=1}^{\infty}  \frac{1}{\left( 2j \atop j\right) } \frac{u^j}{(2j+1)} S_2 
&=& 
  \frac{\theta^3}{6\sin \theta}
- 2 \theta \cot{\tfrac{\theta}{2}} 
- \tfrac{1}{2} \theta^2 
+ 4 \; , 
\\  
\sum_{j=1}^{\infty}  \frac{1}{\left( 2j \atop j\right) }  \frac{u^j}{(2j+1)} \bar{S}_1 
&=& 
\frac{1}{\sin \theta}
\left[
2 \Ls{2}{\pi\!-\!\theta} 
\!-\! 2 \theta L_\theta
\!+\! \Ls{2}{\theta} 
\!+\! \theta l_\theta
\right]
\nonumber \\ && 
+ \tfrac{1}{2} \theta \cot{\tfrac{\theta}{2}} 
- \frac{2}{\sin{\tfrac{\theta}{2}}} \; \Ti{2}{\tan{\tfrac{\theta}{4}}}
- 1 \; , 
\\  
\sum_{j=1}^{\infty}  \frac{1}{\left( 2j \atop j\right) }  \frac{u^j}{(2j+1)} S_1^2  
&=& 
- \tfrac{1}{2} \theta^2 
+ 4 \cot \tfrac{\theta}{2} \left[ \Ls{2}{\pi-\theta} - \theta L_\theta \right]
- 4 
+ 2 \theta \cot \tfrac{\theta}{2}
+ \frac{\theta^3}{6\sin \theta}
\nonumber \\ &&
+ \frac{4}{\sin \theta}
\left[
\Ls{3}{\pi \!-\! \theta} \!-\! \Ls{3}{\pi}
\!+\! \theta L^2_\theta
\!-\! 2 \Ls{2}{\pi-\theta} L_\theta
\right]
\;, 
\\  
\sum_{j=1}^{\infty}  \frac{1}{\left( 2j \atop j\right) }  \frac{u^j}{(2j+1)} S_1 S_2 
&=& 
8 
\!-\! \theta^2 l_\theta
\!+\! 2 \theta \Cl{2}{\pi\!-\!\theta} 
\!-\! 2 \theta \Cl{2}{\theta} 
\!-\! 4 \Cl{3}{\pi\!-\!\theta}
\!-\! 2 \Cl{3}{\theta}
\!-\! \zeta_3 
\nonumber \\ && 
+ 
\cot \tfrac{\theta}{2}  
\left[
\tfrac{1}{6} \theta^3  
\!-\! 4 \theta  
\!+\! 4 \theta L_\theta
\!-\! 4 \Cl{2}{\pi-\theta}
\right]
\nonumber \\ && 
+ \frac{1}{\sin \theta}
\Bigl\{ 
  \theta^2 \left[ \Cl{2}{\pi \!-\! \theta} \!-\! \Cl{2}{\theta} \right]
\!-\! 4 \theta  \left[  \Cl{3}{\pi \!-\! \theta} + \Cl{3}{\theta}  \right] 
\nonumber \\ && \hspace{15mm}
- 8 \Cl{4}{\pi \!-\! \theta}
\!+\! 6 \Cl{4}{\theta}
\!+\! \theta \zeta_3 
\!-\! \tfrac{1}{3} \theta^3 L_\theta
\Bigr\} \;, 
\end{eqnarray}
where $\Ti{N}{z}$ is the inverse tangent integral~(\ref{Ti_N}).
Together with Eqs.~(\ref{S2}) and (\ref{S1barc2}), this provides 
us the result for $R_{2;1}(u)$,
\begin{equation}
R_{2;1}(u_{\theta}) = 4 \Cl{3}{\pi-\theta} - \Cl{3}{\theta}
- 2\theta \Cl{2}{\pi-\theta} + 4\zeta_3 
+  {\tfrac{1}{2}} \theta^2 l_{\theta}
+ {\tfrac{1}{6}}\theta^3\tan{\tfrac{\theta}{2}}
+ {\tfrac{1}{12}}\theta^3\cot{\tfrac{\theta}{2}} \; .
\end{equation}
Finally, using Eq.~(\ref{uniq}) and integrating over $\phi$, we arrive at
\begin{eqnarray}
%%%%%%%%%%%%%%%%%%%%%%%%%%%%%%%%%%%%%%%%%%%%%%%%%%%%%%%%%%%%%%%%%%%%%%%%%%%%%%%%%%%5
\sum_{j=1}^\infty \frac{1}{\left( 2j \atop j\right) } 
\frac{u^j}{j} S_2 \bar{S}_1   &=&  
\tan\tfrac{\theta}{2}  
\Bigl\{ 
  \theta^2 \Cl{2}{\pi-\theta}
- 4 \theta \Cl{3}{\pi-\theta}
- 8 \Cl{4}{\pi-\theta}
- \Cl{4}{\theta}
\nonumber \\  && 
+ \tfrac{1}{6} \theta^3 l_\theta
- \tfrac{1}{3} \theta^3 L_\theta
+ 4 \theta \zeta_3 
\Bigr\} \; . 
\label{S2S1barc1}
\end{eqnarray}  

Using the results for the sums with $c=1$ and
applying Eq.~(\ref{higher_c}), we obtain the following results for the
case $c=2$:
\begin{eqnarray}
%%%%%%%%%%%%%%%%%%%%%%%%%%%%%%%%%%%%%%%%%%%%%%%%%%%%%%%%%%%%%%%%%%%%%%%%%%%%%%%%%%%5
\sum_{j=1}^\infty \frac{1}{\left( 2j \atop j \right) } 
\frac{u^j}{j^2} S_3
 &=&
   \theta^2 \Cl{3}{\theta}
- 6 \theta \Cl{4}{\theta}
- 12 \Cl{5}{\theta}
- \theta^2 \zeta_3
+ 12 \zeta_5 \; , 
\label{S3c2}
\\ 
%%%%%%%%%%%%%%%%%%%%%%%%%%%%%%%%%%%%%%%%%%%%%%%%%%%%%%%%%%%%%%%%%%%%%%%%%%%%%%%%%%%5
\sum_{j=1}^\infty \frac{1}{\left( 2j \atop j \right) } 
\frac{u^j}{j^2} S_1 S_2 
 &=&
- \tfrac{1}{3} \theta^3 \Cl{2}{\pi-\theta}
+ 2 \theta^2 \Cl{3}{\pi-\theta}
+ \theta^2 \Cl{3}{\theta}
+ 8 \theta \Cl{4}{\pi-\theta}
\nonumber \\ &&  
- 6 \theta \Cl{4}{\theta}
- 16 \Cl{5}{\pi-\theta}
- 12 \Cl{5}{\theta} 
+ \tfrac{1}{2} \theta^2 \zeta_3 
- 3 \zeta_5 \; , 
\label{S1S2c2}
\\ 
%%%%%%%%%%%%%%%%%%%%%%%%%%%%%%%%%%%%%%%%%%%%%%%%%%%%%%%%%%%%%%%%%%%%%%%%%%%%%%%%%%%5
\sum_{j=1}^\infty \frac{1}{\left( 2j \atop j \right) } 
\frac{u^j}{j^2} S_2\bar{S}_1   
&=&
- \tfrac{1}{3} \theta^3 \Cl{2}{\pi-\theta}
- \tfrac{1}{6} \theta^3 \Cl{2}{\theta}
+ 2 \theta^2 \Cl{3}{\pi-\theta}
+ 2 \theta^2 \zeta_3 
- \tfrac{1}{2} \theta^2 \Cl{3}{\theta}
\nonumber \\  && 
+ 8 \theta \Cl{4}{\pi-\theta}
+ \theta \Cl{4}{\theta}
- 16 \Cl{5}{\pi-\theta}
+ 2 \Cl{5}{\theta}
- 17 \zeta_5 \; .
\label{S2S1barc2}
\end{eqnarray}

We note that the results for other sums obtained in the previous
sections can be also reproduced by this method. 
Moreover, it is possible to consider not only the combinations
$(\bar{S}_2 + \bar{S}_1^2)$ that correspond to the $J=1$ case
of Eq.~(\ref{expansion}) (given in Eq.~(\ref{expansion2})),
but also the sums containing $\bar{S}_2$ or $\bar{S}_1^2$
separately (which appear for $J\neq 1$). 
As an example, let us consider the sum 
\begin{equation}
\Sigma_{-;2;1}^{-;1}(u)
= \sum_{j=1}^\infty \frac{1}{\left( 2j \atop j \right) } 
\frac{u^j}{j} \bar{S}_2  
= \tan{\tfrac{\theta}{2}} 
\left[
\theta + \int\limits_0^{\theta} {\rm d}\phi\; R_{-;2}(u_{\phi})
\right] \; ,
\label{halfangle1}
\end{equation}
where $R_{-;2}(u)$ corresponds to the case~(\ref{R_b1}).

Substituting 
\begin{equation}
R_{-;2}(u_{\theta})
= 
\frac{2}{\sin{\tfrac{\theta}{2}}}\; 
\Ti{2}{\tan{\tfrac{\theta}{4}}} + {\tfrac{1}{8}}\theta^2 - 1 \; , 
\end{equation}
into Eq.~(\ref{halfangle1}) and integrating over $\phi$,
we arrive at
\begin{equation}
%%%%%%%%%%%%%%%%%%%%%%%%%%%%%%%%%%%%%%%%%%%%%%%%%%%%%%%%%%%%%%%%%%%%%%%%%%%%%%%%%%%5
\sum_{j=1}^\infty \frac{1}{\left( 2j \atop j \right) } 
\frac{u^j}{j} \bar{S}_2   
= \tan\tfrac{\theta}{2}  
\Bigl\{ 4\Ti{3}{\tan\tfrac{\theta}{4}} + \tfrac{1}{24}\theta^3 \Bigl\} \; . 
\end{equation}
Moreover, upon applying Eq.~(\ref{higher_c}) we can perform another
integration and obtain the result for $c=2$,
\begin{equation}
%%%%%%%%%%%%%%%%%%%%%%%%%%%%%%%%%%%%%%%%%%%%%%%%%%%%%%%%%%%%%%%%%%%%%%%%%%%%%%%%%%%5
\sum_{j=1}^\infty \frac{1}{\left( 2j \atop j \right) } 
\frac{u^j}{j^2} \bar{S}_2   
= 4\theta \Ti{3}{\tan\tfrac{\theta}{4}}
- 8 \left[ \Ti{2}{\tan\tfrac{\theta}{4}} \right]^2
+\tfrac{1}{96}\theta^4 \; .
\end{equation}
Using the connections between $\Ti{2}{\tan\tfrac{\theta}{4}}$, 
$\Ti{3}{\tan\tfrac{\theta}{4}}$ and the log-sine functions
(see Eq.~(17) on p.~292 and Eq.~(44) on p.~298 of Ref.~\cite{Lewin}),
we can get other representations,
\begin{eqnarray}
\sum_{j=1}^\infty \frac{1}{\left( 2j \atop j \right) } 
\frac{u^j}{j} \bar{S}_2   
&=&
\tan\tfrac{\theta}{2}  
\Bigl\{ 
\tfrac{1}{24} \theta^3 
+ \tfrac{1}{2} \theta \ln^2\left(\tan\tfrac{\theta}{4}\right)
+ 2 \ln\left(\tan\tfrac{\theta}{4}\right)
\left[ \Ls{2}{\tfrac{\theta}{2}}  + \Ls{2}{\pi - \tfrac{\theta}{2}} \right] 
\nonumber \\ &&
+ 2 \left[ \Ls{3}{\pi - \tfrac{\theta}{2}} - \Ls{3}{\pi} \right]
- 2 \Ls{3}{\tfrac{\theta}{2}}
+ \tfrac{1}{2} \Ls{3}{\theta} 
\Bigr \} \;, 
\label{S2barc1}
\\
\sum_{j=1}^\infty \frac{1}{\left( 2j \atop j \right) } 
\frac{u^j}{j^2} \bar{S}_2   
&\!\!=\!\!&
2 \theta \left[ \Ls{3}{\pi-\tfrac{\theta}{2}}-\Ls{3}{\tfrac{\theta}{2}}
\right]
+\tfrac{1}{2} \theta \Ls{3}{\theta}
-2 \left[ \Ls{2}{\pi-\tfrac{\theta}{2}}+\Ls{2}{\tfrac{\theta}{2}}
\right]^2
\nonumber \\ &&
+\tfrac{1}{6}\pi^3 \theta + \tfrac{1}{96}\theta^4 \; .
\label{S2barc2}
\end{eqnarray}
Note the appearance of log-sine functions of arguments ${\tfrac{\theta}{2}}$
and $\left(\pi -{\tfrac{\theta}{2}}\right)$.

%%%%%%%%%%%%%%%%%%%%%%%%%%%%%%%%%%%%%%%%%%%%%%%%%%%%%%%%%%%%%%%%%%%%%%%%%
\section{Analytical continuation}
\setcounter{equation}{0}

To obtain results valid in other regions of variable $u$ (for
$u<0$ and $u>4$), we will construct
the proper analytical continuation of the expressions presented
in the previous section. 
For generalized log-sine integrals it is described in~\cite{DK01}.
Let us introduce a new variable
\begin{equation}
y \equiv e^{ {\rm i} \sigma \theta}, \hspace{5mm}
\ln(-y-{\rm i}\sigma 0) = \ln{y} - {\rm i} \sigma \pi,
\label{y:def}
\end{equation}
where the choice of the sign $\sigma=\pm 1$ is related to the
causal ``+i0'' prescription for the propagators.
For completeness, we also present  
the inverse relations,
\begin{equation}
\label{y<->u}
u = - \frac{(1-y)^2}{y} \;, \quad  
y = \frac{1-\sqrt{\frac{u}{u-4}}}{1+\sqrt{\frac{u}{u-4}}} \;, \quad
u \frac{{\rm d}}{{\rm d} u} = -\frac{1-y}{1+y} \; y \frac{{\rm d}}{{\rm d}y}  \;, 
\end{equation}
and also expressions for $(1\pm y)$ in terms of $\theta$,
\begin{equation}
1 - y = 2 \sin \tfrac{\theta}{2}\;  e^{ -{\rm i} \sigma (\pi-\theta)/2}  \;, \quad 
1 + y = 2 \cos \tfrac{\theta}{2}\;  e^{  {\rm i} \sigma \theta/2} \;. 
\label{y<->theta}
\end{equation}
In terms of this variable $y$, the analytic continuation of
all generalized log-sine integrals can be expressed in terms 
of Nielsen polylogarithms, whereas for
the function $\Phi(\theta)$ we get
\begin{eqnarray}
\Phi(\theta) & = & \zeta_3 \ln 2 + \tfrac{1}{2} \zeta_4 
-  \H_{-1,0,0,1}(1) 
+ \tfrac{1}{2} \left[ \H_{-1,0,0,1}(y) + \H_{-1,0,0,1}(y^{-1}) \right]
\nonumber \\ && 
- \tfrac{1}{4} \left[\Li{4}{y} + \Li{4}{y^{-1}} \right]
- \tfrac{1}{2} \left[ \ln(1+y) - \tfrac{1}{2} \ln y \right] 
           \left[\Li{3}{y} + \Li{3}{y^{-1}} \right] \;,
\end{eqnarray}
with $\H_{-1,0,0,1}(y)$ defined in Eq.~(\ref{h(-1001)}).

For the cases involving the inverse tangent integrals
$\Ti{N}{\tan \frac{\theta}{2}}$ and
$\Ti{N}{\tan \frac{\theta}{4}}$, the analytic continuation
is straightforward (see Eq.~(\ref{Ti_N})), 
\begin{eqnarray}
\Ti{N}{\tan\tfrac{\theta}{2}} &=& - \frac{\sigma}{2 {\rm i}}
\left[ \Li{N}{\omega}  - \Li{N}{ - \omega} \right] \;, \qquad \;\;
\omega = \frac{1-y}{1+y} = - {\rm i} \sigma \tan\tfrac{\theta}{2} \;. 
\label{def_omega}
\\
\Ti{N}{\tan\tfrac{\theta}{4}} &=& - \frac{\sigma}{2 {\rm i}}
\left[ \Li{N}{\omega_s}  - \Li{N}{ - \omega_s} \right] \;, \qquad
\omega_s = \frac{1-\sqrt{y}}{1+\sqrt{y}} = - {\rm i} \sigma \tan\tfrac{\theta}{4} \;. 
\label{def_omega_s}
\end{eqnarray}

Below we list the most complicated results, corresponding to the
analytical continuation of the results obtained in Section~2:
%%%%%%%%%%%%%%%%%%%%%%%%%%%%%%%%%%%%%%%%%%%%%%%%%%%%%%%%%%%%%%%%%%%%%%%%%%%
\begin{eqnarray}
\label{AN_S1S1_2}
&& \hspace*{-10mm} 
\sum_{j=1}^\infty \frac{1}{\left( 2j \atop j \right) } \frac{u^j}{j^2} S_1^2
 =
- 8 \Snp{1,2}{-y} \ln y
+ 4 \Li{3}{-y} \ln y
- 2 \Li{2}{-y} \ln^2 y
+ 4 \left[ \Li{2}{-y} \right]^2
\nonumber \\ && \hspace{20mm}
- \tfrac{1}{24} \ln^4 y
+ 4  \zeta_2 \Li{2}{-y}
+ \zeta_2 \ln^2 y
+ 4 \zeta_3 \ln y
+ \tfrac{5}{2} \zeta_4 \; ,
\\ && \hspace*{-10mm}
%%%%%%%%%%%%%%%%%%%%%%%%%%%%%%%%%%%%%%%%%%%%%%%%%%%%%%%%%%%%%%%%%%%%%%%%%%%
\label{AN_S1Sb1_2}
\sum_{j=1}^\infty \frac{1}{\left( 2j \atop j \right) } \frac{u^j}{j^2} S_1 \bar{S}_1
 =  
- 10 \Snp{1,2}{-y} \ln y 
+  \Snp{1,2}{y^2} \ln y 
- 2 \Snp{1,2}{y} \ln y 
+ 4 \left[\Li{2}{-y} \right]^2
\nonumber \\ && 
- 2 \Li{2}{y} \Li{2}{-y} 
+ 3 \Li{3}{-y} \ln y 
- \Li{3}{y} \ln y 
- \tfrac{3}{2} \Li{2}{-y} \ln^2 y 
+ \tfrac{1}{2} \Li{2}{y} \ln^2 y 
\nonumber \\ && 
- \tfrac{1}{48} \ln^4 y 
+ 6 \zeta_2 \Li{2}{-y}
-  \zeta_2 \Li{2}{y}
+ \tfrac{5}{4} \zeta_2 \ln^2 y 
+ \tfrac{11}{2} \zeta_3 \ln y 
+ 5 \zeta_4 \; ,
\\ && \hspace*{-10mm} 
%%%%%%%%%%%%%%%%%%%%%%%%%%%%%%%%%%%%%%%%%%%%%%%%%%%%%%%%%%%%%%%%%%%%%%%%%%%
\label{AN_Sb2_2}
\sum_{j=1}^\infty \frac{1}{\left( 2j \atop j \right) } \frac{u^j}{j^2} 
\left( \bar{S}_2 + \bar{S}_1^2 \right)
 =  
- 12 \Snp{1,2}{ -y} \ln y 
+ 2 \Snp{1,2}{y^2} \ln y 
- 6 \Snp{1,2}{y} \ln y 
+ 4 \left[\Li{2}{-y} \right]^2
\nonumber \\ && 
+  \left[\Li{2}{y} \right]^2
- 4 \Li{2}{-y} \Li{2}{y} 
+ 2 \Li{3}{-y} \ln y 
- \Li{3}{y} \ln y 
- \Li{2}{-y} \ln^2 y 
\nonumber \\ && 
+ \tfrac{1}{2} \Li{2}{y} \ln^2 y 
+ 8 \zeta_2 \Li{2}{-y}
- 4 \zeta_2 \Li{2}{y}
+ \zeta_2 \ln^2 y 
+ 8 \zeta_3 \ln y 
+ 10 \zeta_4 \; , 
\\ && \hspace*{-10mm} 
%%%%%%%%%%%%%%%%%%%%%%%%%%%%%%%%%%%%%%%%%%%%%%%%%%%%%%%%%%%%%%%%%%%%%%%%%%%
\label{AN_C1}
\sum_{j=1}^\infty \frac{1}{\left( 2j\atop j\right)} \frac{u^j}{j} {\cal C}_1
=  \frac{1-y}{1+y}
\Bigl[ 
  3 \Li{4}{-y}  
- 3 \Li{4}{y}
+ 28 \Snp{1,3}{-y} 
- 2  \Snp{1,3}{y^2} 
+ 4  \Snp{1,3}{y} 
\nonumber \\ &&
- 14 \Snp{2,2}{-y} 
+ \Snp{2,2}{y^2} 
- 2\Snp{2,2}{y}
+ 28 \Snp{1,2}{-y} \ln(1+y)
- 2 \Snp{1,2}{y^2} \ln(1-y)
\nonumber \\ && 
- 2 \Snp{1,2}{y^2} \ln(1+y)
+ 4 \Snp{1,2}{y} \ln(1-y)
+ 4 \Li{3}{-y} \ln (1-y) 
+ 2 \Li{3}{y} \ln (1-y) 
\nonumber \\ && 
- 10 \Li{3}{-y} \ln (1+y) 
+ 4 \Li{3}{y} \ln (1+y)  
- 2 \Li{2}{-y} \ln^2 (1-y) 
\nonumber \\ && 
- 2 \Li{2}{y^2} \ln (1+y) \ln (1-y) 
+ 12 \Li{2}{-y} \ln^2 (1+y) 
- 2 \Li{2}{y} \ln^2 (1+y) 
\nonumber \\ && 
- 2 \ln y \ln^2 (1-y) \ln (1+y) 
- 2 \ln y \ln (1-y) \ln^2 (1+y) 
+ 4 \ln y \ln^3 (1+y) 
\nonumber \\ && 
+ \tfrac{1}{2} \ln^2 y \ln^2 (1-y) 
+ 2 \ln^2 y \ln (1-y) \ln (1+y)  
- \tfrac{5}{2} \ln^2 y \ln^2 (1+y) 
- \tfrac{1}{2} \ln^3 y \ln (1-y) 
\nonumber \\ && 
+ \tfrac{1}{2} \ln^3 y \ln (1+y) 
+ \tfrac{45}{8} \zeta_4 
- \zeta_3 \ln (1-y) 
- 13 \zeta_3 \ln (1+y) 
+ 7 \zeta_3 \ln y 
+ \tfrac{9}{4} \zeta_2 \ln^2 y 
\nonumber \\ && 
- \zeta_2 \ln^2 (1-y)
+ 2 \zeta_2 \ln (1-y) \ln(1+y)
+ 8 \zeta_2 \ln^2 (1+y) 
- 9 \zeta_2 \ln y \ln (1+y) 
\Bigr] \;, 
%%%%%%%%%%%%%%%%%%%%%%%%%%%%%%%%%%%%%%%%%%%%%%%%%%%%%%%%%%%%%%%%%%%%%%%%%%%%%%
\\ && \hspace*{-10mm}
\label{AN_C2}
\sum_{j=1}^\infty \frac{1}{\left( 2j\atop j\right)} \frac{u^j}{j} {\cal C}_2
=  \frac{1-y}{1+y}
\Bigl[
 2 \Snp{1,3}{y^2}
- 24 \Snp{1,3}{-y}  
- 6 \Snp{1,3}{y}
+ 12 \Snp{2,2}{-y} 
- \Snp{2,2}{y^2} 
\nonumber \\ && 
+ 3 \Snp{2,2}{y} 
- 2 \Li{4}{-y} 
+ \tfrac{5}{2} \Li{4}{y}
- 24 \Snp{1,2}{-y} \ln(1+y) 
- 6 \Snp{1,2}{y} \ln (1-y)  
\nonumber \\ && 
- \Li{3}{y} \ln (1-y) 
+ 2 \Snp{1,2}{y^2} \ln (1-y) 
+ 2 \Snp{1,2}{y^2} \ln (1+y) 
- 4 \Li{3}{-y} \ln (1-y) 
\nonumber \\ && 
+ 8 \Li{3}{-y} \ln (1+y) 
- 4 \Li{3}{y} \ln (1+y) 
+ 2 \Li{2}{-y} \ln^2 (1-y) 
- \Li{2}{y} \ln^2 (1-y) 
\nonumber \\ && 
+ 4 \Li{2}{-y} \ln (1-y) \ln (1+y) 
+ 4 \Li{2}{y} \ln (1-y) \ln (1+y) 
- 10 \Li{2}{-y} \ln^2 (1+y) 
\nonumber \\ && 
+ 2 \Li{2}{y} \ln^2 (1+y) 
- \tfrac{1}{3} \ln y \ln^3 (1-y) 
+ 2 \ln y \ln^2 (1-y) \ln (1+y) 
\nonumber \\ && 
+ 2 \ln y \ln (1-y) \ln^2 (1+y) 
- \tfrac{10}{3} \ln y \ln^3 (1+y) 
- \tfrac{1}{4} \ln^2 y \ln ^2 (1-y) 
\nonumber \\ && 
- 2 \ln^2 y \ln (1-y) \ln (1+y) 
+ 2 \ln^2 y \ln^2 (1+y) 
+ \tfrac{5}{12} \ln^3 y \ln (1-y) 
- \tfrac{1}{3} \ln^3 y \ln (1+y) 
\nonumber \\ && 
- \tfrac{1}{96} \ln^4 y 
- \tfrac{9}{4} \zeta_4 
+ 2 \zeta_3 \ln (1-y) 
+ 11 \zeta_3 \ln (1+y) 
- \tfrac{13}{2} \zeta_3 \ln y 
- 7 \zeta_2 \ln^2 (1+y) 
\nonumber \\ && 
+ 2 \zeta_2 \ln^2 (1-y) 
- 2 \zeta_2 \ln (1-y) \ln (1+y) 
- \tfrac{7}{4} \zeta_2 \ln^2 y 
- \zeta_2 \ln y \ln (1-y) 
\nonumber \\ && 
+ 8 \zeta_2 \ln y \ln (1+y) 
\Bigr] \;, 
\\ && \hspace*{-10mm}
%%%%%%%%%%%%%%%%%%%%%%%%%%%%%%%%%%%%%%%%%%%%%%%%%%%%%%%%%%%%%%%%%%%%%%%%%%%
\label{AN_S1_3}
\sum_{j=1}^\infty \frac{1}{\left( 2j \atop j \right) } \frac{u^j}{j^3} S_1
 =  
 4 H_{-1,0,0,1}(-y) 
+ \Snp{2,2}{y^2}
- 4 \Snp{2,2}{y} 
- 4 \Snp{2,2}{-y} 
- 6 \Li{4}{-y} 
\nonumber \\ && 
- 2 \Li{4}{y}
+ 4 \Snp{1,2}{-y}  \ln y 
+ 4 \Snp{1,2}{ y}  \ln y 
- 2 \Snp{1,2}{ y^2}  \ln(y)
+ 4 \Li{3}{-y} \ln (1-y) 
\nonumber \\ && 
+ 2 \Li{3}{-y} \ln y
+ 2 \Li{3}{y}  \ln y
- \Li{2}{y} \ln^2 y 
- 4 \Li{2}{-y} \ln y \ln (1-y) 
\nonumber \\ && 
- \tfrac{1}{3} \ln^3 y \ln (1-y) 
+ \tfrac{1}{24} \ln^4 y 
+ 2 \zeta_2 \Li{2}{y}
- \tfrac{1}{2} \zeta_2 \ln^2 y 
+ 2 \zeta_2 \ln y \ln (1-y) 
\nonumber \\ && 
+ 6 \zeta_3 \ln (1-y) 
- 3 \zeta_3 \ln y 
- 4 \zeta_4 \; , 
\\ && \hspace*{-10mm}
%%%%%%%%%%%%%%%%%%%%%%%%%%%%%%%%%%%%%%%%%%%%%%%%%%%%%%%%%%%%%%%%%%%%%%%%%%%
\label{AN_Sb1_3}
\sum_{j=1}^\infty \frac{1}{\left( 2j \atop j \right) } \frac{u^j}{j^3} \bar{S}_1
 =  
 4 H_{-1,0,0,1}(-y) 
+ \Snp{2,2}{y^2}
- 8 \Snp{2,2}{y} 
- 4 \Snp{2,2}{-y} 
- 6 \Li{4}{-y} 
\nonumber \\ && 
+ 2 \Li{4}{y}
- \left[ \Li{2}{y} \right]^2
+ 4 \Snp{1,2}{-y}  \ln y 
+ 8 \Snp{1,2}{ y}  \ln y 
- 2 \Snp{1,2}{ y^2} \ln y 
+ \tfrac{1}{48} \ln^4 y 
\nonumber \\ && 
+ 4 \Li{3}{-y} \ln (1-y) 
- 4 \Li{3}{y} \ln (1-y) 
+ 2 \Li{3}{-y} \ln y 
- 4 \Li{2}{-y} \ln y \ln (1-y) 
\nonumber \\ && 
+ 2 \Li{2}{y} \ln y \ln (1-y) 
- \tfrac{1}{2} \Li{2}{y} \ln^2 y 
- \tfrac{1}{6} \ln^3 y \ln (1-y) 
+ 4 \zeta_2 \ln y \ln (1-y) 
\nonumber \\ && 
- \zeta_2 \ln^2 y 
+ 10 \zeta_3 \ln (1-y) 
- 5 \zeta_3 \ln y 
+ 4 \zeta_2 \Li{2}{y}
- \tfrac{19}{2} \zeta_4 \; , 
\\ && \hspace*{-10mm}
%%%%%%%%%%%%%%%%%%%%%%%%%%%%%%%%%%%%%%%%%%%%%%%%%%%%%%%%%%%%%%%%%%%%%%%%%%%
\label{AN_S1^3_1}
\sum_{j=1}^\infty \frac{1}{\left( 2j \atop j \right) } \frac{u^j}{j} S_1^3
=  \frac{1-y}{1+y}
\Bigl[
- 48 \Snp{1,2}{-y} \ln (1+y) 
- 48 \Snp{1,3}{-y}
+ 24 \Snp{2,2}{-y}
\nonumber \\ && 
- 12 \zeta_2 \ln^2 (1+y) 
- 24 \ln^2 (1+y) \Li{2}{-y}
+ 24 \zeta_3 \ln (1+y) 
+ 24 \ln(1+y) \Li{3}{-y}
\nonumber \\ && 
- 8 \ln y \ln^3 (1+y) 
+ 12 \zeta_2 \ln y \ln (1+y) 
+ 6 \ln^2 y \ln^2 (1+y) 
- \ln^3 y \ln(1+y)
\nonumber \\ && 
+ \tfrac{1}{24} \ln^4 y 
- \tfrac{3}{2} \zeta_2 \ln^2 y 
+ 3 \ln^2 y \Li{2}{-y}
+ \ln^2 y \Li{2}{y}
 - 5 \zeta_3 \ln y 
- 12 \ln y \Li{3}{-y}
\nonumber \\ && 
- 4 \ln y \Li{3}{y}
+ \tfrac{3}{2} \zeta_4 
+ 12 \Li{4}{-y}
+ 6 \Li{4}{y}
\Bigr] \;, 
\\ && \hspace*{-10mm}
%%%%%%%%%%%%%%%%%%%%%%%%%%%%%%%%%%%%%%%%%%%%%%%%%%%%%%%%%%%%%%%%%%%%%%%%%%%
\label{AN_S3_2}
\sum_{j=1}^\infty \frac{1}{\left( 2j \atop j \right) } \frac{u^j}{j^2} S_3 
=  
- 12 \Li{5}{y} 
+ 6 \ln y \Li{4}{y}
- \ln^2 y \Li{3}{y}
- \tfrac{1}{120} \ln^5 y 
\nonumber \\ && 
+ \zeta_3 \ln^2 y 
+ 6 \zeta_4 \ln y 
+ 12 \zeta_5 
\;, 
\\ && \hspace*{-10mm}
%%%%%%%%%%%%%%%%%%%%%%%%%%%%%%%%%%%%%%%%%%%%%%%%%%%%%%%%%%%%%%%%%%%%%%%%%%%
\label{AN_S1S2_2}
\sum_{j=1}^\infty \frac{1}{\left( 2j \atop j \right) } \frac{u^j}{j^2} S_1 S_2  
=  
- 12 \Li{5}{y} 
- 16 \Li{5}{-y} 
+ 6 \ln y \Li{4}{y}
+ 8 \ln y \Li{4}{-y}
\nonumber \\ && 
- \ln^2 y \Li{3}{y}
- 2 \ln^2 y \Li{3}{-y}
+ \tfrac{1}{3} \ln^3 y \Li{2}{-y}
+ \tfrac{1}{120} \ln^ 5 y 
- \tfrac{1}{6} \zeta_2 \ln^3 y 
\nonumber \\ && 
- \tfrac{1}{2} \zeta_3 \ln^2 y 
- \zeta_4 \ln y 
- 3 \zeta_5
\;,
\\ && \hspace*{-10mm}
%%%%%%%%%%%%%%%%%%%%%%%%%%%%%%%%%%%%%%%%%%%%%%%%%%%%%%%%%%%%%%%%%%%%%%%%%%%
\label{AN_S2Sb1_2}
\sum_{j=1}^\infty \frac{1}{\left( 2j \atop j \right) } \frac{u^j}{j^2} S_2 \bar{S}_1  
=   
  2 \Li{5}{y}
- 16 \Li{5}{-y}
- \ln y \Li{4}{y}
+ 8 \ln y \Li{4}{-y}
\nonumber \\ && 
+ \tfrac{1}{2} \ln^2 y \Li{3}{y}
- 2 \ln^2 y \Li{3}{-y}
+ \tfrac{1}{3} \ln^3 y \Li{2}{-y}
- \tfrac{1}{6} \ln^3 y \Li{2}{y}
+ \tfrac{1}{240} \ln^ 5 y 
\nonumber \\ && 
- \tfrac{1}{3} \zeta_2 \ln^3 y 
- 2 \zeta_3 \ln^2 y 
- 8 \zeta_4 \ln y 
- 17 \zeta_5
\;,
\end{eqnarray}
%%%%%%%%%%%%%%%%%%%%%%%%%%%%%%%%%%%%%%%%%%%%%%%%%%%%%%%%%%%%%%%%%%%%%%%%%%%
where we have used the relations~(\ref{Hrel1}) and (\ref{Hrel2}).

For the results involving $\Lsc{2,3}{\theta}$ we can use 
Eq.~(\ref{Lsc-Ti}) to express 
them in terms of $\mbox{Ti}$ function, and then employ
Eq.~(\ref{def_omega}). For example, starting from 
Eq.~(\ref{S1S1Sb1_1}) we obtain
\begin{eqnarray}
&& \hspace*{-10mm}
%%%%%%%%%%%%%%%%%%%%%%%%%%%%%%%%%%%%%%%%%%%%%%%%%%%%%%%%%%%%%%%%%%%%%%%%%%%
\label{AN_S1S1Sb1_1}
\sum_{j=1}^\infty \frac{1}{\left( 2j \atop j \right) } \frac{u^j}{j} S_1^2 \bar{S}_1  
=  \frac{1-y}{1+y}
\Biggl\{
    \Li{4}{y}
+ 8 \Li{4}{-y}
- 4 \Li{4}{\omega}
+ 4 \Li{4}{-\omega}
\nonumber \\ && 
+ 2 \Snp{1,3}{y^2}
- 8 \Snp{1,3}{y}
- 48 \Snp{1,3}{-y}
- \Snp{2,2}{y^2}
+ 4 \Snp{2,2}{y}
+ 24 \Snp{2,2}{-y}
\nonumber \\ && 
+ 2 \ln (1+y) 
\left[ 
  \Snp{1,2}{y^2} 
- 24 \Snp{1,2}{-y} 
+ 10 \Li{3}{-y}
- 2  \Li{3}{y} 
\right]
\nonumber \\ && 
+ 2 \ln (1-y) 
\left[ 
   \Snp{1,2}{y^2} 
- 4 \Snp{1,2}{y}  
- 2 \Li{3}{-y}
\right]
- 8 \ln y \Li{3}{-y} 
\nonumber \\ && 
+ 2 \Li{2}{-y} 
\left[ 
    \ln^2 (1-y) 
+ 2 \ln (1-y) \ln (1+y) 
- 11 \ln^2(1+y) 
+ \ln^2 y 
\right]
\nonumber \\ && 
+  2 \Li{2}{y} 
\left[ 
    \ln^2 (1+y) 
-   \ln^2 (1-y) 
+ 2 \ln (1-y) \ln (1+y) 
\right]
+ \tfrac{1}{48} \ln^4 y 
\nonumber \\ && 
+ 2 \ln y \left[ 
  \ln (1+y) \ln^2 (1-y) 
+ \ln (1-y) \ln^2 (1+y) 
- \tfrac{1}{3} \ln^3 (1-y) 
- \tfrac{11}{3} \ln^3 (1+y) 
\right]
\nonumber \\ && 
- 2 \ln^2 y \ln (1-y) \ln (1+y) 
+ 5 \ln^2 y \ln^2 (1+y) 
+ \tfrac{1}{6} \ln^3 y \ln (1-y) 
- \tfrac{2}{3} \ln^3 y \ln (1+y) 
\nonumber \\ && 
\!+\! \zeta_2 \left[ 
  3  \ln^2 (1\!-\!y) 
\!-\! 2  \ln (1\!-\!y) \ln (1\!+\!y) 
\!-\! 13 \ln^2 (1\!+\!y) 
\!-\! 2  \ln y \ln (1\!-\!y) 
\!+\! 14 \ln y \ln (1\!+\!y) 
\!-\! \tfrac{3}{2} \ln^2 y
\right]
\nonumber \\ && 
\!+\! \zeta_3 \left[ 
  3  \ln (1-y) 
- 6  \ln y 
+ 23 \ln (1+y) 
\right]
+ \tfrac{33}{4} \zeta_4 
\Biggr\} \;,
\end{eqnarray}
where $\omega=(1-y)/(1+y)$, see Eq.~(\ref{def_omega}).
We note that the analytic continuation of $\Lsc{2,3}{\theta}$
can also be obtained directly, using the integral
representation (\ref{Lsc}). This procedure is described 
in Appendix~A, Eqs.~(\ref{begin_AN_Lsc})--(\ref{AC_of_Lsc}).

The analytic continuation of the sum involving $\bar{S}_2$ 
can be presented in terms of the
variable $\omega_s$ given in Eq.~(\ref{def_omega_s}),
\begin{equation}
%%%%%%%%%%%%%%%%%%%%%%%%%%%%%%%%%%%%%%%%%%%%%%%%%%%%%%%%%%%%%%%%%%%%%%%%%%%
\label{AN_S2b_c2}
\sum_{j=1}^\infty \frac{1}{\left( 2j \atop j \right) } \frac{u^j}{j^2} \bar{S}_2
=  
2 \ln y  \Bigl[ \Li{3}{\omega_s}  - \Li{3}{-\omega_s} \Bigr] 
+ 2 \Bigl[ \Li{2}{\omega_s}  - \Li{2}{-\omega_s} \Bigr]^2 
+ \tfrac{1}{96} \ln^4 y \;. 
\end{equation}

The results for 
lower values of $c$ can be deduced using
\begin{equation}
\sum_{j=1}^\infty  \frac{u^j}{j^c} f(j) =
u \frac{{\rm d}}{{\rm d}u} \sum_{j=1}^\infty \frac{u^j}{j^{c+1}} f(j) \;. 
\label{derivative}
\end{equation}
For example, 
%%%%%%%%%%%%%%%%%%%%%%%%%%%%%%%%%%%%%%%%%%%%%%%%%%%%%%%%%%%%%%%%%%%%%%%%%%
\begin{eqnarray} 
\label{AN_S1_2}
\sum_{j=1}^\infty \frac{1}{\left( 2j \atop j\right) } \frac{u^j}{j^2} S_1
&=& 
4 \Li{3}{-y} - 2 \Li{2}{-y} \ln y - \tfrac{1}{6} \ln^3 y
+ 3 \zeta_3  + \zeta_2 \ln y \; ,
%%%%%%%%%%%%%%%%%%%%%%%%%%%%%%%%%%%%%%%%%%%%%%%%%%%%%%%%%%%%%%%%%%%%%%%%%%%
\\  
\label{AN_Sb1_2}
\sum_{j=1}^\infty \frac{1}{\left( 2j \atop j \right) } \frac{u^j}{j^2} \bar{S}_1
&=& 
-2 \Li{3}{y} + 4 \Li{3}{-y} - 2 \Li{2}{-y} \ln y + \Li{2}{y}\ln y
\nonumber \\ && 
- \tfrac{1}{12} \ln^3 y + 2 \zeta_2 \ln y + 5 \zeta_3 \;.
\end{eqnarray}
The results for these two sums can be extracted from Ref.~\cite{FKV99}.
Further results for the sums are collected in Appendix~C.

%%%%%%%%%%%%%%%%%%%%%%%%%%%%%%%%%%%%%%%%%%%%%%%%%%%%%%%%%%%%%%%%%%%%%%%%%%%
\section{Application to Feynman diagrams}
\setcounter{equation}{0}

Below  we present results for the $\ep$-expansion of one- and two-loop 
master integrals shown in Fig.~1. 
In the rest of this paper we use the notation $u=p^2/m^2$.

\begin{figure}[th]
\begin{center}
\centerline{\vbox{\epsfysize=90mm \epsfbox{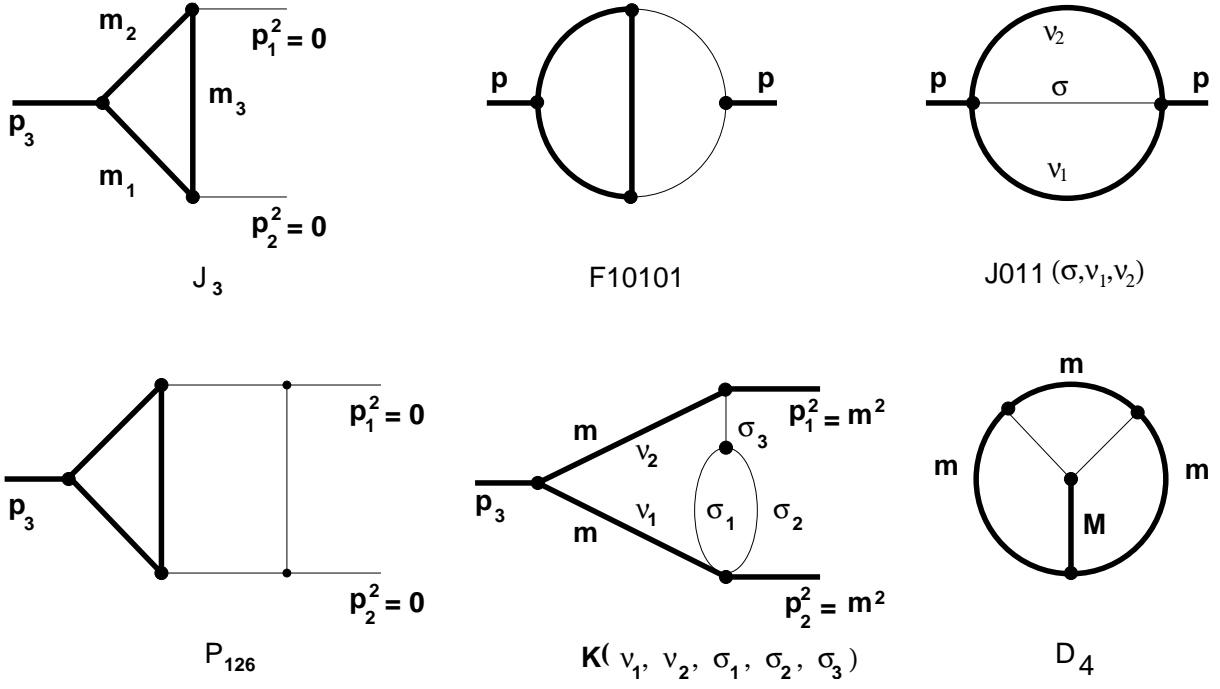}}}
\caption{\label{propagator} One- two- and three-loop diagrams
considered in the paper.
Bold and thin lines correspond to massive and
massless propagators, respectively.}
\end{center}
\end{figure}

%%%%%%%%%%%%%%%%%%%%%%%%%%%%%%%%%%%%%%%%%%%%%%%%%%%%%%%%%%%%%%%%%%%%%%%%%%%%%%%%%%%%%%%%%
\subsection{One-loop vertex}

Let us consider a one-loop triangle diagram with
$m_1=m_2=m_3\equiv m$, $p_1^2=p_2^2=0$, with an arbitrary (off-shell)
value of $p_3^2\equiv p^2$. Such diagrams occur, for example, in Higgs decay
into two photons or two gluons via a massive quark loop.
Following the notation of Ref.~\cite{BD-TMF}, we will denote
this integral (with unit powers of propagators) as
$J_3(1,1,1; m)$. According to Eq.~(40) of Ref.~\cite{BD-TMF}, the result
in an arbitrary space-time dimension $n=4-2\ep$ is
\begin{equation}
J_3(1,1,1; m) \left. \right|_{p_1^2 = p_2^2 = 0} =
- \tfrac{1}{2} \mbox{i} \pi^{2-\ep} (m^2)^{-1-\ep} \Gamma(1+\ep)\;
{}_{3}F_2 \left(\begin{array}{c|} 1, 1, 1+\ep \\
\frac{3}{2}, 2 \end{array} ~\frac{p^2}{4m^2} \right) \; .
\end{equation}
For this integral a number of one-fold integral representations are 
available, see Eqs.~(3.10)--(3.11) in Ref.~\cite{DK01}.
Expanding in $\ep$ we get
\begin{eqnarray}
\label{J3sums}
&& \hspace{-5mm}
J_3(1,1,1; m) \left. \right|_{p_1^2 = p_2^2 = 0} 
= - \mbox{i} \pi^{2-\ep} \Gamma(1+\ep)\; \frac{(m^2)^{-\ep}}{p^2}
\sum_{j=1}^{\infty} \frac{1}{\left( 2j \atop j\right) } \frac{u^j}{j^2} 
\left[ 1 + \ep S_1
+ \tfrac{1}{2}\ep^2 \left( S_1^2 -S_2 \right)
\right.
\nonumber \\  && \hspace{8mm}
+ \tfrac{1}{6}\ep^3 \left( S_1^3 -3 S_1 S_2 + 2 S_3 \right)
+ \tfrac{1}{24}\ep^4 \left( S_1^4 -6 S_1^2 S_2 + 8 S_1 S_3 + 3 S_2^2 - 6 S_4 \right)
\left. 
+ {\cal O} (\ep^5) 
\right] .
\hspace*{9mm}
\end{eqnarray}
Substituting results for the occurring inverse binomial sums, we
obtain
\begin{eqnarray}
J_3(1,1,1; m) \left. \right|_{p_1^2 = p_2^2 = 0} 
&\!\!=\!\!& - \mbox{i} \pi^{2-\ep} \Gamma(1+\ep)\; \frac{(m^2)^{-\ep}}{p^2}
\Bigl\{ 
 - \tfrac{1}{2} \ln^2 y 
%%%%%%%%%%%%%%%%%%%%%%%%%%%%
\nonumber \\ &&
+ \ep \left[ 
4 \Li{3}{-y}
- 2 \Li{2}{-y} \ln y 
- \tfrac{1}{6} \ln^3 y 
+ \zeta_2 \ln y 
+ 3 \zeta_3 
\right]
%%%%%%%%%%%%%%%%%%%%%%%%%%%%
\nonumber \\ && 
+ \ep^2 \Bigl[ 
2 \left[\Li{2}{-y} \right]^2
+ 2 \Li{3}{-y} \ln y 
- 4 \Snp{1,2}{-y} \ln y 
- \Li{2}{-y} \ln^2 y 
\nonumber \\ && 
+ 2 \zeta_2 \Li{2}{-y}
+ \tfrac{1}{2}\zeta_2 \ln^2 y 
+ 2 \zeta_3 \ln y 
- \tfrac{1}{24} \ln^4 y 
+ \tfrac{5}{4} \zeta_4
\Bigl]
+ {\cal O}(\ep^3) 
\Bigl\} \; . 
\label{Hgg1}
\end{eqnarray}
These results correspond to the analytic continuation of
Eqs.~(3.13), (3.14) and (3.16) from Ref.~\cite{DK01}.

The $\ep^3$ term is Eq.~(\ref{J3sums}) contains the same
combination of sums as ${\cal C}_0$ in Eq.~(\ref{combinations2}).
However, the sum in Eq.~(\ref{J3sums}) contains $j^2$
in the denominator, i.e., it corresponds to the case $c=2$.
Using the result~(\ref{combination:1}) (with $c=1$) 
together with (\ref{higher_c}), we obtain a one-fold
integral representation for the $\ep^3$-term in Eq.~(\ref{J3sums}). 
Analyzing it, we see that we get only one new non-trivial integral,  
\begin{equation}
\label{nontrivial5}
\int\limits_0^\theta {\mbox d} \phi \; 
\Ls{2}{\phi} \; \ln^2 \left| 2\sin\tfrac{\phi}{2} \right| \; ,
\end{equation}
while all other terms can be expressed in terms of known functions.
For $\theta=\tfrac{2\pi}{3}$ the integral~(\ref{nontrivial5}) 
is connected with the new element $\chi_5$ of the {\em odd} basis,
whereas for $\theta=\tfrac{\pi}{2}$ it is related to
the new element $\widetilde{\chi}_5$ 
of the {\em even} basis, see Section~3.3 in Ref.~\cite{DK01} 
and Eqs.~(14)--(15) in Ref.~\cite{review}.

At the same time, the $\ep^4$-term of Eq.~(\ref{J3sums})
(and all other even powers of $\ep$)
can be calculated in terms of log-sine functions,
so that their analytic continuation can be expressed in terms of 
Nielsen polylogarithms.
Namely, for the $\ep^4$-term, combining 
Eqs.~(3.14), (3.19) from Ref.~\cite{DK01} and Eq.~(\ref{S4G}),
we obtain the result for the sum 
\[
\sum_{j=1}^{\infty} \frac{1}{\left( 2j \atop j\right) } \frac{u^j}{j^c} 
\left( S_1^4 -6 S_1^2 S_2 + 8 S_1 S_3 - 3 S_2^2\right)
\]
with $c=2$. Using Eq.~(\ref{derivative}), we also obtain
the result for the case $c=1$. 
%%%%%%%%%%%%%%%%%%%%%%%%%%%%%%%%%%%%%%%%%%%%%%%%%%%%%%%%%%%

\subsection{Two-loop self-energy integral $F_{10101}$}

This integral is a good illustration of the
application of general expressions given in Section~3.
The off-shell result for this integral in arbitrary
dimension was presented in~\cite{BFT93} (where it was called
$\widetilde{I}_3$, see Eq.~(22) of \cite{BFT93}).
For unit powers of propagators, the result reads~\footnote{
In given normalization each loop is divided by $\pi^{2-\ep} \Gamma(1+\ep)$.}
\begin{eqnarray}
&& \hspace*{-15mm}
m^{2+4\ep}
(1-2\ep) F_{10101}(p^2,m) =
\frac{1}{(1-\ep^2)(1+2\ep)}
{}_{4}F_3 \left(\begin{array}{c|} 1, 1+\ep,1+\ep, 1+2\ep \\
\frac{3}{2} +\ep ,2+\ep, 2-\ep \end{array} ~\frac{p^2}{4m^2} \right)
\nonumber\\ && \hspace*{-15mm}
-\frac{1}{2 \ep (1+\ep)}
{}_{3}F_2 \left(\begin{array}{c|} 1, 1\!+\!\ep, 1\!+\!\ep \\
\frac{3}{2}, 2+\ep \end{array} ~\frac{p^2}{4m^2} \right)
+ \frac{1}{2\ep}
\frac{\Gamma^2(1-\ep)}{\Gamma(1\!-\!2\ep)} \left(-\frac{m^2}{p^2}\right)^\ep
{}_{3}F_2 \left(\begin{array}{c|} 1, 1, 1+\ep \\
\frac{3}{2}, 2 \end{array} ~\frac{p^2}{4m^2} \right). 
\end{eqnarray}
Expanding in $\ep$, we obtain
\begin{eqnarray}
&& \hspace*{-10mm}
p^2 (m^2)^{2\ep}
(1-2\ep) F_{10101}(p^2,m) 
\nonumber\\ && \hspace*{-10mm}
=
\sum_{j=1}^\infty \frac{1}{\left( 2j \atop j\right) } \frac{u^j}{j^2} 
\Biggl\{ 
\frac{3}{j} - \ln(-u)
+ \ep \Biggl[
- \frac{1}{j^2}
+ \frac{11}{j} S_1 
- \frac{4}{j} \bar{S}_1 
- \ln (-u) S_1 
+ \tfrac{1}{2} \ln^2 (-u) 
- \zeta_2 
\Biggr]
+ {\cal O} (\ep^2)\Biggr \}
\nonumber\\ && \hspace*{-10mm}
=  6 \Li{3}{y} - 6 \Li{2}{y} \ln y  -2 \ln^2 y \ln (1-y) - 6 \zeta_3 
%%%%%%%%%%%%%%%%%%%%%%%%%%%%%%%%%%%
\nonumber\\ && \hspace*{-10mm}
+ \ep \Bigl\{ 
  28  H_{-1,0,0,1}(-y) 
+ 7 \Snp{2,2}{y^2}
- 28 \Snp{2,2}{-y}
- 16 \Snp{2,2}{y}
- 42 \Li{4}{-y} 
- 26 \Li{4}{y}
\nonumber\\ && \hspace*{-10mm}
+ 4 \left[\Li{2}{y} \right]^2
+ 16 \Snp{1,2}{y} \ln y 
- 14 \Snp{1,2}{y^2} \ln y 
+ 28 \Snp{1,2}{-y} \ln y 
+ 18 \Li{3}{-y} \ln y 
\nonumber\\ && \hspace*{-10mm}
+ 20 \Li{3}{y} \ln y 
+ 20 \Li{3}{-y} \ln (1-y) 
+ 12 \Li{3}{y} \ln (1-y) 
- 2 \Li{2}{-y} \ln^2 y 
- 9 \Li{2}{y} \ln^2 y 
\nonumber\\ && \hspace*{-10mm}
- 24 \Li{2}{-y} \ln y \ln (1-y) 
- 4 \Li{2}{y} \ln y \ln (1-y)
- 2 \ln^3 y \ln (1-y) 
+ 6 \zeta_2 \Li{2}{y}
\nonumber\\ && \hspace*{-10mm} 
+ 4 \zeta_2 \ln y \ln (1-y) 
- 12 \zeta_3 \ln y 
+ 24 \zeta_3 \ln (1-y) 
- 9 \zeta_4 
\Bigr\}
+ {\cal O} (\ep^2) \; .
\label{f10101}
\end{eqnarray}
The result for the finite part coincides with~\cite{B90},
whereas the result for the $\ep$-term is new.
%%%%%%%%%%%%%%%%%%%%%%%%%%%%%%%%%%%%%%%%%%%%%%%%%%%%%%%%%%%%%%%%%%%%%%%%%%%%%%%%%%%%%%%%%%%%%%%%
%
%J011
%
%%%%%%%%%%%%%%%%%%%%%%%%%%%%%%%%%%%%%%%%%%%%%%%%%%%%%%%%%%%%%%%%%%%%%%%%%%%%%%%%%%%%%%%%%%%%%%%

\subsection{Two-loop sunset-type diagram $J_{011}$}

Let us consider sunset-type diagrams with two equal masses and 
one zero mass (see Fig.~1). The off-shell result for the sunset-type 
integral $J_{011}$ with arbitrary powers of propagators 
has been obtained~\footnote{The imaginary part of the sunset diagrams in an arbitrary dimension  
is presented in \cite{sunset:im}.} 
in Refs.~\cite{BFT93,D91}
%~\footnote{The integral
%$J_{011}(\sigma,\nu_1,\nu_2;p^2,m)$ is related with integral 
%$L(\sigma,\nu_1,\nu_2;p^2,m)$ defined in Ref.~\cite{D91} 
%as
%$$
%J_{011}(\sigma,\nu_1,\nu_2)  
%\equiv \frac{(-1)^{1\!+\!\sigma\!+\!\nu_1\!+\!\nu_2}}{(\pi^{2\!-\!\ep})^2 \Gamma^2(1\!+\!\ep)} 
%                            L(\sigma,\nu_1,\nu_2;p^2,m).
%$$
%}, 
by using the 
Mellin--Barnes technique \cite{BD-TMF}:
\begin{eqnarray}
J_{011}(\sigma,\nu_1,\nu_2;p^2,m) \!\! &=& \!\!
(m^2)^{n-\sigma-\nu_1-\nu_2}
\frac{
\Gamma(\nu_1\!+\!\nu_2\!+\!\sigma\!-\!n)
\Gamma\left(\frac{n}{2}\!-\!\sigma\right)
\Gamma\left(\nu_2\!+\!\sigma\!-\!\frac{n}{2}\right)
\Gamma\left(\nu_1\!+\!\sigma\!-\!\frac{n}{2}\right)}
{\Gamma(\nu_1)\Gamma(\nu_2)\Gamma\left(\frac{n}{2}\right)
\Gamma(\nu_1+\nu_2+2\sigma-n) \Gamma^2 \left(3-\frac{n}{2}\right)}
\nonumber\\ && \times
{}_{4}F_3 \left( \begin{array}{c|}
\sigma,\nu_1+\nu_2+\sigma-n,
\nu_2+\sigma-\tfrac{n}{2}, \nu_1+\sigma-\tfrac{n}{2} \\
\tfrac{n}{2}, \sigma + \tfrac{1}{2}(\nu_1+\nu_2-n),
\sigma+ \tfrac{1}{2}(\nu_1+\nu_2+1-n)
\end{array}~\frac{p^2}{4 m^2} \right) .
\label{j011_off-shell}
\end{eqnarray}
For simplicity,  in the definition of the integral $J_{011}$
we will omit the arguments $p^2$ and $m$
$$
J_{011}(\sigma,\nu_1,\nu_2)  \equiv J_{011}(\sigma,\nu_1,\nu_2;p^2,m).
$$
Let us remind that for the integrals $J_{011}$ with 
different integer values of $\sigma$ and $\nu_i$
there are two master integrals~\cite{T97a} of this type,
$J_{011}(1,1,1)$ and $J_{011}(1,1,2)$.
However, two other independent combinations
of the integrals of this type happen to be more suitable for 
constructing the $\ep$-expansion, $J_{011}(1,2,2)$ and
$[J_{011}(1,2,2)+2J_{011}(2,1,2)$] (see also in Ref.~\cite{thresholds}). 
The latter combination corresponds to the integral $J_{011}(1,1,1)$ in 
$2-2\ep$ dimensions \cite{T96}.
In Ref.~\cite{single} 
$J_{011}(1,1,3)$ has been used as the second integral.
To construct the $\ep$-expansion of the integrals 
$J_{011}(1,1,1)$ and $J_{011}(1,1,2)$ up to 
order $\ep^2$, the integral $J_{011}(1,2,2)$ should also 
be expanded up to $\ep^2$, whereas
$[J_{011}(1,2,2)+2J_{011}(2,1,2)]$ or 
$J_{011}(1,1,3)$ up to the order $\ep$ only. 
In Appendix~C we give an example of a realistic calculation
which demonstrates the required orders of the $\ep$-expansion
for the integrals involved.

For $J_{011}(1,2,2)$ we have obtained
%%%%%%%%%%%%%%%%%%%%%%%%%%%%%%%%%%%%%%%%%%%%%%%%%%%%%%%%%%%%
\begin{eqnarray}
&& \hspace{-10mm}
J_{011}(1,2,2)  = 
\frac{(m^2)^{-1-2\ep}}{(1-\ep)(1+2\ep)}
{}_{3}F_2 \left( \begin{array}{c|}
1,  1+\ep, 1+2\ep \\
\tfrac{3}{2} + \ep, 2-\ep
\end{array}~\frac{p^2}{4 m^2} \right)  
\nonumber \\ && \hspace*{-5mm}
= 2 \frac{(m^2)^{-2\ep}}{p^2}
\sum_{j=1}^{\infty} \frac{1}{\left( 2j \atop j\right) } \frac{u^j}{j^2} 
\Biggl\{ 
1 + \ep \Biggl[ \frac{1}{j} + 5 S_1 - 2 \bar{S}_1
\Biggr]
\nonumber \\ &&
+ \ep^2 \Biggl[ 
  \tfrac{1}{j^2} 
+ \tfrac{5}{j}S_1 
- \tfrac{2}{j} \bar{S}_1
+ \tfrac{25}{2} S_1^2 
- 10 S_1 \bar{S}_1 
- \tfrac{5}{2} S_2
+ 2 \bar{S}_1^2 
+ 2 \bar{S}_2 
\Biggr]
\nonumber \\ &&
+ \ep^3 \left[
  \tfrac{1}{j^3}  
+ \tfrac{5}{j^2}S_1 
- \tfrac{2}{j^2} \bar{S}_1
+ \tfrac{25}{2 j} S_1^2 
- \tfrac{10}{j} S_1 \bar{S}_1 
- \tfrac{5}{2 j} S_2
+ \tfrac{2}{j}  \left( \bar{S}_1^2 + \bar{S}_2 \right) 
+ \tfrac{125}{6} S_1^3 
- 25 S_1^2 \bar{S}_1 
\right.
\nonumber \\ &&
\left.
- \tfrac{25}{2} S_1 S_2 
+ \tfrac{11}{3} S_3  
+ 5 S_2 \bar{S}_1 
+ 10 S_1 \Biggl( \bar{S}_1^2 + \bar{S}_2 \right) 
- \tfrac{4}{3} \left( \bar{S}_1^3 + 3 \bar{S}_1 \bar{S}_2 + 2 \bar{S}_3  \Biggr)
\right]
+ {\cal O} (\ep^4) 
\Biggr\}
%%%%%%%%%%%%%%%%%%%%%%%%%%%%%
\nonumber \\ && \hspace*{-5mm}
= 2 \frac{(m^2)^{-2\ep}}{p^2} 
\left( 1-y \right)^{ 2 \ep}
y^{ 2\ep}
\Biggl\{ 
- \tfrac{1}{2} \ln^2 y 
\nonumber \\ &&
+ \ep 
\Bigl[ 
 \tfrac{1}{2} \ln^3 y 
+ \zeta_2 \ln y 
-6 \ln y  \Li{2}{-y} 
-4 \ln y  \Li{2}{y} 
+ 3 \zeta_3 
+ 12 \Li{3}{-y} 
+ 6 \Li{3}{y}
\Bigr]
\nonumber \\ &&
+ \ep^2 
\Bigl[  
 12 H_{-1,0,0,1} (-y)
-12 \ln y  \Snp{1,2}{-y} 
-12 \ln y \Snp{1,2}{y^2} 
+ 8 \ln y  \Snp{1,2}{y} 
- 12 \Snp{2,2}{-y}
\nonumber \\ &&
+ 3 \Snp{2,2}{y^2}
- 12 \ln (1-y) \Li{3}{-y} 
-\tfrac{7}{24} \ln^4 y 
- \tfrac{1}{2} \zeta_2 \ln^2 y 
+ 2 \zeta_3 \ln y 
+ 6 \ln y  \Li{3}{-y} 
\nonumber \\ && 
+ 8 \ln y  \Li{3}{y} 
+ 6 \zeta_2 \Li{2}{-y} 
+ 4 \zeta_2 \Li{2}{y}
+ \tfrac{13}{4} \zeta_4 
+ 18 \left[ \Li{2}{-y} \right]^2
\nonumber \\ && 
+ 12 \Li{2}{y} \Li{2}{-y}
+ 4 \left[ \Li{2}{y} \right]^2
- \tfrac{9}{4} \Li{4}{y^2}
\Bigr]
+ {\cal O} (\ep^3) 
\Biggr\} \; ,
\label{j011_122}
\end{eqnarray}
where the result for the $\ep^2$-term is new.

For another combination, we get
%%%%%%%%%%%%%%%%%%%%%%%%%%%%%%%%%%%%%%%%%%%%%%%%%%%%%%%%%%%%%%%%%%%%%%%%%%%%%%%%%%%%%%%%%%%
\begin{eqnarray}
&& \hspace{-10mm}
J_{011}(1,2,2) + 2 J_{011}(2,1,2)
= - \frac{(m^2)^{-1-2\ep}}{\ep(1+2\ep)}
{}_{3}F_2 \left( \begin{array}{c|}
1,  1+\ep, 1+2\ep \\
\tfrac{3}{2} + \ep, 1-\ep
\end{array}~\frac{p^2}{4 m^2} \right)  
\nonumber \\ && \hspace*{-10mm}
= - \frac{2(m^2)^{-2\ep}}{p^2 } 
\sum_{j=1}^{\infty} 
\frac{1}{\left( 2j \atop j\right) } \frac{u^j}{j}
\Biggl\{ \frac{1}{\ep} 
+ \left[ 5 S_1 - 2 \bar{S}_1 \right]
+ \ep \left[
\tfrac{25}{2} S_1^2 
- 10 S_1 \bar{S}_1 
- \tfrac{5}{2} S_2
+ 2 \bar{S}_1^2
+ 2 \bar{S}_2
\right]
\nonumber \\ && \hspace*{10mm}
+ \ep^2 \left[
\tfrac{125}{6} S_1^3 
- 25 S_1^2 \bar{S}_1 
- \tfrac{25}{2} S_1 S_2 
+ \tfrac{11}{3} S_3  
+ 5 S_2 \bar{S}_1 
+ 10 S_1 \Biggl( \bar{S}_1^2 + \bar{S}_2 \right) 
\nonumber \\ && \hspace*{20mm}
- \tfrac{4}{3} \left( \bar{S}_1^3 + 3 \bar{S}_1 \bar{S}_2 + 2 \bar{S}_3  \Biggr)
\right]
\!+\! {\cal O} (\ep^3) 
\Biggr\}
\nonumber \\ && \hspace*{-10mm}
= - \frac{2(m^2)^{-2\ep}}{p^2}\;
\frac{(1-y)^{1-2\ep}}{(1+y)^{1+6\ep}}  
\Biggl\{ 
\frac{1}{\ep}\ln y 
+ 
2 \ln^2 y 
- \zeta_2 
- 6 \Li{2}{-y} 
- 2 \Li{2}{y}
\nonumber \\ && 
+ \ep \Bigl[ 
24 \Snp{1,2}{-y} 
\!+\! 6 \Snp{1,2}{y^2}
\!-\! 8 \Snp{1,2}{y} 
\!+\! \tfrac{5}{3} \ln^3 y 
\!-\! 4 \zeta_2 \ln y 
\!-\! 11 \zeta_3 
\!-\! 24 \Li{3}{-y} 
\!-\! 8 \Li{3}{y}
\Bigr] 
\nonumber \\ && 
+ \ep^2 \Bigl[ 
  24 \Li{4}{\omega}
- 24 \Li{4}{-\omega}
- 24 \Snp{1,3}{y^2} 
+ 12 \Snp{2,2}{y^2}
+ 64 \Snp{1,3}{y} 
\nonumber \\ && \hspace{10mm}
+ 96 \Snp{2,2}{-y}
- 32 \Snp{2,2}{y}
- 192 \Snp{1,3}{-y}
+ 52 \Li{4}{y}
+ 48 \Li{4}{-y}
\nonumber \\ && \hspace{10mm}
- 6 \ln\omega 
\left[ 
2 \Snp{1,2}{y^2} - 8 \Snp{1,2}{y} - 8 \Snp{1,2}{-y}  + 7 \zeta_3 
\right]
- 72 \ln y \Li{3}{-y} 
- 48 \ln y \Li{3}{y}
\nonumber \\ && \hspace{10mm}
+ 6 \ln^2 \omega 
\left[ 
  2 \Li{2}{y}
- 2 \Li{2}{-y}
- 3 \zeta_2 
\right]
+ 12 \ln^2 y \Li{2}{y}
+ 18 \ln^2 y \Li{2}{-y}
\nonumber \\ && \hspace{10mm}
+ 4 \ln y \ln^3 \omega
+ \tfrac{11}{12} \ln^4 y 
- 5 \zeta_2 \ln^2 y
- 26 \zeta_3 \ln y 
- 57 \zeta_4 
\Bigr] 
+ {\cal O} (\ep^3) 
\Biggr\} \;,
\label{j011_c}
\end{eqnarray}
where $\omega=(1-y)/(1+y)$, see Eq.~(\ref{def_omega}).
The $\ep$-term can be related to that of the result for
$J_{011}(1,1,3)$ presented in Ref.~\cite{single}, whereas
the result for the $\ep^2$ term is new. 

%%%%%%%%%%%%%%%%%%%%%%%%%%%%%%%%%%%%%%%%%%%%%%%%%%%%%%%%%%%%%%%%%%%%%%%%%%%%%%%
%%%%%%%%%%%%%%%%%%%%%%%%%%%%%%%%%%%%%%%%%%%%%%%%%%%%%%%%%%%%%%%%%%%%%%%%%%%%%%%
\subsection{Two-loop vertex diagrams } 

Consider the two-loop vertex-type diagram $P_{126}$ given 
in Ref.~\cite{FKV99}\footnote{The $q^2$ from Ref.~\cite{FKV99} 
corresponds to our $p^2$, whereas their $z$ corresponds to our $u$. 
For $P_{126}$ we keep the normalization used in in Ref.~\cite{FKV99} 
for two-loop vertices: each loop integral is divided by 
$\pi^{2-\ep}m^{-2\ep} e^{-\gamma_E\ep}$, where $\gamma_E$ is Euler's 
constant. Note that it is different from the normalization used in 
Ref.~\cite{Hgg:num}, where each loop is divided by 
${\mbox i} \pi^{2-\ep} \Gamma(1+\ep) \mu^{-2\ep}$, 
where $\mu$ is the scale parameter of dimensional regularization \cite{dimreg}. 
Furthermore, in numerical results presented in Table~1 of 
Ref.~\cite{Hgg:num} an extra common 
factor $m^4\Gamma(1-2\ep)/\Gamma^2(1-\ep)$ was extracted.}
(see Fig.~1),
\begin{eqnarray} 
(p^2)^2 P_{126}  &=& 
    \sum_{j=1}^{\infty} \frac{1}{\left( 2j \atop j\right) } \frac{u^j}{j^2}
      \Biggl\{
      \frac{1}{\varepsilon^2}
    + \frac{1}{\varepsilon} \Bigl[
           - S_1
           - \log(-u)
             \Bigr]
\nonumber \\ &&
      + \tfrac{1}{2} \log^2(-u)
      - S_1 \log(-u)
      - \tfrac{3}{2} S_2
      - \tfrac{15}{2} S_1^2
      + 4 S_1 \bar{S}_1  
      + 2 \frac{S_1}{j}
+ {\cal O} (\ep)
      \Biggr\} \; .
\end{eqnarray}
It corresponds to one of the two-loop contributions to 
a boson decay into two massless particles, 
with a massive triangle subloop. Diagrams of such type  
have been intensively studied for Higgs boson production 
via gluon fusion~\cite{Hgg}.

Using our approach, we obtain~\footnote{Recently this result 
was checked numerically by G.~Passarino and S.~Uccirati
with the help of their approach~\cite{passarino}. 
After fixing a typo in an earlier version of our result,
the results are in full agreement.
The corrected result below was also confirmed 
(according to a private communication by R.~Bonciani)
by R.~Bonciani, P.~Mastrolia and E.~Remiddi
\cite{confirnation_BMR}, using their approach~\cite{vertex_BMR}.
}
\begin{eqnarray} 
(p^2)^2 P_{126}  &\!\!=\!\!& 
     - \frac{1}{2\ep^2} \ln^2 y 
+ \frac{1}{\ep} \Bigl[ 
2 \Li{2}{-y} \ln y 
\!-\! 4 \Li{3}{-y}
\!+\! \ln^2 y \ln (1-y) 
\!-\! \tfrac{1}{3} \ln^3 y 
\!-\! \zeta_2 \ln y 
\!-\! 3 \zeta_3 
\Bigr]
\nonumber \\ &&
+  8 H_{-1,0,0,1}(-y) 
+ 2 \Snp{2,2}{y^2}
- 8  \Snp{2,2}{-y}
- 8  \Snp{2,2}{y}
- 4 \Li{4}{y}
- 12 \Li{4}{-y} 
\nonumber \\ &&
- 8 \Li{2}{y} \Li{2}{-y} 
- 14 \left[ \Li{2}{-y} \right]^2
+ 28 \Snp{1,2}{-y} \ln y 
- 10 \Li{3}{-y} \ln y 
\nonumber \\ &&
+ 7 \Li{2}{-y} \ln^2 y 
- 4 \Li{2}{-y} \ln y \ln (1-y) 
- \ln^2 y \ln^2 (1-y) 
- \tfrac{1}{6} \ln^4 y 
\nonumber \\ &&
+ \tfrac{2}{3} \ln^3 y \ln (1-y) 
+ 2 \zeta_2 \ln y \ln (1-y) 
- \tfrac{5}{2} \zeta_2 \ln^2 y 
+ 6 \zeta_3 \ln (1-y) 
- 11 \zeta_3 \ln y
\nonumber \\ &&
- 6 \Li{2}{-y} \zeta_2 
- \tfrac{27}{4} \zeta_4 
+ {\cal O} (\ep) \; .
\label{Hgg2}
\end{eqnarray}
Here, the result for the finite part is new.
Alternatively, results of such type
can be obtained in a different way, using 
a technique based on Mellin--Barnes contour integrals~\cite{S_Bha}. 

%%%%%%%%%%%%%%%%%%%%%%%%%%%%%%%%%%%%%%%%%%%%%%%%%%%%%%%%%%%%%%%%%%%%%%%%%
%\subsection{Two-loop vertex diagram J}

Let us consider another two-loop vertex-type diagram shown in Fig.~1, 
$K(\nu_1,\nu_2,\sigma_1,\sigma_2,\sigma_3)$, where two external momenta 
are on shell ($p_1^2=p_2^2=m^2$) and we also put $p_3^2=p^2$.
Note that $K(1,1,1,1,0)$ appears as one of the master integrals 
in Ref.~\cite{vertex_BMR} (see Eqs.~(102)--(106) of~\cite{vertex_BMR}).

First of all, integration over the momentum of the massless loop 
(with powers of the propagators equal to $\sigma_1$ and $\sigma_2$) 
can be easily performed,
yielding the corresponding massless propagator to the power 
$\sigma_1+\sigma_2-\tfrac{n}{2}$, times a well-known factor
containing $\Gamma$ functions. The resulting integral is nothing but
a one-loop triangle function, 
$J_2(\nu_1, \nu_2, \sigma_1+\sigma_2+\sigma_3-\tfrac{n}{2}; m)$
in the notation of Ref.~\cite{BD-TMF}, 
with two external momenta ($p_1$ and $p_2$) on shell
and $\nu_3=\sigma_1+\sigma_2+\sigma_3-\tfrac{n}{2}$.
Using the results of Ref.~\cite{BD-TMF}, one can see that in this
on-shell limit the three-point integral $J_2$ reduces to 
a two-point function $J^{(2)}$ with two massive propagators
(with the powers $\nu_1$ and $\nu_2$), but with a shifted value
of the space-time dimension ($n\to n-2\nu_3$),
\begin{equation}
\label{3to2}
J_2(\nu_1, \nu_2, \nu_3; m)\big|_{p_1^2=p_2^2=m^2,\; p_3\equiv p}
= \pi^{\nu_3}\; {\rm i}^{-2\nu_3}\;
\frac{\Gamma(n-\nu_1-\nu_2-2\nu_3)}{\Gamma(n-\nu_1-\nu_2-\nu_3)}\;
J^{(2)}(n-2\nu_3; \nu_1, \nu_2) \; .
\end{equation}
This is a generalization to arbitrary values of $\nu_i$ of the relation
found in Ref.~\cite{DOS1,DK01} for $\nu_1=\nu_2=\nu_3=1$.

Using Eq.~(\ref{3to2}), we arrive at the following 
result for the diagram $K(\nu_1,\nu_2,\sigma_1,\sigma_2,\sigma_3;p^2,m)$
(which we normalize by dividing each loop by 
${\rm i}\pi^{n/2}\Gamma(3-\tfrac{n}{2})$):
\begin{eqnarray}
&& \hspace*{-15mm}
K(\nu_1,\nu_2,\sigma_1,\sigma_2,\sigma_3; p^2, m)
\nonumber \\
&=&  
\frac{(-1)^{\nu_1\!+\!\nu_2\!+\!\sigma_1\!+\!\sigma_2\!+\!\sigma_3}}
{(m^2)^{\nu_1\!+\!\nu_2\!+\!\sigma_1\!+\!\sigma_2\!+\!\sigma_3\!-\!n}}
\frac{\Gamma(\nu_1\!+\! \nu_2 \!+\! \sigma_1 \!+\! \sigma_2\!+\!\sigma_3\!-\!n)
     \Gamma\left(\frac{n}{2}\!-\!\sigma_1\right)
     \Gamma\left(\frac{n}{2}\!-\!\sigma_2\right)
     \Gamma\left(\sigma_1\!+\! \sigma_2\!-\!\frac{n}{2}\right)
}
{\Gamma(\sigma_1)\Gamma(\sigma_2)\Gamma(\nu_1\!+\!\nu_2)
 \Gamma(n\!-\!\sigma_1\!-\!\sigma_2) \Gamma^2(3-\tfrac{n}{2})} 
\nonumber\\ && \times
\frac{\Gamma(2n\!-\!\nu_1\!-\!\nu_2\!-\!2\sigma_1\!-\! 2 \sigma_2\!-\! 2\sigma_3)}
    { \Gamma \left(\frac{3n}{2}\!-\!\nu_1\!-\!\nu_2\!-\!\sigma_1\!-\!\sigma_2
       \!-\!\sigma_3\right) }
{}_{3}F_2 \left( \begin{array}{c|}
\nu_1,\nu_2, \nu_1\!+\!\nu_2\!+\!\sigma_1\!+\!\sigma_2\!+\!\sigma_3\!-\!n \\
\tfrac{1}{2}(\nu_1+\nu_2), \tfrac{1}{2}(\nu_1+\nu_2+1)
\end{array}~\frac{u}{4} \right) ,
\hspace*{10mm}
\end{eqnarray}
where $u=p^2/m^2$.

For $\nu_1=\nu_2=\sigma_1=\sigma_2=1$ and $\sigma_3=0$, 
the parameter $\nu_3$ 
in Eq.~(\ref{3to2}) gets equal to $\ep$, and the function
corresponds to a one-loop two-point function in $4-4\ep$
dimensions (i.e., $\ep\to 2\ep$),
\begin{equation}
K(1,1,1,1,0; p^2, m)
= (m^2)^{-2\ep}
\frac{1}{2\ep^2} \
\frac{\Gamma^2(1-\ep) \Gamma (1+2\ep) \Gamma(2-4\ep)}
{\Gamma(2-2\ep) \Gamma(2-3\ep) \Gamma(1+\ep)}\;
{}_{2}F_1 \left( \begin{array}{c|}
1, 2\ep \\
\tfrac{3}{2}
\end{array}~\frac{u}{4} \right) \; . 
\end{equation}
As in the one-loop case (cf.\ Eqs.~(C29)--(C30) of Ref.~\cite{DOS1}),
the $_2F_1$ function can be reduced to the type considered in
section~2 by means of one of the Kummer relations,
\[
(1-4\ep)\;
{}_{2}F_1 \left( \begin{array}{c|}
1, 2\ep \\
\tfrac{3}{2}
\end{array}~z \right)
= 1- 4\ep (1-z)\;
{}_{2}F_1 \left( \begin{array}{c|}
1, 1+2\ep \\
\tfrac{3}{2}
\end{array}~z \right) \; .
\]
All orders of the $\ep$-expansion of the resulting $_2F_1$ function
are given in  (see also in Refs.~\cite{D-ep,DK01}).
For another special case, $K(1,1,1,1,1)$,
the parameter $\nu_3$ 
in Eq.~(\ref{3to2}) is equal to $1+\ep$, and the function
corresponds to a one-loop two-point function in $2-4\ep$
dimensions. In this case, we directly get the $_2F_1$ function 
of the type~(\ref{basis_II}), with $\ep\to 2\ep$.

Analytic continuation in terms of the Nielsen polylogarithms was
discussed in section~2.2 of Ref.~\cite{DK01}.
In the case of $K(1,1,1,1,0)$, we get
\begin{eqnarray}
K(1,1,1,1,0; p^2, m)
&=& 
(m^2)^{-2\ep}
\frac{1}{2\ep^2} \
\frac{\Gamma^2(1-\ep) \Gamma (1+2\ep) \Gamma(1-4\ep)}
{\Gamma(1-2\ep) \Gamma(1-3\ep) \Gamma(1+\ep)}
\frac{1}{(1-2\ep)(1-3\ep)}
\nonumber \\ && 
\hspace*{-20mm} 
\times
\Biggl\{ 
1 - \frac{(1+y)^{1-4\ep}}{2(1-y)} (1-y^{4\ep}) 
- 2 \ep \frac{(1+y)^{1-4\ep}}{y^{-2\ep}(1-y)}
\sum_{j=0}^{\infty} (-4\ep)^j
\nonumber \\ && 
\hspace*{-20mm} 
\times
\sum_{p=0}^{j-1} \frac{\ln^p y}{2^p p!} 
\sum_{k=1}^{j-p} (-2)^{-k}
\left[ \Snp{k,j+1-k-p}{-y} - (-1)^p \Snp{k,j+1-k-p}{-y^{-1}} \right] 
\Biggr\} \; ,
\hspace*{8mm}
\label{K11110_Snp}
\end{eqnarray}
where the generalized polylogarithms $\Snp{k,p}{-y^{-1}}$ 
can be expressed in terms of inverse
argument by means of the standard formulae given in Ref.~\cite{Nielsen}.
In this way, we have obtained results for {\em all} coefficients
of the $\ep$-expansion of $K(1,1,1,1,0)$.
The first three coefficients 
coincide with those given in Eqs.~(104)--(106) of Ref.~\cite{vertex_BMR}.

%%%%%%%%%%%%%%%%%%%%%%%%%%%%%%%%%%%%%%%%%%%%%%%%%%%%%%%%%%%%%%%%%%%%%%%%%%%%%%%
%%%%%%%%%%%%%%%%%%%%%%%%%%%%%%%%%%%%%%%%%%%%%%%%%%%%%%%%%%%%%%%%%%%%%%%%%%%%%%%
\subsection{Three-loop vacuum diagram $D_4$ } 
%
%D4
%

Consider a three-loop vacuum diagram with two different masses
shown in Fig.~({\ref{propagator}}). Such an integral with 
equal masses ($M=m$) enters as a master integral in Avdeev's 
package~\cite{leo96} and {\sf MATAD}~\cite{matad}. 
Although there is no similar package for three-loop vacuum 
diagrams with two different masses, 
it is clear that such configuration (with unit powers 
of the propagators) will remain one of the master integrals
in this more general case.
 
This integral can be calculated by integrating over the momentum $p$
the off-shell two-loop diagram $F_{10101}$ 
(also shown in Fig.~{\ref{propagator}}) 
with a massive propagator 
containing an arbitrary mass $M$ (see Eq.~(4.7) in \cite{DK01}),
\begin{eqnarray}
&& \hspace*{-20mm}
{D_4}(1,1,1,1,1,1;u) = 
\frac{1}{{\rm i} \pi^{n/2}} \int \frac{{\rm d}^n p}{p^2-M^2}\; F_{10101}(p^2,m) \; ,
\end{eqnarray}
where $u=M^2/m^2$.
The result of this integration can be presented in the following form:
\begin{eqnarray}
\label{D4_hypergeom}
&& \hspace*{-20mm}
(m^2)^{3 \ep} (1\!-\!\ep) (1\!-\!2\ep) {D_4}(1,1,1,1,1,1;u) 
\nonumber \\ && \hspace*{-10mm}  
= 
-\frac{u^{1\!-\!\ep}}{\ep(1\!-\!\ep)(1\!+\!\ep)(1\!+\!2\ep)}
~{}_{4}F_3 \left(\begin{array}{c|}
1, 1\!+\!\ep, 1\!+\!\ep, 1\!+\!2\ep \\
\frac{3}{2}\!+\!\ep, 2\!+\!\ep, 2\!-\!\ep\end{array} ~\frac{u}{4} \right) 
\nonumber \\ && \hspace*{-10mm}
%%%%%%%%%%%%%%%%%%%%%%%%%%%%%%%%%%%%%%%%%%%%%%%%%%%%%%%%%%%%%%%%%%%%%%
+ \frac{u \Gamma(1\!-\!\ep) \Gamma^2(1\!+\!2\ep)\Gamma(1\!+\!3\ep)}
{\ep (1\!+\!2\ep)(1\!+\!4\ep)\Gamma^2(1\!+\!\ep) \Gamma(1\!+\!4\ep)}
~{}_{4}F_3 \left(\begin{array}{c|}
1, 1\!+\!2\ep, 1\!+\!2\ep, 1\!+\!3\ep \\
\frac{3}{2}\!+\!2\ep, 2, 2\!+\!2\ep\end{array} ~\frac{u}{4} \right)
\nonumber \\ && \hspace*{-10mm}
%%%%%%%%%%%%%%%%%%%%%%%%%%%%%%%%%%%%%%%%%%%%%%%%%%%%%%%%%%%%%%%%%%%%%%
+ \frac{u^{1\!-\!\ep}}{2\ep^2 (1\!+\!\ep)}
~{}_{3}F_2 \left(\begin{array}{c|}
1, 1\!+\!\ep, 1\!+\!\ep \\
\frac{3}{2}, 2\!+\!\ep\end{array} ~\frac{u}{4} \right)
%%%%%%%%%%%%%%%%%%%%%%%%%%%%%%%%%%%%%%%%%%%%%%%%%%%%%%%%%%%%%%%%%%%%%%
- \frac{u}{2\ep^2 (1\!+\!2\ep)^2}
~{}_{3}F_2 \left(\begin{array}{c|}
1, 1\!+\!2\ep, 1\!+\!2\ep \\
\frac{3}{2}\!+\!\ep, 2\!+\!2\ep\end{array} ~\frac{u}{4} \right)
\nonumber \\ && \hspace*{-10mm}
%%%%%%%%%%%%%%%%%%%%%%%%%%%%%%%%%%%%%%%%%%%%%%%%%%%%%%%%%%%%%%%%%%%%%%
- u^{1\!-\!2\ep} \frac{\Gamma(1\!-\!\ep) \Gamma(1\!+\!2\ep)}{4\ep^2 \Gamma(1\!+\!\ep)}
~{}_{3}F_2 \left(\begin{array}{c|} 1, 1, 1\!+\!\ep \\
\frac{3}{2}, 2 \end{array} ~\frac{u}{4} \right)
%%%%%%%%%%%%%%%%%%%%%%%%%%%%%%%%%%%%%%%%%%%%%%%%%%%%%%%%%%%%%%%%%%%%%%
+ \frac{\Gamma(1\!-\!\ep) \Gamma^2(1\!+\!2\ep) \Gamma(1\!+\!3\ep)}
{4\ep^4 \Gamma^2(1\!+\!\ep)\Gamma(1\!+\!4\ep)}
\nonumber \\ && \hspace*{-10mm}
%%%%%%%%%%%%%%%%%%%%%%%%%%%%%%%%%%%%%%%%%%%%%%%%%%%%%%%%%%%%%%%%%%%%%%
+ \frac{u \Gamma(1\!-\!\ep) \Gamma^2(1\!+\!2\ep)\Gamma(1\!+\!3\ep)}
{4\ep^2 (1\!+\!2\ep)(1\!+\!4\ep)\Gamma^2(1\!+\!\ep) \Gamma(1\!+\!4\ep)}
~{}_{3}F_2 \left(\begin{array}{c|}
1, 1\!+\!2\ep, 1\!+\!3\ep \\
\frac{3}{2}\!+\!2\ep, 2\!+\!2\ep\end{array} ~\frac{u}{4} \right)
- \frac{1}{4\ep^4} \; ,
\end{eqnarray}
where, as before, $u = 4 z = 4 \sin^2\tfrac{\theta}{2}$.
All hypergeometric functions occurring in Eq.~(\ref{D4_hypergeom})
belong to the type considered in section~2. 
Applying the results for the $\ep$-expansion of these functions, 
we arrive at 
\begin{eqnarray}
&& \hspace*{-13mm}
(m^2)^{3 \ep} (1-\ep) (1-2\ep) {D_4}(1,1,1,1,1,1;u) 
\nonumber \\ && \hspace*{-10mm}
=  
\frac{2 \zeta_3}{\ep} - 9 \zeta_4
+ \sum_{j=1}^{\infty} \frac{u^j}{\left( 2j \atop j\right) } 
\Biggl\{ 
- \frac{1}{2 j^2} \ln^2 u 
+  \frac{3}{j^3} \ln u
- \frac{5}{j^4}
- \frac{1}{j^2} \zeta_2
\nonumber \\ && 
+ \frac{4}{j^3} S_1
- \frac{4}{j^3} \bar{S}_1
+ \frac{2}{j^2} S_1^2
- \frac{4}{j^2} S_1 \bar{S}_1
- \frac{1}{j^2} S_2
+ \frac{2}{j^2} \left(\bar{S}_1^2 + \bar{S}_2  \right)
+ {\cal O} (\ep) 
\Biggr\}
\nonumber \\ && \hspace*{-10mm}
= \! 
\frac{2 \zeta_3}{\ep} 
\!+\! 2 \LS{4}{1}{\theta}
\!+\! 8 l_\theta \left[\Cl{3}{\theta}  \!-\! \zeta_3 \right]
\!-\! 2 \theta \Ls{3}{\theta}
\!-\! 6 \left[ \Ls{2}{\theta} \right]^2
\!+\! \tfrac{1}{12} \theta^4
\!-\! \tfrac{1}{2} \zeta_2 \theta^2 
\!-\! 9 \zeta_4
+ {\cal O} (\ep) . \hspace*{4mm}
\label{D4_theta}
\end{eqnarray}
Using analytic continuation described in section~3, we can also
present this result in terms of the variable $y$ 
defined in (\ref{y<->u}),
\begin{eqnarray}
&& \hspace*{-15mm}
(m^2)^{3 \ep} (1-\ep) (1-2\ep) {D_4}(1,1,1,1,1,1;u) 
\nonumber \\ && \hspace*{-10mm}
=  
\frac{2 \zeta_3}{\ep} 
+ \tfrac{1}{4} \ln^2 u \ln^2 y 
+ \ln u \biggl[
  6 \Li{3}{y} 
- 6 \ln y \Li{2}{y} 
+ \tfrac{1}{2} \ln^3 y 
- 3 \ln^2 y \ln (1-y) 
- 6 \zeta_3 
\biggr]
\nonumber \\ && 
+ 4 \Li{4}{y}
- 4 \Snp{2,2}{y}
+ 6 \left[\Li{2}{y} \right]^2
- 4 \ln (1-y) \Li{3}{y} 
+ 12 \ln y \ln (1-y) \Li{2}{y}
\nonumber \\ &&
- 3 \ln^2 y \Li{2}{y}
+ 5 \ln^2 y \ln^2 (1-y) 
- \tfrac{7}{3} \ln^3 y \ln (1-y) 
+ \tfrac{1}{4} \ln^4 y
\nonumber \\ && 
- 12 \zeta_2 \Li{2}{y}
- 8 \zeta_2 \ln y \ln (1-y)           
+ \tfrac{3}{2} \zeta_2 \ln^2 y 
+ 4 \zeta_3 \ln (1-y) 
+ 3 \zeta_4 
+ {\cal O}(\ep).
\label{D4_y}
\end{eqnarray}
One should remember however, that in this case the variable $y$
becomes complex (although all imaginary parts should cancel 
for real masses), whereas Eq.~(\ref{D4_theta}) is explicitly real. 

As a non-trivial check on these results we consider two 
particular values, $M^2=m^2$ and $M^2=0$. 
In the first case ($\theta=\tfrac{\pi}{3}$), 
we reproduce the known result (see in Ref.~\cite{B99}) 
for the master integral ${\rm \bf D}_4\equiv{D_4}(1,1,1,1,1,1;u)|_{u \to 1}$,
\begin{equation}
(m^2)^{3 \ep} (1-\ep) (1-2\ep) \left. {D_4}(1,1,1,1,1,1;u) \right|_{u \to 1}
=\frac{2 \zeta_3 }{\ep} 
- \tfrac{77}{12}\zeta_4 
- 6 \left[ \Ls{2}{\tfrac{\pi}{3}} \right]^2 + {\cal O}(\ep) \;. 
\end{equation}
In the second case, $M^2=0$ ($\theta=0$, $y\to 1$), 
the result for the master integral $B_M$ is reproduced
(which was first calculated in Ref.~\cite{B92}),  
\begin{equation}
(m^2)^{3 \ep} (1\!-\!\ep) (1\!-\!2\ep) 
\left. {D_4}(1,1,1,1,1,1;u) \right|_{u \to 0}
= (m^2)^{3 \ep} (1\!-\!\ep) (1\!-\!2\ep) B_M 
= \frac{2 \zeta_3 }{\ep} - 9 \zeta_4 + {\cal O}(\ep)  \;. 
\end{equation}

In a similar manner, analytical results can be deduced for other 
diagrams with two different mass scales,
like $D_3$ and $E_3$ (for notations, see in Ref.~\cite{DK01}). 
To our knowledge, this is the first example of the calculation of the finite part 
of a three-loop vacuum diagram with six internal lines with two different mass scales. 
In previous publications~\cite{three-different}, only the divergent parts of some 
three-loop vacuum integrals have been analyzed. 

%%%%%%%%%%%%%%%%%%%%%%%%%%%%%%%%%%%%%%%%%%%%%%%%%%%%%%%%%%%%%%%%%%%%%%%%%%%%%%%
\section{Conclusion}

In this paper, we have studied the
{\it multiple inverse binomial sums} of the type~(\ref{binsum}), 
for arbitrary values of the variable~$u$.
Our analysis was based on the connection between 
these sums and terms of the expansion of certain
hypergeometric functions with respect to the parameter~$\ep$.
Using known results for the hypergeometric functions,
together with the integral representation~(\ref{higher_c}),
we have obtained a number of new analytical results
up to the weight {\bf 4}, Eqs.~(\ref{S1a})--(\ref{last4}),
(\ref{S1c2})--(\ref{S2bar_2}) and (\ref{S1_3})--(\ref{S1bar_3}).
Moreover, in some cases like (\ref{S2G})--(\ref{S4G})
the results can be obtained for an arbitrary weight.
Constructing analytical continuation of the obtained results, 
we have expressed them in terms of the generalized
polylogarithms (\ref{AN_S1S1_2})--(\ref{AN_Sb1_3}).
In the cases considered, only one new function was needed,
in addition to the basis of Nielsen polylogarithms.
As such function one can take, e.g., the harmonic polylogarithm 
$H_{-1,0,0,1}(-y)$, where the variable $y$ is given in
Eq.~(\ref{y<->u}).

This approach allowed us to construct some terms of the 
$\ep$-expansion of the generalized hypergeometric function 
${}_{P+1}F_P$ (\ref{expansion}) and 
obtain new analytical results for higher terms of the 
$\ep$-expansion of some one- and two-loop 
propagator-type (see Eqs.~(\ref{f10101}), (\ref{j011_122}) and (\ref{j011_c})) 
and vertex-type (see Eqs.~(\ref{Hgg1}) and (\ref{Hgg2})) 
diagrams depending on one dimensionless variable $u=p^2/m^2$.
As a by-product, we have analytically proven the 
earlier published results for the three-loop vacuum integrals, 
the finite part of ${\bf D_4}$ and 
the $\ep$-part of ${\bf D_3}$ and ${\bf E_3}$,
corresponding to the particular value of $u=1$ ($z=\tfrac{1}{4}$),
see Sections~4.3, 4.5 and 4.6 in Ref.~\cite{DK01} and Refs.~\cite{B99,odd,d_3}.
These integrals enter as master integrals in Avdeev's 
package~\cite{leo96} and {\sf MATAD}~\cite{matad}. 
Moreover, the developed technique is useful in 
the calculation of two-loop vertex-type 
diagrams~\cite{vertex_DS,vertex_BMR}.

\vspace{5mm}

\noindent
{\bf Acknowledgements.}
We are grateful to F.~Jegerlehner, A.~Kotikov, 
M.~Spira, V.A.~Smirnov and O.V.~Tarasov for useful discussions.
We would like to thank R.~Delbourgo and J.~Fleischer for their
interest in our work, and
J.~Gluza for checks of some formulae.
We are very thankful to G.~Passarino and S.~Uccirati for
an independent numerical check of some expressions from sections~3
and 4 which helped to identify typos in earlier versions of
Eqs.~(3.11) and (\ref{Hgg2}). 
This research was supported in part by INTAS-CERN grant No.~99-0377
and by the Australian Research Council grant No.~A00000780.

%%%%%%%%%%%%%%%%%%%%%%%%%%%%%%%%%%%%%%%%%%%%%%%%%%%%%%%%%%%%%%%%%%%%%%%%%%%%%%%
\appendix

%%%%%%%%%%%%%%%%%%%%%%%%%%%%%%%%%%%%%%%%%%%%%%%%%%%%%%%%%%%%%%%%
\section{Harmonic polylogarithms of complex arguments}
\label{HPCA}
\setcounter{equation}{0}

Here we collect some properties 
of the harmonic polylogarithms introduced in~\cite{RV00}. 
Let us define a $w$-dimensional vector $\vec{a}=(a,\vec{b})$, where
$a$ is the leftmost component of $\vec{a}$, while $\vec{b}$ stands
for the vector of the remaining $(w-1)$ components. The harmonic
polylogarithms of weight $w$ are then defined as follows:
\begin{equation}
\H_{\vec{a}}(y) = \int\limits_0^y \d x \ \f(a;x) \ \H_{\vec{b}}(x) \ ,
\label{eq:def}
\end{equation}
where the three rational fractions $\f(a;x)$ are given by
\[
   \f(+1;x) = \frac{1}{1-x} \;, \quad  
   \f(0;x)  = \frac{1}{x} \;, \quad  
   \f(-1;x) = \frac{1}{1+x} \; .
\]
The derivatives can be written in a compact form
\begin{equation}
\frac{\d}{\d y} \H_{\vec{a}}(y) = \f(a;y) \H_{\vec{b}}(y) \; .
\label{eq:derive}
\end{equation}

Let us put $y=e^{{\rm i}\theta}$, so that 
\begin{equation}
\H_{\vec{a}}(e^{{\rm i}\theta})  
= \H_{\vec{a}}(1) 
+ {\rm i} \int\limits_0^\theta \d \phi \ e^{{\rm i}\phi}\;
\f(a;e^{{\rm i}\phi}) 
\ \H_{\vec{b}}(e^{{\rm i}\phi})  \; ,
\end{equation}
where 
\begin{eqnarray}
   e^{{\rm i}\phi} \f(+1;e^{{\rm i}\phi}) &=& 
   \frac{e^{{\rm i}(\pi+\phi)/2}}{2 \sin \tfrac{\phi}{2} } 
   = - \frac{1}{2} 
\left( 1 - {\rm i} \cot \tfrac{\phi}{2}  \right)  
\; ,
\nonumber\\
   e^{{\rm i}\phi} \f(0;e^{{\rm i}\phi})  &=&  1 \; , 
\nonumber\\
   e^{{\rm i}\phi} \f(-1;e^{{\rm i}\phi}) &=& 
   \frac{e^{{\rm i}\phi/2}}{2 \cos \tfrac{\phi}{2} } 
   = \frac{1}{2} 
\left( 1 + {\rm i} \tan \tfrac{\phi}{2} \right) 
\; .
\end{eqnarray}

Consider, for example, the harmonic polylogarithm $H_{-1,0,0,1}(y)$
given in Eq.~(\ref{h(-1001)}).
Using the decomposition of 
$\Li{3}{e^{{\rm i}\phi}}$ into 
the real and imaginary parts 
\cite{Lewin} and integrating by parts, we obtain 
expressions in terms of Clausen's and generalized log-sine functions,
\begin{eqnarray}
\H_{-1,0,0,1}(e^{{\rm i}\theta})&\!\!=\!\!& 
\H_{-1,0,0,1}(1)  
+ \frac{{\rm i}}{2} \int\limits_0^\theta {\rm d} \phi \; \Li{3}{e^{{\rm i}\phi}}
\left( 1 + {\rm i} \frac{\sin \tfrac{\phi}{2}}{\cos \tfrac{\phi}{2} }  \right) 
\nonumber \\ 
&\!\!=\!\!& \H_{-1,0,0,1}(1)   
- \tfrac{1}{96} \theta^2 (2\pi-\theta)^2
+ \Phi(\theta) 
+ \ln \left(2 \cos \tfrac{\theta}{2} \right) \Cl{3}{\theta} - \zeta_3 \ln2 
\nonumber \\ &&
+ {\rm i} \ln \left(2 \cos \tfrac{\theta}{2} \right) \Gl{3}{\theta}
+ \tfrac{1}{2} {\rm i} \left[ \Cl{4}{\theta} + \Cl{4}{\pi-\theta} \right]
\nonumber \\ && 
- {\rm i} \left[ 
\Gl{2}{\theta}
\Cl{2}{\pi-\theta}
+ \tfrac{1}{2} (\pi- \theta) \Cl{3}{\pi-\theta}
- \tfrac{1}{2}\pi \Cl{3}{\pi}
\right] \;, 
\end{eqnarray}
where 
$$
\Cl{3}{\pi} = -\tfrac{3}{4} \zeta_3 \;, \quad
\Gl{2}{\theta} =  \zeta_2 - \tfrac{1}{2} \pi \theta + \tfrac{1}{4}\theta^2 \;,
\quad 
\Gl{3}{\theta} = \tfrac{1}{12}\theta \left(\pi-\theta \right) 
\left(2\pi-\theta \right) \; ,
$$
and $\Phi(\theta)$ is defined in Eq.~(\ref{def_Phi}).

In this paper, we have also used the following relations:
\begin{eqnarray}
\label{Hrel1}
H_{-1,0,0,1}\left( -y^{-1} \right) &=& H_{-1,0,0,1}(-y)
+ \Li{4}{-y^{-1}}
+ \Li{4}{y} 
- \Li{3}{y} \ln y 
- \zeta_2 \Li{2}{1-y} 
\nonumber \\ && 
+ \tfrac{1}{2} \ln^2 y \Li{2}{y} 
+ \tfrac{1}{6} \ln^3 y \ln (1-y)
- \tfrac{1}{8} \zeta_4
- {\rm i} \sigma \pi \tfrac{3}{4} \zeta_3 \; 
\end{eqnarray}
and 
\begin{equation}
\label{Hrel2}
H_{-1,0,0,1}(y) +  H_{-1,0,0,1}(-y) 
= \Li{2}{y} \Li{2}{-y}
+ \ln (1+y) \Li{3}{y}
+ \ln (1-y) \Li{3}{-y} \; .
\end{equation}
%%%%%%%%%%%%%%%%%%%%%%%%%%%%%%%%%%%%%%%%%%%
%         int_li3
%%%%%%%%%%%%%%%%%%%%%%%%%%%%%%%%%%%%%%%%%%%%
The following two representations of $H_{-1,0,0,1}(y)$ are also useful:
\begin{eqnarray}
H_{-1,0,0,1}(y)
& = & \Li{4}{y}-{\textstyle{3\over4}} \zeta_3 \ln(1+y)
+ {\textstyle{1\over2}}
\int\limits_0^1 \frac{\mbox{d} x\; \ln^2{x}\; \ln(1-xy)}{1+x} \; , 
\\
H_{-1,0,0,1}(y)
& = & 
-{\textstyle{3\over4}} \zeta_3 \ln(1+y)
+ {\textstyle{1\over16}} \sum\limits_{l=1}^{\infty}
\frac{y^l}{l} \left[ \psi''\left(\frac{l+1}{2}\right)
-\psi''\left(\frac{l}{2}\right) \right] .
\end{eqnarray}

Instead of $\Phi(\theta)$, 
one could also introduce another function. One of such possibilities
is to consider a generalization of the Glashier function.
Let us recall that the real part of 
$\Li{j}{e^{{\rm i}\theta}}$ can be presented as\footnote{We use
the standard notation $\Li{j}{r,\theta}=
{\rm Re}\left[\Li{j}{re^{{\rm i}\theta}}\right]$
(see in Ref.~\cite{Lewin}).}
\begin{equation}
\Li{j}{1,\theta} =
\frac{(-1)^{j-1}}{2(j-2)!}\int\limits_0^1
\frac{{\rm d}\xi}{\xi}\; \ln^{j-2}\xi\; 
\ln\left(1-2\xi\cos\theta+\xi^2\right)
=
\Biggl\{
\begin{array}{l}
\Cl{j}{\theta}, \quad j\;\;{\rm odd}\\
\Gl{j}{\theta}, \quad j\;\;{\rm even}
\end{array}
\end{equation}
where $\Cl{j}{\theta}$ and $\Gl{j}{\theta}$ are
Clausen and Glashier functions, respectively.
In particular, $\Gl{j}{\theta}$ is just a polynomial
in $\theta$.

A possible non-trivial generalization of $\Gl{j}{\theta}$
(for even $j$) could be
\begin{equation}
\Gl{j}{\theta; a} =
-\frac{1}{2(j-2)!}\int\limits_0^1
\frac{{\rm d}\xi}{\xi+a}\; \ln^{j-2}\xi\; 
\ln\left(1-2\xi\cos\theta+\xi^2\right), \quad
j\;\;{\rm even},
\end{equation}
so that $\Gl{j}{\theta; 0}=\Gl{j}{\theta}$.
In particular,
\begin{equation}
\label{Gl2_gen}
\Gl{2}{\theta; a} = -\ln{b}\;\ln\frac{a+1}{a}
+ \Li{2}{\frac{a+1}{b},\widetilde{\theta}}
- \Li{2}{\frac{a}{b},\widetilde{\theta}}\; ,
\end{equation}
where
\begin{equation}
\cos{\widetilde{\theta}}
=\frac{a+\cos\theta}{\sqrt{1+2a\cos\theta+a^2}}, 
\end{equation}
so that
\begin{equation}
a = \frac{\sin(\theta-{\widetilde{\theta}})}
         {\sin{\widetilde{\theta}}}, \quad
b = \sqrt{1+2a\cos\theta+a^2} 
= \frac{\sin\theta}{\sin{\widetilde{\theta}}} \; .
\end{equation}
Using Eqs.~(17) and (18) on p.~293 of~\cite{Lewin},
we can see that the general result (\ref{Gl2_gen})
simplifies in the case $a=1$ (${\widetilde{\theta}}=\tfrac{\pi}{2}$),
\begin{equation}
\Gl{2}{\theta; 1} = 
-\tfrac{1}{4}\; \Li{2}{\cos^2{\tfrac{\theta}{2}}}
+ \tfrac{1}{8}(\pi-\theta)^2 
- \tfrac{1}{2} \ln^2{2} \; .
\end{equation}

Then, let us consider
\begin{equation}
\label{Gl4(theta,1)}
\Gl{4}{\theta; 1} = -\frac{1}{4}
\int\limits_0^1 \frac{{\rm d}\xi}{\xi+1}\;
\ln^2\xi\; \ln\left(1-2\xi\cos\theta+\xi^2\right) \; . 
\end{equation}
For $\theta=0$ we get
\begin{equation}
\Gl{4}{0; 1} = U_{3,1} - \zeta_4 \; ,
\end{equation}
where $U_{3,1}$ is the alternating two-fold Euler sum 
considered in~\cite{euler,B99} (see also Ref.~\cite{GKP}), 
$$
U_{3,1} = 
- 2 \Li{4}{\tfrac{1}{2}} + \tfrac{1}{2} \zeta_4
- \tfrac{1}{12} \ln^4 2 + \tfrac{1}{2} \zeta_2 \ln^2 2 \;. 
$$

For general $\theta$,
one can see that $\Gl{4}{\theta;1}$ is related to the real part
of a harmonic polylogarithm,
\begin{equation} 
{\rm Re}H_{-1,0,0,1}\big(e^{{\rm i}\theta}\big)
= \Gl{4}{\theta} - \Gl{4}{\theta; 1}
- \tfrac{3}{4} \zeta_3 \ln\left(2\cos\tfrac{\theta}{2}\right).
\end{equation}
In particular, the function $\Phi(\theta)$
can be presented as
\begin{equation}
\Phi(\theta) = -\Gl{4}{\theta; 1} + \Gl{4}{0; 1}
+\tfrac{7}{4}\zeta_3\ln{2} - \tfrac{1}{96}\theta^2(2\pi-\theta)^2
-\left[\Cl{3}{\theta}-\Cl{3}{\pi}\right]
\ln\left(2\cos\tfrac{\theta}{2}\right) ,
\end{equation}
where $\Cl{3}{\pi}=-\tfrac{3}{4}\zeta_3$.

Using the symmetry property of $\Phi(\theta)$, Eq.~(\ref{sym_Phi}), 
we get
\begin{eqnarray}
\Gl{4}{\theta; 1} + \Gl{4}{\pi-\theta; 1} &=&
\tfrac{11}{8}\zeta_4 - \tfrac{1}{48}\theta^2(\pi-\theta)^2
-\Cl{2}{\theta}\;\Cl{2}{\pi-\theta}
\nonumber \\ &&
-\left[\Cl{3}{\theta}-\Cl{3}{\pi}\right]
\ln\left(2\cos\tfrac{\theta}{2}\right)
\nonumber \\ &&
-\left[\Cl{3}{\pi-\theta}-\Cl{3}{\pi}\right]
\ln\left(2\sin\tfrac{\theta}{2}\right).
\label{sym_Gl4}
\end{eqnarray}
For special values of $\theta$, Eq.~(\ref{sym_Gl4}) yields
\begin{eqnarray}
\Gl{4}{\pi; 1} &=& -U_{3,1} + \tfrac{19}{8}\zeta_4
-\tfrac{7}{4}\zeta_3 \ln{2} ,
\\
\Gl{4}{\tfrac{\pi}{2}; 1} &=& \tfrac{161}{256}\zeta_4
-\tfrac{21}{64}\zeta_3\ln{2} - G^2 , 
\\
\Gl{4}{\tfrac{\pi}{3}; 1}+\Gl{4}{\tfrac{2\pi}{3}; 1}
&=& \tfrac{277}{216}\zeta_4 - \tfrac{13}{24}\zeta_3\ln{3}
-\tfrac{2}{3}\left[ \Cl{2}{\tfrac{\pi}{3}} \right]^2 \; ,
\end{eqnarray}
where $G$ is the Catalan constant.
Moreover, with the help of {\sf PSLQ} algorithm \cite{PSLQ} one can
obtain results for $\Gl{4}{\tfrac{\pi}{3}; 1}$ and 
$\Gl{4}{\tfrac{2\pi}{3}; 1}$ separately,
\begin{eqnarray}
\Gl{4}{\tfrac{\pi}{3}; 1}
&=& \tfrac{259}{108} \zeta_4
- \tfrac{13}{24} \zeta_3 \ln{3}
-\tfrac{1}{3}\left[ \Cl{2}{\tfrac{\pi}{3}} \right]^2
+ \tfrac{1}{4} \pi \Ls{3}{\tfrac{2\pi}{3}}
- \tfrac{3}{8} \LS{4}{1}{\tfrac{2\pi}{3}} \; ,
\\
\Gl{4}{\tfrac{2\pi}{3}; 1}
&=& -\tfrac{241}{216} \zeta_4
-\tfrac{1}{3}\left[ \Cl{2}{\tfrac{\pi}{3}} \right]^2
- \tfrac{1}{4} \pi \Ls{3}{\tfrac{2\pi}{3}}
+ \tfrac{3}{8} \LS{4}{1}{\tfrac{2\pi}{3}} \; .
\end{eqnarray}

We would also like to discuss the analytical continuation 
of $\Lsc{2,3}{\theta}$ (see Eq.~(\ref{Lsc})),
\[
\Lsc{2,3}{\theta}=-\int\limits_0^{\theta}
{\rm d}\phi\; \ln \Bigl|2\sin\tfrac{\phi}{2}\Bigr| \;
\ln^2 \Bigl|2\cos\tfrac{\phi}{2}\Bigr| \; .
\]
If we introduce a variable $z=e^{{\rm i}\sigma \phi}$, we see that
\begin{equation}
\ln \left( 2\sin \tfrac{\phi}{2} \right)
\leftrightarrow \ln(1-z) - \tfrac{1}{2} \ln z 
+ \tfrac{1}{2}{\rm i}\sigma\pi, \quad
\ln \left( 2\cos \tfrac{\phi}{2} \right)
\leftrightarrow \ln(1+z) - \tfrac{1}{2} \ln z,
\label{begin_AN_Lsc}
\end{equation}
and the analytical continuation of $\Lsc{2,3}{\theta}$
is given by
\begin{eqnarray}
{\rm i} \sigma \Lsc{2,3}{\theta}
\!= \! \int\limits_y^1 \frac{{\rm d}z}{z}
\Bigl\{ 
 \ln (1\!-\!z) 
\!-\! \tfrac{1}{2} \ln z
\!+\! \tfrac{1}{2} {\rm i} \sigma \pi
\Bigr\} 
\Bigl[ \ln^2 (1 \!+\!z) 
\!-\! \ln z \ln (1 \!+\! z) 
\!+\! \tfrac{1}{4} \ln^2 z
\Bigr] \;,
\nonumber 
\end{eqnarray}
with $y$ defined in (\ref{y:def}). 
This integral can be calculated by using the following relations
\begin{eqnarray} 
\int\limits_0^y \frac{{\rm d}z}{z} \ln(1+z) \ln (1-z) \ln z   
&=& 
  \Snp{2,2}{y} 
+ \Snp{2,2}{-y}
- \tfrac{1}{4} \Snp{2,2}{y^2} 
\nonumber \\ && 
+ \ln y \Bigl[ \tfrac{1}{2} \Snp{1,2}{y^2} 
- \Snp{1,2}{y} 
- \Snp{1,2}{-y} 
        \Bigr ] 
\;,  
\\ 
\int\limits_0^y \frac{{\rm d}z}{z} \ln^2 (1\mp z) \ln (1\pm z)    
&=&
2 \ln(1\pm y) \Snp{1,2}{\pm y} - 2 \H_{-1,0,1,1}(\pm y) ,
\end{eqnarray} 
where
\begin{equation}
\H_{-1,0,1,1}(y) \equiv \int\limits_0^y \frac{{\rm d}z}{1+z}\;
\Snp{1,2}{z}
\end{equation}
is another harmonic polylogarithm \cite{harmonic}. We note that
\begin{equation}
\H_{-1,0,1,1}(\pm y) = 
\tfrac{1}{4} \Snp{1,3}{y^2}-\Snp{1,3}{\mp y}+\ln(1\pm y)\Snp{1,2}{\pm y}
\pm \tfrac{1}{12}\int\limits_0^y \frac{{\rm d}z}{z}\; 
\ln^3\left(\frac{1-z}{1+z}\right) ,
\label{new_H_via_Li}
\end{equation}
\begin{eqnarray}
\H_{-1,0,1,1}(y) + \H_{-1,0,1,1}(-y) 
&=&
\tfrac{1}{2} \Snp{1,3}{y^2} - \Snp{1,3}{y} - \Snp{1,3}{-y}
\nonumber \\ &&
+ \ln(1+y) \Snp{1,2}{y} + \ln(1-y) \Snp{1,2}{-y} \; .
\end{eqnarray}

Therefore, the analytic continuation of $\Lsc{2,3}{\theta}$ reads
\begin{eqnarray}
\label{AC_of_Lsc}
{\rm i}\sigma \Lsc{2,3}{\theta} &=&
\Snp{2,2}{y}-\tfrac{1}{4}\Snp{2,2}{y^2}
-2\Snp{1,3}{y}+\tfrac{1}{2}\Snp{1,3}{y^2}
+\tfrac{1}{2}\Li{4}{y}+\Li{4}{-y}
\nonumber \\ &&
+\ln{y} \left[ \tfrac{1}{2}\Snp{1,2}{y^2} - \Snp{1,2}{y}
-\tfrac{1}{2}\Li{3}{y}-\Li{3}{-y} \right]
\nonumber \\ &&
+ \ln^2{y} \left[ \tfrac{1}{4} \Li{2}{y}+\tfrac{1}{2}\Li{2}{-y} \right]
+\tfrac{1}{32}\ln^4{y}
-\tfrac{3}{16}\zeta_4
-\tfrac{1}{6} \int\limits_0^y \frac{{\rm d}z}{z}\; 
\ln^3\left(\frac{1-z}{1+z}\right)
\nonumber \\ &&
-{\rm i}\sigma\pi \left\{ 
\Snp{1,2}{-y} 
- \tfrac{1}{2}\Li{3}{-y}
+\tfrac{1}{2}\ln{y} \Li{2}{-y} 
+\tfrac{1}{24}\ln^3{y}
-\tfrac{1}{2}\zeta_3 \right\} \; .
\end{eqnarray}

Note that the integral occurring in Eqs.~(\ref{new_H_via_Li}) and 
(\ref{AC_of_Lsc}) is directly related to the integral in 
Eq.~(\ref{Lsc23_via_int}), and it
can be calculated in terms of polylogarithms,
\begin{eqnarray}
\label{I(z)viaLi4}
\int\limits_0^y \frac{{\rm d}z}{z}\; 
\ln^3\left(\frac{1-z}{1+z}\right)
&=&
-\tfrac{45}{4}\zeta_4
+ 6 \left[ \Li{4}{\omega} - \Li{4}{-\omega} \right]
- 6 \ln{\omega} \left[ \Li{3}{\omega} - \Li{3}{-\omega} \right]
\nonumber \\ &&
+3 \ln^2{\omega} \left[ \Li{2}{\omega} - \Li{2}{-\omega} \right]
+ \ln^3{\omega} \ln{y} \; ,
\hspace*{8mm}
\end{eqnarray}
where $\omega=(1-y)/(1+y)$ (see Eq.~(\ref{def_omega})).
Therefore, $\H_{-1,0,1,1}(\pm y)$ reduces to polylogarithms, as well
as the analytic continuation of $\Lsc{2,3}{\theta}$. We note that
the imaginary part on the r.h.s.\ of Eq.~(\ref{AC_of_Lsc}) gets 
cancelled in the results for the corresponding sums.

The following relations are useful for the transformations:
\begin{eqnarray}
%%%%%%%%%%%%%%%%%%%%%%%%%%%%%%%%%%%%%%%%%%%%%%%%%%%%%%%%%%%%%%%%%%%%%% 
\Li{2}{\omega} - \Li{2}{- \omega}
&=& -\Li{2}{y}+\Li{2}{-y}-\ln\omega \ln{y} + \tfrac{3}{2}\zeta_2 \; , 
\\
\Li{3}{\omega} - \Li{3}{- \omega}
&=& 
 \tfrac{1}{2} \Snp{1,2}{y^2} 
- 2  \Snp{1,2}{y} 
- 2  \Snp{1,2}{-y} 
- \ln \omega \left[ \Li{2}{y} - \Li{2}{- y} \right] 
\nonumber \\ && 
- \tfrac{1}{2} \ln^2 \omega  \ln y
+ \tfrac{3}{2} \zeta_2 \ln \omega
+ \tfrac{7}{4} \zeta_3 \;.
%%%%%%%%%%%%%%%%%%%%%%%%%%%%%%%%%%%%%%%%%%%%%%%%%%%%%%%%%%%%%%%%%%%%%%
\end{eqnarray}

%%%%%%%%%%%%%%%%%%%%%%%%%%%%%%%%%%%%%%%%%%%%%%%%%%%%%%%%%%%%%%%%%%%%%%%%%%%%%%%%%%%%%%%%%
\section{Additional identities between inverse binomial sums}
\label{AIBIBN}
\setcounter{equation}{0}
%=====================================================================
As we have seen, the $\ep$-expansion of the
hypergeometric functions
produces series of the type (\ref{binsum}).
One can use certain properties of hypergeometric functions
to get relations between the sums (\ref{binsum}) involved
in the $\ep$-expansion. 

Let us consider Eq.~(B.18) of Ref.~\cite{DK01}
(which follows from Eq.~(22) on p.~498 of~\cite{PBM3}),
\begin{equation}
{}_{3}F_2 \left(\begin{array}{c|}
1+a_1 \ep, 1+ a_2 \ep, 1+ \tfrac{1}{2}(a_1+a_2)\ep  \\
\tfrac{3}{2}+ \tfrac{1}{2}(a_1+a_2)\ep, 2+(a_1+a_2)\ep \end{array} ~z
\right) =
(1\!-\!z) \left[
{}_{2}F_1 \left(\begin{array}{c|} 1+\tfrac{1}{2} a_1 \ep,
1+\tfrac{1}{2} a_2 \ep \\
\tfrac{3}{2}+\tfrac{1}{2}(a_1+a_2)\ep \end{array} ~z \right)
\right]^2 .
\label{less_trivial}
\end{equation}
It reduces the given $_3F_2$ function to a square of the $_2F_1$
function.
Substituting the $\ep$-expansions of the $_3F_2$ and $_2F_1$ functions 
into (\ref{less_trivial}), we
obtain the following relations between the sums:
\begin{eqnarray}
&&
\Sigma_{-;-;2}^{-;-}(u) = \frac{4-u}{2u} 
\left[ \Sigma_{-;-;1}^{-;-}(u)\right]^2 \; ,
\\ &&
\Sigma_{1;-;2}^{1;-}(u) - \Sigma_{-;1;2}^{-;1}(u) - \Sigma_{-;-;3}^{-;-}(u)
= \frac{4-u}{u} \; \Sigma_{-;-;1}^{-;-}(u)
\left[
\Sigma_{1;-;1}^{1;-}(u) - \Sigma_{-;1;1}^{-;1}(u) \right] \; ,
\\ &&
\Sigma_{2;-;2}^{1;-}(u) = \frac{4-u}{4u} \; \Sigma_{-;-;1}^{-;-}(u)
\Sigma_{2;-;1}^{1;-}(u) \; ,
\\ &&
\Sigma_{1;-;2}^{2;-}(u) - 2 \Sigma_{1;1;2}^{1;1}(u)
+\Sigma_{-;1;2}^{-;2}(u) + \Sigma_{-;2;2}^{-;1}(u) 
-2\Sigma_{1;-;3}^{1;-}(u) + 2\Sigma_{-;1;3}^{-;1}(u)
+2\Sigma_{-;-;4}^{-;-}(u) 
\nonumber \\ && \hspace*{10mm}
= \frac{4-u}{u}
\Bigl[
\Sigma_{-;-;1}^{-;-}(u)
\left( \Sigma_{1;-;1}^{2;-}(u) - 2 \Sigma_{1;1;1}^{1;1}(u)
+\Sigma_{-;1;1}^{-;2}(u) + \Sigma_{-;2;1}^{-;1}(u) 
-\tfrac{3}{8} \Sigma_{2;-;1}^{1;-}(u)\right)
\nonumber \\ && \hspace*{24mm}
+ \left( \Sigma_{1;-;1}^{1;-}(u) - \Sigma_{-;1;1}^{-;1}(u)\right)^2 
\Bigl] \; ,
\end{eqnarray}
with $u=4z$, so that
\[
\frac{4-u}{u} = \frac{1-z}{z} 
= \cot^2\tfrac{\theta}{2} \; . 
\]

Another interesting relation, Eq.~(B.19) of Ref.~\cite{DK01} 
(which follows from Eq.~(20) on p.~498 of~\cite{PBM3}), reads
\begin{eqnarray}
&& \hspace*{-10mm}
{}_{3}F_2 \left(\begin{array}{c|}
1+a_1 \ep, 1+ a_2 \ep, 1+ \tfrac{1}{2}(a_1+a_2)\ep  \\
\tfrac{3}{2}+ \tfrac{1}{2}(a_1+a_2)\ep, 1+(a_1+a_2)\ep \end{array} ~z
\right) =
{}_{2}F_1 \left(\begin{array}{c|} 1+\tfrac{1}{2}a_1\ep,
1+\tfrac{1}{2} a_2 \ep \\
\tfrac{3}{2}+\tfrac{1}{2}(a_1+a_2)\ep \end{array} ~z \right)
\nonumber \\ &&     
\times \left\{ 1 + \frac{a_1 a_2 \ep^2 z}
{2\left[1+(a_1+a_2)\ep\right]}\;
{}_{3}F_2 \left(\begin{array}{c|}
1+\tfrac{1}{2} a_1 \ep, 1+\tfrac{1}{2} a_2 \ep, 1 \\
\tfrac{3}{2}+ \tfrac{1}{2}(a_1+a_2)\ep, \; 2 \end{array} ~z \right)
\right\} \; .
\label{even_less_trivial}
\end{eqnarray}
In this way, we get the following relations:
\begin{eqnarray}
&&
\Sigma_{2;-;1}^{1;-}(u) = \tfrac{1}{3}
\Sigma_{-;-;1}^{-;-}(u) \Sigma_{-;-;2}^{-;-}(u) \; ,
\nonumber \\
&& 
3\Sigma_{1,2;-;1}^{1,1;-}(u)
-3\Sigma_{2,1;1}^{1;1}(u) 
- \tfrac{7}{2} \Sigma_{3;-;1}^{1;-}(u)
\nonumber \\ && \hspace*{20mm}
= \Sigma_{-;-;1}^{-;-}(u)
\left[ \Sigma_{1;-;2}^{1;-}(u) - \Sigma_{-;1;2}^{-;1}(u) \right]
+ \Sigma_{-;-;2}^{-;-}(u)
\left[ \Sigma_{1;-;1}^{1;-}(u) - \Sigma_{-;1;1}^{-;1}(u) \right] \; .
\hspace*{7mm}
\end{eqnarray}
We have checked that all above equations are satisfied by
the explicit results for the sums 
listed in this paper.

Let us also present some relations for higher-order sums:
\begin{eqnarray}
\label{higher1}
&& \hspace{-5mm}
\Sigma_{3;-;2}^{1;-}(u) 
+ \Sigma_{2;-;3}^{1;-}(u) 
- \Sigma_{2,1;-;2}^{1,1;-}(u) 
+ \Sigma_{2;1;2}^{1;1}(u) 
= 
\frac{4\!-\!u}{4u}\! 
\left\{
\Sigma_{2;-;1}^{1;-}(u)
\left[ 
  \Sigma_{-;1;1}^{-;1}(u) 
- \Sigma_{1;-;1}^{1;-}(u) 
\right]
\right.
\nonumber \\ && 
\left.
+ 
\Sigma_{-;-;1}^{-;-}(u) 
\left[ 
  \Sigma_{2;1;1}^{1;1}(u) 
- \Sigma_{2,1;-;1}^{1,1;-}(u) 
+ \tfrac{1}{2} \Sigma_{3;-;1}^{1;-}(u) 
\right] 
\right\} , 
\\ 
\label{higher2}
&& \hspace{-5mm}
  \Sigma_{2;-;2}^{2;-}(u) 
- \Sigma_{4;-;2}^{1;-}(u) 
=
\frac{4-u}{16u} 
\left\{  
\left[   \Sigma_{2;-;1}^{1;-}(u) \right]^2
+  \Sigma_{-;-;1}^{-;-}(u)  \left[
  \Sigma_{2;-;1}^{2;-}(u) 
- \Sigma_{4;-;1}^{1;-}(u)   \right]
\right\} \; ,
\nonumber \\ 
\label{higher3}
&& \hspace{-5mm}
\Sigma_{2;-;1}^{2;-}(u) 
-\Sigma_{4;-;1}^{1;-}(u) 
= \tfrac{2}{45} \Sigma_{-;-;1}^{-;-}(u) 
\left\{
\left[ \Sigma_{-;-;2}^{-;-}(u) \right]^2
+ 3 \Sigma_{2;-;2}^{1;-}(u) 
\right\} \;, 
\\
\label{higher4}
&& \hspace{-5mm}
2 \left[\Sigma_{-;-;2}^{-;-}(u)  \right]^2
\left[ 
  \Sigma_{1;-;1}^{1;-}(u) 
- \Sigma_{-;1;1}^{-;1}(u) 
\right]
+ 4  \Sigma_{-;-;2}^{-;-}(u)  \Sigma_{-;-;1}^{-;-}(u) 
\left[ 
  \Sigma_{1;-;2}^{1;-}(u) 
- \Sigma_{-;1;2}^{-;1}(u) 
\right] 
\nonumber \\ && \hspace{-5mm}
+ 6 \Sigma_{2;-;2}^{1;-}(u) 
\left[ 
  \Sigma_{1;-;1}^{1;-}(u) 
- \Sigma_{-;1;1}^{-;1}(u) 
\right]
+ 6  \Sigma_{-;-;1}^{-;-}(u) 
\left[ 
  \Sigma_{2,1;-;2}^{1,1;-}(u) 
- \Sigma_{2;1;-}^{1;1}(u) 
\right] 
\nonumber \\ && \hspace{-5mm}
+ 45  \Sigma_{2;1;1}^{2;1}(u) 
- 45 \Sigma_{2,1;-;1}^{2,1;-}(u) 
+ 45 \Sigma_{4,1;-;1}^{1,1;-}(u) 
- 45 \Sigma_{4;1;1}^{1;1}(u) 
+ 93 \Sigma_{3,2;-;1}^{1,1;-}(u) 
- 93 \Sigma_{5;-;1}^{1;-}(u) 
\nonumber \\ && \hspace{-5mm}
+ 4 \Sigma_{3;-;1}^{1;-}(u)  \Sigma_{-;-;2}^{-;-}(u) 
- 3 \Sigma_{3;-;2}^{1;-}(u)  \Sigma_{-;-;1}^{-;-}(u)  
= 0 \; , 
\end{eqnarray}
where Eqs.~(\ref{higher1}) and (\ref{higher2}) follow 
from Eq.~(\ref{less_trivial}),
while Eqs.~(\ref{higher3}) and (\ref{higher4}) follow 
from Eq.~(\ref{even_less_trivial}).

One more relation can be derived from Eq.~(47) 
on p.~456 of~\cite{PBM3},
\begin{equation}
{}_{2}F_1 \left(\begin{array}{c|}
1+a_1 \ep, 1+ a_2 \ep \\
\tfrac{3}{2}+ \tfrac{1}{2}(a_1+a_2)\ep 
\end{array} ~\sin^2\tfrac{\theta}{2}
\right) =
\cos\theta \;\;
{}_{2}F_1 \left(\begin{array}{c|} 1+\tfrac{1}{2} a_1 \ep,
1+\tfrac{1}{2} a_2 \ep \\
\tfrac{3}{2}+\tfrac{1}{2}(a_1+a_2)\ep \end{array} ~\sin^2\theta \right) .
\label{less_trivial2}
\end{equation}
We can compare the $\ep$-expansions of these $_2F_1$ functions,
considering the coefficients of $\ep^k$ as functions of $\theta$.
Introducing
\[
u=4\sin^2\tfrac{\theta}{2}, \qquad
{\widetilde{u}}=4\sin^2\theta, \qquad
A_1=a_1+a_2, \qquad A_2=a_1^2+a_2^2 \; ,
\]
we get the following relations at orders $\ep^0$ and $\ep^1$:
\begin{eqnarray}
\cot \tfrac{\theta}{2}
\sum_{j=1}^{\infty} \frac{1}{\left(2j\atop j\right)}\frac{u^j}{j}
&=& \tfrac{1}{2} 
\cot \theta
\sum_{j=1}^{\infty} \frac{1}{\left(2j\atop j\right)}
\frac{{\widetilde{u}}^j}{j} \; ,
\\
\cot \tfrac{\theta}{2}
\sum_{j=1}^{\infty} \frac{1}{\left(2j\atop j\right)}\frac{u^j}{j}
\left( \tfrac{3}{2}S_1 - \bar{S}_1 \right)
&=&\tfrac{1}{2} 
\cot \theta
\sum_{j=1}^{\infty} \frac{1}{\left(2j\atop j\right)}
\frac{{\widetilde{u}}^j}{j} 
\left( S_1 - \bar{S}_1 \right) \; .
\end{eqnarray}
At order $\ep^2$, comparison of the coefficients 
of $A_2$ and $A_1^2$ yields
\begin{eqnarray}
&&
\cot \tfrac{\theta}{2}
\sum_{j=1}^{\infty} \frac{1}{\left(2j\atop j\right)}\frac{u^j}{j} S_2
=\tfrac{1}{8} 
\cot \theta
\sum_{j=1}^{\infty} \frac{1}{\left(2j\atop j\right)}
\frac{{\widetilde{u}}^j}{j} S_2 \; ,
\\
&&
\cot \tfrac{\theta}{2}
\sum_{j=1}^{\infty} \frac{1}{\left(2j\atop j\right)}\frac{u^j}{j}
\left[ \tfrac{1}{2}\left(\bar{S}_2+\bar{S}_1^2\right) 
-\tfrac{3}{2} S_1 \bar{S}_1 
-\tfrac{1}{8}S_2 +\tfrac{9}{8}S_1^2 \right]
\nonumber \\ && \hspace*{30mm}
=\tfrac{1}{2} 
\cot \theta
\sum_{j=1}^{\infty} \frac{1}{\left(2j\atop j\right)}
\frac{{\widetilde{u}}^j}{j} 
\left[ \tfrac{1}{2}\left(\bar{S}_2+\bar{S}_1^2\right) 
-S_1 \bar{S}_1 
-\tfrac{1}{8}S_2 +\tfrac{1}{2}S_1^2 \right] \; .
\end{eqnarray}
All these equations are satisfied by analytic expressions
for these sums in terms of $\theta$ from Section~2 of this paper. 
Remember that one needs to substitute $\theta\to 2\theta$ 
as an argument on the r.h.s.

At order $\ep^3$, we have two independent structures,
$A_1^3$ and $A_1 A_2$. It is more convenient, however, 
to compare the coefficients of $A_1^3$ and $A_1(A_1^2-A_2)$.
The first equation coming from the coefficients of $A_1^3$ yields
\begin{equation}
\cot \tfrac{\theta}{2}
\sum_{j=1}^{\infty} \frac{1}{\left(2j\atop j\right)}\frac{u^j}{j}
\left(
\tfrac{3}{16}{\cal C}_0+\tfrac{3}{4}{\cal C}_1+\tfrac{1}{2}{\cal C}_2
\right)
=\tfrac{1}{2} 
\cot \theta
\sum_{j=1}^{\infty} \frac{1}{\left(2j\atop j\right)}
\frac{{\widetilde{u}}^j}{j} 
\left(
\tfrac{1}{24}{\cal C}_0 + \tfrac{1}{2}{\cal C}_1 + \tfrac{1}{2}{\cal C}_2
\right) \; , 
\end{equation}
where ${\cal C}_j$ are the combinations of the harmonic sums
defined in Eqs.~(\ref{combinations2}).
This equation is also satisfied, if we use analytic expressions
for these sums in terms of $\theta$ given in Section~2.

The second equation at order $\ep^3$ comes from the
coefficient of $A_1(A_1^2-A_2)$,
\begin{equation}
\cot \tfrac{\theta}{2}
\sum_{j=1}^{\infty} \frac{1}{\left(2j\atop j\right)}\frac{u^j}{j}
\left( -\tfrac{3}{4}S_1 S_2 + \tfrac{1}{2}S_2 \bar{S}_1
+\tfrac{1}{2}S_3 \right)
=\tfrac{1}{2} 
\cot \theta
\sum_{j=1}^{\infty} \frac{1}{\left(2j\atop j\right)}
\frac{{\widetilde{u}}^j}{j} 
\left( -\tfrac{1}{8}S_1 S_2 + \tfrac{1}{8}S_2 \bar{S}_1
+\tfrac{1}{16}S_3 \right)\; .
\end{equation}
If we introduce two functions 
\begin{eqnarray}
\Psi_1(\theta) &=& \cot\tfrac{\theta}{2}
\sum_{j=1}^{\infty} \frac{1}{\left(2j\atop j\right)}\frac{u^j}{j}
\left(S_1S_2-S_3\right) \; ,
\\
\Psi_2(\theta) &=& \cot\tfrac{\theta}{2}
\sum_{j=1}^{\infty} \frac{1}{\left(2j\atop j\right)}\frac{u^j}{j}
\left(S_3-2S_2\bar{S}_1\right) \; ,
\end{eqnarray}
we obtain an interesting relation between them,
\begin{equation}
3\Psi_1(\theta) + \Psi_2(\theta) 
= \tfrac{1}{4}\Psi_1(2\theta) + \tfrac{1}{8}\Psi_2(2\theta) \; .
\end{equation}
This relation is satisfied
by the explicit results for the sums involved given in 
Eqs.~(\ref{S3c1}), (\ref{S1S2c1}), (\ref{S2S1barc1}).

%%%%%%%%%%%%%%%%%%%%%%%%%%%%%%%%%%%%%%%%%%%%%%%%%%%%%%%%%%%%%%%%%%%%%%%%%%%%%%%%%%
\section{Further results for the inverse binomial sums}
\setcounter{equation}{0}
%=====================================================================

For completeness, in this Appendix we collect some results for 
the {\it multiple inverse binomial sums} of lower weights. 
They can be obtained by applying the operator 
$u ({\rm d}/{\rm d}u)$ to our results presented in 
sections~2 and 3. 
These sums occur in lower terms of the $\ep$-expansion 
of hypergeometric functions given in Eq.~(\ref{expansion}).

First of all, we list explicit results for some particular cases of
the general formulae (\ref{KV}) and (\ref{S2G}), in terms
of the angular variable $\theta$ (\ref{def_theta}): 
%%%%%%%%%%%%%%%%%%%%%%%%%%%%%%%%%%%%%%%%%%%%%%%%%%%%%%%%%%%%%%%%%%%%%%%%%%%
%%%%%%%%%%%%%%%%%%%%%%%%%%%%%%%%%%%%%%%%%%%%%%%%%%%%%%%%%%%%%%%%%%%%%%%%%%%
\begin{eqnarray}
\sum_{j=1}^\infty \frac{1}{\left( 2j \atop j\right) } \frac{u^j}{j}  
&=& 
 \theta \tan\tfrac{\theta}{2} \; ,
\label{j^1}
\\  
%%%%%%%%%%%%%%%%%%%%%%%%%%%%%%%%%%%%%%%%%%%%%%%%%%%%%%%%%%%%%%%%%%%%%%%%%%%
\sum_{j=1}^\infty \frac{1}{\left( 2j \atop j\right) } \frac{u^j}{j^2} 
&=& 
\tfrac{1}{2} \theta^2 \;, 
\label{j^2}
\\ 
%%%%%%%%%%%%%%%%%%%%%%%%%%%%%%%%%%%%%%%%%%%%%%%%%%%%%%%%%%%%%%%%%%%%%%%%%%%
\sum_{j=1}^\infty \frac{1}{\left( 2j \atop j \right) } \frac{u^j}{j^4} 
&=& 
- 2 \LS{4}{1}{\theta}
+ 4 l_{\theta}
\left[\Cl{3}{\theta} + \theta \Cl{2}{\theta} - \zeta_3 \right] 
+\theta^2 l_{\theta}^2 \;,
\label{j^4}
\\ 
%%%%%%%%%%%%%%%%%%%%%%%%%%%%%%%%%%%%%%%%%%%%%%%%%%%%%%%%%%%%%%%%%%%%%%%%%%%
\sum_{j=1}^\infty \frac{1}{\left( 2j \atop j \right) } \frac{u^j}{j^2} S_2 
&=& 
\tfrac{1}{24} \theta^4 \;. 
\label{S2_2}
\end{eqnarray}

Then, let us present a few more complicated examples 
of the results in terms of $\theta$, 
\begin{eqnarray}
&&
%%%%%%%%%%%%%%%%%%%%%%%%%%%%%%%%%%%%%%%%%%%%%%%%%%%%%%%%%%%%%%%%%%%%%%%%%%%
\sum_{j=1}^\infty \frac{1}{\left( 2j \atop j \right) } \frac{u^j}{j} S_1^2 \bar{S}_1
=  
\tan\tfrac{\theta}{2} 
\Biggl\{ 
\Cl{2}{\pi-\theta} 
\left[ 28 L^2_\theta - 8 L_\theta l_\theta + \tfrac{5}{2} \theta^2 \right]
+ \Cl{2}{2 \theta} \left[ 2 L^2_\theta + \tfrac{1}{4} \theta^2 \right]
\nonumber \\ && \hspace{15mm}
+ 4 L_\theta \Ls{3}{ \theta}
- 2 L_\theta \Ls{3}{2 \theta}
+ 4 \Lsc{2,3}{\theta}
- \tfrac{1}{2} \theta^2 \Cl{2}{\theta}
- \Cl{4}{\theta}
\nonumber \\ && \hspace{15mm}
+  \left[\Ls{3}{\pi-\theta} - \Ls{3}{\pi} \right] 
   \left[ 4 l_\theta - 28 L_\theta \right]
- 8 \theta \left[ \Cl{3}{\pi-\theta} - \Cl{3}{\pi} \right]
\nonumber \\ && \hspace{15mm}
+ 8 \left[\Ls{4}{\pi-\theta} - \Ls{4}{\pi}\right]
- 16 \left[\Cl{4}{\pi-\theta} - \Cl{4}{\pi} \right]
\nonumber \\ && \hspace{15mm}
+ \tfrac{1}{6} \theta^3 l_\theta
- \tfrac{2}{3} \theta^3 L_\theta
- 8 \theta L^3_\theta
+ 13 \zeta_3 \theta 
+ 4 \theta l_\theta  L^2_\theta
\Biggr\} 
\label{S1S1Sb1_1}
\\ && 
%%%%%%%%%%%%%%%%%%%%%%%%%%%%%%%%%%%%%%%%%%%%%%%%%%%%%%%%%%%%%%%%%%%%%%%%%%%
\sum_{j=1}^\infty \frac{1}{\left( 2j \atop j \right) } \frac{u^j}{j} S_1 
\left[ \bar{S}_1^2 + \bar{S}_2 \right] =  
\tan\tfrac{\theta}{2} 
\Biggl\{ 
\Cl{2}{\pi-\theta}
\left[ 2 \theta^2 + 2 l^2_\theta - 20 L_\theta l_\theta + 32 L^2_\theta \right]
- \tfrac{1}{3} \Ls{4}{2\theta}
\nonumber \\ && \hspace{15mm}
+ \Cl{2}{2\theta} \left[ \tfrac{1}{4} \theta^2 - 2 L_\theta l_\theta + 4 L^2_\theta \right]
+ \Ls{3}{2 \theta} \left[ l_\theta - 4 L_\theta \right]
- 2 \Ls{3}{\theta} \left[ l_\theta - 5 L_\theta \right]
\nonumber \\ && \hspace{15mm}
+ \left[\Ls{3}{\pi-\theta} - \Ls{3}{\pi} \right] 
  \left[ 10 l_\theta - 32 L_\theta \right]
- 6 \theta \left[\Cl{3}{\pi-\theta} - \Cl{3}{\pi} \right] 
- \Cl{4}{\theta}
\nonumber \\ && \hspace{15mm}
+ \frac{22}{3} \left[\Ls{4}{\pi-\theta} - \Ls{4}{\pi} \right] 
+ 10 \Lsc{2,3}{\theta}
- 12 \Cl{4}{\pi-\theta}
+ \tfrac{2}{3} \Ls{4}{ \theta}
\nonumber \\ && \hspace{15mm}
- \tfrac{1}{2} \theta^2 \Cl{2}{\theta}
+ 10 \zeta_3 \theta 
+ \tfrac{1}{6} \theta^3 l_\theta
- \tfrac{1}{2} \theta^3 L_\theta
- 8 \theta L^3_\theta
+ 8 \theta L^2_\theta l_\theta
- 2 \theta  L_\theta l^2_\theta
\Biggr\} 
\label{S1S2b_1}
\\ && 
%%%%%%%%%%%%%%%%%%%%%%%%%%%%%%%%%%%%%%%%%%%%%%%%%%%%%%%%%%%%%%%%%%%%%%%%%%%
\sum_{j=1}^\infty \frac{1}{\left( 2j \atop j \right) } \frac{u^j}{j} 
\left[ \bar{S}_1^3 + 3 \bar{S}_1 \bar{S}_2 + 2 \bar{S}_3 \right] =  
\tan\tfrac{\theta}{2} 
\Biggl\{ 
9 \Cl{2}{\pi-\theta}
\left[ \tfrac{1}{4} \theta^2 + (2 L_\theta - l_\theta)^2 \right]
\nonumber \\ && \hspace{15mm}
+ \tfrac{3}{2} \Cl{2}{2\theta} \left[ \tfrac{1}{4} \theta^2 + ( 2 L_\theta - l_\theta )^2  \right]
+ 18 \Lsc{2,3}{\theta}
- \tfrac{3}{2} \Cl{4}{\theta}
- \tfrac{3}{4} \theta^2 \Cl{2}{\theta}
\nonumber \\ && \hspace{15mm}
- 18 ( 2 L_\theta - l_\theta ) 
  \left[ \Ls{3}{\pi-\theta} - \Ls{3}{\pi} - \tfrac{1}{2} \Ls{3}{\theta} + \tfrac{1}{6} \Ls{3}{2 \theta} \right]
+ \tfrac{21}{2} \zeta_3 \theta
\nonumber \\ && \hspace{15mm}
\!+\! 6 \left[ \Ls{4}{\pi \!-\! \theta} \!-\! \Ls{4}{\pi} \right]
\!-\! 6 \theta \left[ \Cl{3}{\pi \!-\! \theta} \!-\! \Cl{3}{\pi} \right]
\!-\! \Ls{4}{2\theta}
\!+\! 3 \Ls{4}{\theta}
\nonumber \\ && \hspace{15mm}
\!-\! 12 \Cl{4}{\pi \!-\! \theta}
\!+\! \tfrac{1}{4} \theta^3 l_\theta
\!-\! \tfrac{1}{2} \theta^3 L_\theta
\!+\! 8 \theta l^3_\theta
\!-\! 6 \theta l^2_\theta L_\theta
\!+\! 12 \theta l_\theta L^2_\theta
\!-\! 8 \theta L^3_\theta
\!-\! 7 \theta l^3_\theta
\Biggr\} \;. 
\label{S3b_1}
\end{eqnarray}
%%%%%%%%%%%%%%%%%%%%%%%%%%%%%%%%%%%%%%%%%%%%%%%%%%%%%%%%%%%%%%%%%%%%%%%%%%%
%XXXXXX

Using the analytic continuation procedure described in Section~3,
these results can be rewritten in terms of
the conformal variable $y$ defined in Eq.~(\ref{y<->u}). 
Below we list these analytically-continued results for
Eqs.~(\ref{j^1})--(\ref{S2_2}), as well as for those 
sums from Section~2 whose analytical continuations 
were not presented in Section~3:
%%%%%%%%%%%%%%%%%%%%%%%%%%%%%%%%%%%%%%%%%%%%%%%%%%%%%%%%%%%%%%%%%%%%%%%%%%%
\begin{eqnarray}
\sum_{j=1}^\infty \frac{1}{\left( 2j \atop j\right) } \frac{u^j}{j}  & = &
\frac{1-y}{1+y} \ln y,
\label{j^1_y}
\\
\sum_{j=1}^\infty \frac{1}{\left( 2j \atop j\right) } \frac{u^j}{j^2} & = &
-\tfrac{1}{2} \ln^2 y,
\label{j^2_y}
\\
\sum_{j=1}^\infty \frac{1}{\left( 2j \atop j \right) } \frac{u^j}{j^3} & = &
2 \Li{3}{y}
- 2 \ln y \Li{2}{y}
- \ln^2 y \ln (1-y) 
+ \tfrac{1}{6} \ln^3 y
- 2 \zeta_3 \;, 
\label{j^3_y}
\\
\sum_{j=1}^\infty \frac{1}{\left( 2j \atop j \right) } \frac{u^j}{j^4} & = & 
  4 \Snp{2,2}{y}
- 4 \Li{4}{y} 
- 4 \Snp{1,2}{y} \ln y 
+ 4 \Li{3}{y} \ln (1-y) 
+ 2 \Li{3}{y} \ln y 
\nonumber \\ && 
- 4 \Li{2}{y} \ln y \ln (1-y) 
- \ln^2 y \ln^2 (1-y) 
+ \tfrac{1}{3} \ln^3 y \ln (1-y) 
- \tfrac{1}{24} \ln^4 y 
\nonumber \\ && 
- 4 \ln (1-y)  \zeta_3 
+ 2 \ln y  \zeta_3 
+ 3 \zeta_4 \; ,
\label{j^4_y}
%%%%%%%%%%%%%%%%%%%%%%%%%%%%%%%%%%%%%%%%%%%%%%%%%%%%%%%%%%%%%%%%%%%%%%%%%%%
\\
\sum_{j=1}^\infty \frac{1}{\left( 2j \atop j\right) } \frac{u^j}{j} S_1
&=&
\frac{1-y}{1+y}
\left[ -2 \Li{2}{-y} - 2 \ln y \ln (1+y) + \tfrac{1}{2} \ln^2 y - \zeta_2 \right] ,
\label{S1_y}
%%%%%%%%%%%%%%%%%%%%%%%%%%%%%%%%%%%%%%%%%%%%%%%%%%%%%%%%%%%%%%%%%%%%%%%%%%%
\\
\sum_{j=1}^\infty \frac{1}{\left( 2j \atop j \right) } \frac{u^j}{j} \bar{S}_1
&=&
\frac{1-y}{1+y} \biggl[
\Li{2}{y} - 2 \Li{2}{-y} - 2 \ln y \ln(1+y)
+ \ln y \ln (1-y) 
\nonumber \\ && 
+ \tfrac{1}{4} \ln^2 y 
- 2 \zeta_2   \biggr] ,
\label{Sb1_y}
%%%%%%%%%%%%%%%%%%%%%%%%%%%%%%%%%%%%%%%%%%%%%%%%%%%%%%%%%%%%%%%%%%%%%%%%%%%
\\ 
\sum_{j=1}^\infty \frac{1}{\left( 2j \atop j \right) } \frac{u^j}{j} S_1^2
&=&
\frac{1-y}{1+y}
\biggl[
8 \Snp{1,2}{-y}
- 4 \Li{3}{-y}
+ 8 \Li{2}{-y} \ln(1+y)
+ 4 \ln^2 (1+y) \ln y
\nonumber \\ && 
- 2 \ln(1+y) \ln^2 y
+ \tfrac{1}{6} \ln^3 y
+ 4 \zeta_2 \ln(1+y) 
- 2 \zeta_2 \ln y 
- 4 \zeta_3
\biggr] \; ,
\label{S1S1_y}
\\ 
%%%%%%%%%%%%%%%%%%%%%%%%%%%%%%%%%%%%%%%%%%%%%%%%%%%%%%%%%%%%%%%%%%%%%%%%%%%
%%%%%%%%%%%%%%%%%%%%%%%%%%%%%%%%%%%%%%%%%%%%%%%%%%%%%%%%%%%%%%%%%%%%%%%%%%%
\sum_{j=1}^\infty \frac{1}{\left( 2j \atop j \right) } \frac{u^j}{j} S_1 \bar{S}_1
&\!\!=\!\!&
 \frac{1-y}{1+y}
\biggl[
10 \Snp{1,2}{-y}
- \Snp{1,2}{y^2}
+ \Li{3}{y}
+ 2 \Snp{1,2}{y}
- 3 \Li{3}{-y}
\nonumber \\ &&
-   2 \ln (1-y) \Li{2}{-y}
+ 8 \ln(1+y)  \Li{2}{-y}
- 2 \ln(1+y)  \Li{2}{y}
\nonumber \\ &&
- 2 \ln y \ln (1-y) \ln (1+y)
- \ln y \ln^2 (1-y)
+ 4 \ln y \ln^2 (1+y)
\nonumber \\ &&
+ \tfrac{1}{2} \ln^2 y \ln(1-y)
- \tfrac{3}{2} \ln^2 y \ln(1+y)
+ \tfrac{1}{12} \ln^3 y
+ 6 \zeta_2 \ln(1+y) 
\nonumber \\ &&
- \zeta_2 \ln(1-y)
- \tfrac{5}{2} \zeta_2 \ln y 
- \tfrac{11}{2} \zeta_3
\biggr] \; ,
\label{S1Sb1_y}
\\
%%%%%%%%%%%%%%%%%%%%%%%%%%%%%%%%%%%%%%%%%%%%%%%%%%%%%%%%%%%%%%%%%%%%%%%%%%%%%%%%%%%%%%%%%
\sum_{j=1}^\infty \frac{1}{\left( 2j \atop j \right) } \frac{u^j}{j} S_2
& = &  - \frac{1-y}{6(1+y)}\; \ln^3 y \; ,
\label{S2_1_y}
\\ 
%%%%%%%%%%%%%%%%%%%%%%%%%%%%%%%%%%%%%%%%%%%%%%%%%%%%%%%%%%%%%%%%%%%%%%%%%%%%%%%%%%%%%%%%%
\sum_{j=1}^\infty \frac{1}{\left( 2j \atop j \right) } \frac{u^j}{j^2} S_2
& = &  \tfrac{1}{24} \ln^4 y \; ,
\label{S2_2_y}
\\
%%%%%%%%%%%%%%%%%%%%%%%%%%%%%%%%%%%%%%%%%%%%%%%%%%%%%%%%%%%%%%%%%%%%%%%%%%%
\label{AN_S3_1}
\sum_{j=1}^\infty \frac{1}{\left( 2j \atop j \right) } \frac{u^j}{j} S_3 
&=&  \frac{1-y}{1+y}
\Bigl[
\tfrac{1}{24} \ln^4 y 
+ 6 \Li{4}{y}
+ \ln^2 y \Li{2}{y}
- 2 \zeta_3 \ln y 
- 4 \ln y \Li{3}{y} 
 - 6 \zeta_4 
\Bigr] \;, 
\nonumber \\
&& \hspace{10mm} 
\\ 
%%%%%%%%%%%%%%%%%%%%%%%%%%%%%%%%%%%%%%%%%%%%%%%%%%%%%%%%%%%%%%%%%%%%%%%%%%%
\label{AN_S1S2_1}
\sum_{j=1}^\infty \frac{1}{\left( 2j \atop j \right) } \frac{u^j}{j} S_1 S_2  
&\!\!=\!\!&  \frac{1\!-\!y}{1\!+\!y}
\Bigl[
\tfrac{1}{3} \ln^3 y \ln(1\!+\!y) 
- \! \tfrac{1}{24} \ln^4 y 
+ \! \tfrac{1}{2} \zeta_2 \ln^2 y 
+ \ln^2 y \Li{2}{-y}
+ \ln^2 y \Li{2}{y}
\nonumber \\ && 
+ \zeta_3 \ln y 
\!-\! 4 \ln y \Li{3}{-y}
\!-\! 4 \ln y \Li{3}{y}
\!+\! \zeta_4 
+ \! 8 \Li{4}{-y}
+ \! 6 \Li{4}{y}
\Bigr] , 
\\
%%%%%%%%%%%%%%%%%%%%%%%%%%%%%%%%%%%%%%%%%%%%%%%%%%%%%%%%%%%%%%%%%%%%%%%%%%%
\label{AN_S2Sb1_1}
\sum_{j=1}^\infty \frac{1}{\left( 2j \atop j \right) } \frac{u^j}{j} S_2 \bar{S}_1
&\!\!=\!\!&  \frac{1-y}{1+y}
\Bigl[
  \tfrac{1}{3} \ln^3 y \ln (1+y) 
- \tfrac{1}{6} \ln^3 y \ln (1-y) 
- \tfrac{1}{48} \ln^4 y 
+ \zeta_2 \ln^2 y 
+ 4 \zeta_3 \ln y 
\nonumber \\ && 
+ \ln^2 y \Li{2}{-y}
- 4 \ln y \Li{3}{-y}
+ 8 \zeta_4 
+ 8 \Li{4}{-y}
- \Li{4}{y}
\Bigr] \;. 
\end{eqnarray}

For the combination~(\ref{S2S1bar}) involving $(\bar{S}_2 + \bar{S}_1^2)$
we get
\begin{eqnarray}
&& \hspace*{-10mm}
%%%%%%%%%%%%%%%%%%%%%%%%%%%%%%%%%%%%%%%%%%%%%%%%%%%%%%%%%%%%%%%%%%%%%%%%%%%%%%%%%%%%%%%%%
%%%%%%%%%%%%%%%%%%%%%%%%%%%%%%%%%%%%%%%%%%%%%%%%%%%%%%%%%%%%%%%%%%%%%%%%%%%%%%%%%%%%%%%%%
\sum_{j=1}^\infty \frac{1}{\left( 2j \atop j \right) } \frac{u^j}{j}
\left( \bar{S}_2 + \bar{S}_1^2 \right)
 =
 \frac{1-y}{1+y}
\biggl\{
12 \Snp{1,2}{-y}
- 2 \Snp{1,2}{y^2}
+ \Li{3}{y}
+ 6 \Snp{1,2}{y}
- 8 \zeta_3
\nonumber \\ &&
- 2 \Li{3}{-y}
- 4 \ln (1-y^2) \Li{2}{y}
+ 2 \ln (1-y) \Li{2}{-y}
+ 8 \ln(1+y)  \Li{2}{-y}
- 2 \zeta_2 \ln y 
\nonumber \\ &&
+ \ln y \left[ \ln  (1-y) - 2 \ln (1+y) \right]^2
+ \left( \tfrac{1}{2} \ln^2 y - 4 \zeta_2 \right) 
\left[ \ln  (1-y) - 2 \ln (1+y) \right]
\biggr\} \; ,
\label{Sb2comb_y}
\end{eqnarray}
%%%%%%%%%%%%%%%%%%%%%%%%%%%%%%%%%%%%%%%%%%%%%%%%%%%%%%%%%%%%%%%%%%%%%%%%%%%
For a separate term of this sum involving $\bar{S}_2$, we get 
\begin{equation}
\label{AN_S2b_c1}
\sum_{j=1}^\infty \frac{1}{\left( 2j \atop j \right) } \frac{u^j}{j} \bar{S}_2
=  \frac{1-y}{1+y}
\Bigl[
  2 \Li{3}{-\omega_s}
- 2 \Li{3}{\omega_s}
- \tfrac{1}{24} \ln^3 y 
\Bigr] \; , 
\end{equation}
where $\omega_s$ is defined in Eq.~(\ref{def_omega_s}). 

The results for the {\em multiple inverse binomial sums}
presented in this paper are summarized in Table~1.
When two equation numbers are present, the first one refers to
the result in terms of the angular variable $\theta$, whereas
the second one corresponds to its analytical continuation. 
An asterisk means that the corresponding equation holds for general $c$.
The symbol $\dagger$ means that the results for the sums 
involving the combinations
$S_1 \left( \bar{S}_2 + \bar{S}_1^2 \right)$
and $\bar{S}_1^3 + 3\bar{S}_1 \bar{S}_2 + 2 \bar{S}_3$ 
can be extracted from the expressions (\ref{AN_C1}) and (\ref{AN_C2}) 
given for the sums involving ${\cal C}_1$ and ${\cal C}_2$, 
respectively,
using the definitions~(\ref{combinations}).

\begin{table}
$$
%%%%%%%%%%%%%%%%%%%%%%%%%%%%%%%%%%%%%%%%%%%%%%%%%%%%%%%%%%%%%%%%%%%%%%%%%%%%%%%%%%%%%%%%%
\begin{tabular}[h]{|c||c|c|c|c|}
\hline
{}& $c=1$ & $c=2$ & $c=3$ & $c=4$ \\ 
\hline 
1 & (\ref{j^1}), (\ref{j^1_y}) & 
    (\ref{j^2}), (\ref{j^2_y}) & 
    (\ref{j^3}), (\ref{j^3_y}) & 
    (\ref{j^4}), (\ref{j^4_y}) \\
$S_1$ & (\ref{S1a}), (\ref{S1_y}) & 
        (\ref{S1c2}), (\ref{AN_S1_2}) & 
        (\ref{S1_3}), (\ref{AN_S1_3}) & 
        {} \\
$\bar{S}_1$ & (\ref{S1bar}), (\ref{Sb1_y}) & 
              (\ref{S1barc2}), (\ref{AN_Sb1_2}) & 
              (\ref{S1bar_3}), (\ref{AN_Sb1_3}) & 
              {} \\ 
$S_2$ & (\ref{S2}),  (\ref{S2_1_y}) & 
        (\ref{S2_2}), (\ref{S2_2_y}) & 
        (\ref{S2G})${}^{\star}$ & 
        (\ref{S2G})${}^{\star}$ \\ 
$S_1^2$ & (\ref{S1S1}), (\ref{S1S1_y}) & 
          (\ref{S1S1c2}), (\ref{AN_S1S1_2}) & 
          {} & 
          {} \\ 
$S_1 \bar{S}_1$ & (\ref{S1S1bar}), (\ref{S1Sb1_y}) & 
                  (\ref{S1S1barc2}), (\ref{AN_S1Sb1_2}) & 
                  {} & 
                  {} \\ 
$\bar{S}_2$  & (\ref{S2barc1}),  (\ref{AN_S2b_c1}) & (\ref{S2barc2}),(\ref{AN_S2b_c2})  & {} & {} \\ 

$\bar{S}_2+\bar{S}_1^2$ & (\ref{S2S1bar}), (\ref{Sb2comb_y}) & 
                          (\ref{S2bar_2}), (\ref{AN_Sb2_2}) & 
                          {} & 
                          {} \\ 
$S_3$ & (\ref{S3c1}), (\ref{AN_S3_1}) & 
        (\ref{S3c2}), (\ref{AN_S3_2}) & 
        {} & 
        {} \\ 
$S_1 S_2$ & (\ref{S1S2c1}), (\ref{AN_S1S2_1}) & 
            (\ref{S1S2c2}), (\ref{AN_S1S2_2}) & 
            {} & 
            {} \\ 
$S_1^3$ & (\ref{S1S1S1c1}), (\ref{AN_S1^3_1}) & {} & {} & {} \\ 
$S_2 \bar{S}_1$ & (\ref{S2S1barc1}), (\ref{AN_S2Sb1_1}) & 
                  (\ref{S2S1barc2}), (\ref{AN_S2Sb1_2}) & 
                  {} & 
                  {} \\ 
$S_1^2 \bar{S}_1$ & (\ref{S1S1Sb1_1}), (\ref{AN_S1S1Sb1_1}) & {} & {} & {} \\ 
$S_1 \left( \bar{S}_2 + \bar{S}_1^2 \right)$ & (\ref{S1S2b_1}), (\ref{AN_C1})${}^{\dagger}$ 
& {} & {} & {} \\ 
$\bar{S}_1^3 + 3\bar{S}_1 \bar{S}_2 + 2 \bar{S}_3 $ 
& (\ref{S3b_1}), (\ref{AN_C2})${}^{\dagger}$  & {} & {} & {} \\ 
$S_2^2 - S_4 $ & (\ref{S4}) & (\ref{S4G})${}^{\star}$ 
& (\ref{S4G})${}^{\star}$ & (\ref{S4G})${}^{\star}$ \\ 
\hline
\end{tabular}
$$
\caption{Equation index for the inverse binomial sums}
\end{table}

%%%%%%%%%%%%%%%%%%%%%%%%%%%%%%%%%%%%%%%%%%%%%%%%%%%%%%%%%%%%%%%%%%%%%%%%%%%%%%%%%%%%%%%%%
\section{Connection between binomial, harmonic and inverse binomial sums}
\setcounter{equation}{0}
%=====================================================================

Using the procedure described in section~2.2, one can also construct 
the $\ep$-expansion 
of hypergeometric functions of the following types:
\[
{}_{P+1}F_P\left(\begin{array}{c|}
\tfrac{3}{2} \!+\!b_1\ep, \ldots , 
\tfrac{3}{2} \!+\!b_{J} \ep, \;
1\!+\!a_1\ep,\ldots ,  
1\!+\!a_K\ep,
2\!+\!d_1\ep, \ldots , 
2\!+\!d_L\ep  \\
\tfrac{3}{2} \!+\! f_1 \ep,\ldots ,  
\tfrac{3}{2} \!+\! f_{J-1} \ep, \;
1\!+\!e_1\ep, \ldots ,  
1\!+\!e_R\ep,
2\!+\!c_1\ep, \ldots ,   
2\!+\!c_{K+L-R}\ep 
\end{array} ~  u \right) \;, 
\]
%where $u=4z$ and
\[
{}_{P+1}F_P\left(\begin{array}{c|}
\tfrac{3}{2} \!+\! b_1\ep, \ldots ,
\tfrac{3}{2} \!+\! b_{J} \ep, \;
1\!+\!a_1\ep,\ldots ,
1\!+\!a_K\ep,
2\!+\!d_1\ep, \ldots , 
2\!+\!d_L\ep  \\
\tfrac{3}{2} \!+\! f_1 \ep,\ldots ,  
\tfrac{3}{2} \!+\! f_{J} \ep, \;
1\!+\!e_1\ep, \ldots ,  
1\!+\!e_R\ep,
2\!+\!c_1\ep, \ldots ,
2\!+\!c_{K+L-R-1}\ep 
\end{array} ~  z \right) \;,
\]
where $P=K+L+J-1$ and $u=4z$.
By analogy with Eq.~(\ref{PFQ2}),
the $\ep$-expansion of these functions can be written in the following form:
\begin{eqnarray}
&& \hspace*{-7mm}
_{P+1}F_P\left(\begin{array}{c|}
\{ \tfrac{3}{2} +b_i\ep\}^{J}, \;
\{ 1+a_i\ep\}^K, \; \{ 2+d_i\ep\}^L  \\
\{ \tfrac{3}{2} + f_i\ep\}^{J-1}, \;
\{ 1+e_i\ep \}^R,
\{ 2+c_i\ep \}^{K+L-R}
\end{array} ~  u \right)
\nonumber \\ && \hspace*{5mm}
= \frac{2}{u}
\frac{ \Pi_{s=1}^{K+L-R} (1+c_s\ep) 
      \Pi_{k=1}^{J-1} (1 + 2 f_k\ep)
     }{
      \Pi_{i=1}^{L}   (1+d_i\ep)
      \Pi_{r=1}^{J} (1 + 2 b_r\ep)
       }
\sum_{j=1}^\infty \left( 2j \atop j \right)   \frac{z^j}{j^{K-R-1}}
\Delta \; ,
\label{binomial}
%%%%%%%%%%%%%%%%%%%%%%%%%%%%%%%%%%%%%%%%%%%%%%%%%%%%%%%%%%%%%%%%%%%
\\ && \hspace*{-7mm}
_{P+1}F_P\left(\begin{array}{c|}
\{ \tfrac{3}{2} +b_i\ep\}^{J}, \;
\{ 1+a_i\ep\}^K, \; \{ 2+d_i\ep\}^L  \\
\{ \tfrac{3}{2} + f_i\ep\}^{J}, \;
\{ 1+e_i\ep \}^R,
\{ 2+c_i\ep \}^{K+L-R-1}
\end{array} ~  z \right)
\nonumber \\ && \hspace*{5mm}
= \frac{1}{z}
\frac{ \Pi_{s=1}^{K+L-R-1} (1+c_s\ep)
      \Pi_{k=1}^{J} (1 + 2 f_k\ep)
     }{
\Pi_{i=1}^{L} (1+d_i\ep)
\Pi_{r=1}^{J} (1 + 2 b_r\ep)
       }
\sum_{j=1}^\infty \frac{z^j}{j^{K-R-1}}
\Delta \;,
\label{harmonic}
\end{eqnarray}
where the function $\Delta$ is defined in the same way as in 
Eq.~(\ref{expansion}). One should only remember that 
the upper summation limits for the coefficients 
$B_k$ and $C_k$ are changed, since the numbers of 
the parameters $b_i$ and $c_i$ in Eqs.~(\ref{binomial}) 
and (\ref{harmonic}) are different.

The sums appearing in Eq.~(\ref{binomial}) are expressible in terms 
of the {\it multiple binomial sums} \cite{poleII},
whereas the sums of Eq.~(\ref{harmonic})  are reduced to 
the  {\it multiple harmonic sums}.
Using relations between hypergeometric function of different arguments, 
it is possible to express one type of sums in terms of the another one, 
plus some trivial part. 
Let us illustrate this on the example of $_2F_1$ function~(\ref{2F1_def}).
Using the standard formula of analytic continuation to the argument $1/z$,
we obtain a combination of two $_2F_1$ functions. To bring them to the form
of~(\ref{binomial}), we need to shift some of the parameters using
\[
\left. {}_2F_1\left(\begin{array}{c} a,\; b\\ c \end{array} \right| z \right)
= 1 + \frac{abz}{c} \; 
\left. {}_3F_2\left(\begin{array}{c} 1,\; a+1,\; b+1\\2,\; c+1 \end{array} 
\right| z \right) \; .
\] 
Finally, using the duplication formula for the argument of
the $\Gamma$-function, we arrive at the following relation:
\begin{eqnarray}
&& \hspace*{-16mm}
2 z( a_1-a_2)\ep \frac{\Gamma(1+b\ep)}{\Gamma(2+2b\ep)}
\left. _2F_1\left( \begin{array}{c} 1+a_1 \ep, \; 1 + a_2 \ep \\
                  \tfrac{3}{2} + b \ep \end{array} \right| z  \right) 
\nonumber \\ 
&=& \frac{1}{(-4z)^{a_1\ep}}
\frac{\Gamma(1+(a_2-a_1)\ep) \Gamma( 1+(b-a_1)\ep)}{\Gamma(1+a_2\ep) \Gamma(1+2(b-a_1)\ep)}
\nonumber \\ && \times 
\Biggl\{ 1 + \frac{(1+a_1\ep) (1+2(a_1-b)\ep) }{2 z (1+(a_1-a_2)\ep) }
\left. _3F_2\left( \begin{array}{c} 1, \; 2 + a_1\ep, \; \tfrac{3}{2} + (a_1 - b) \ep \\
                  2, \; 2 + (a_1-a_2) \ep \end{array} \right| \frac{1}{z}  \right) 
\Biggr \}
\nonumber \\ && 
- 
\frac{1}{(-4z)^{a_2\ep}}
\frac{\Gamma(1+(a_1-a_2)\ep) \Gamma( 1+(b-a_2)\ep)}{\Gamma(1+a_1\ep) \Gamma(1+2(b-a_2)\ep)}
\nonumber \\ && \times
\Biggl\{ 1 + \frac{(1+a_2\ep) (1+2(a_2-b)\ep) }{2 z (1+(a_2-a_1)\ep) }
\left. _3F_2\left( \begin{array}{c} 1, \; 2 + a_2\ep, \; \tfrac{3}{2} + (a_2 - b) \ep \\
                  2, \; 2 + (a_2-a_1) \ep \end{array} \right| \frac{1}{z}  \right) 
\Biggr\}
\;. 
\end{eqnarray}
Here, the $\ep$-expansion of the hypergeometric function on the l.h.s.\ 
can be expressed in terms of the {\it multiple inverse binomial sums} 
(see Eq.~(\ref{PFQ2})), whereas the functions on the r.h.s. 
yield the {\it multiple binomial sums} (see Eq.~(\ref{binomial}) ).
 
Furthermore, the following three quadratic relations for $_2F_1$ functions
(see, e.g., in~\cite{PBM3}) allow us to connect some {\it multiple binomial sums} 
(appearing in Eq.~(\ref{binomial}))
with  {\it multiple harmonic sums} (see Eq.~(\ref{harmonic})):
\begin{eqnarray}
%%%%%%%%%%%%%%%%%%%%%%%%%%%%%%%%%%%%%%%%%%%%%%%%%%%%%%%%%%%%%%%%%%%%%%%%%
&& \hspace*{-8mm} 
\left. _2F_1\left( \begin{array}{c} 1 \!+\! b\ep, \; \tfrac{3}{2} \!+\! a \ep \\
                  2 \!+\! (a\!+\!b) \ep \end{array} \right|u  \right) 
= \frac{1}{\sqrt{1-u}} 
(1+\chi)^{2 + 2 a \ep}\;
\left. _2F_1\left( \begin{array}{c}  2 \!+\! 2 a\ep, \; 1 \!+\! (a \!-\! b) \ep \\
                  2 \!+\! (a\!+\!b) \ep \end{array} \right|
-\chi \right) ,
\hspace*{10mm} 
%%%%%%%%%%%%%%%%%%%%%%%%%%%%%%%%%%%%%%%%%%%%%%%%%%%%%%%%%%%%%%%%%%%%%%%%%
\\ 
&& \hspace*{-8mm} 
\left. _2F_1\left( \begin{array}{c} 1 \!+\! b\ep, \; \tfrac{3}{2} \!+\! a \ep \\
                  2 \!+\! 2 b \ep \end{array} \right|u  \right) 
= 
(1+\chi)^{3 + 2 a \ep}\;
\left. _2F_1\left( \begin{array}{c}  \tfrac{3}{2} + a\ep, \; 1 + (a - b) \ep \\
                  \tfrac{3}{2} + b \ep \end{array} \right|  
                  ~\chi \right) 
\;, 
%%%%%%%%%%%%%%%%%%%%%%%%%%%%%%%%%%%%%%%%%%%%%%%%%%%%%%%%%%%%%%%%%%%%%%%%%
\\ 
&& \hspace*{-8mm} 
~{}_{2}F_1 \left(\begin{array}{c|} 1, \frac{1}{2}\\ 
                                2-\ep \end{array} ~u\right)
= \frac{1}{1\!-\!2\ep} \Biggl[
2(1 \!-\! \ep) 
- (1\!-\!u) ~{}_{2}F_1 \left(\begin{array}{c|} 1, \frac{3}{2}\\ 
                                2-\ep \end{array} ~u\right) \Biggr]
= 
(1\!+\!\chi)\;
~{}_{2}F_1 \left(\begin{array}{c|} 1, \ep \\ 
                                2-\ep \end{array} 
~\chi\right) ,
\nonumber \\ 
&& \hspace*{10mm} 
\label{all-order}
\end{eqnarray}
where 
\begin{equation}
\label{def_chi}
\chi = \frac{1-\sqrt{1-u}}{1+\sqrt{1-u}}, \quad u = \frac{4\chi}{(1+\chi)^2},
\end{equation}
and all orders of the $\ep$-expansion of the last $_2F_1$ function on the r.h.s.\ 
of Eq.~(\ref{all-order}) are known through 
Eq.~(2.14) of Ref.~\cite{DK01}. 

Combining all these relations together with the results for the 
{\em multiple inverse binomial sums} presented in this paper,
we have reproduced several known results for {\em multiple binomial
sums}, including those obtained in Ref.~\cite{poleII}.
For completeness, we list a number of such results,
including trivial ones, which could be extracted from Ref.~\cite{poleII}
(although were not explicitly listed there), 
\begin{eqnarray}
&&
%\label{sum000}
\sum\limits_{j=1}^\infty \left( 2j \atop j\right) z^j =
\frac{2 \chi}{1-\chi} , 
\nonumber \\ &&
%\label{sum10000}
\sum\limits_{j=1}^\infty \left( 2j \atop j\right) \frac{z^j}{j} =
    2\ln(1+\chi) ,
\nonumber \\ &&
\sum\limits_{j=1}^\infty \left( 2j \atop j\right) \frac{z^j}{j^2}  =
    -2\Li{2}{-\chi} - 2\ln^2(1+\chi) , 
\nonumber \\ &&
\sum\limits_{j=1}^\infty \left( 2j \atop j\right)  \frac{z^j}{j^3}  =
    4 \Snp{1,2}{-\chi} - 2 \Li{3}{-\chi} + 4 \Li{2}{-\chi} \ln(1+\chi)
    + \tfrac{4}{3} \ln^3(1+\chi) , 
\nonumber \\ &&
\sum\limits_{j=1}^\infty \left( 2j \atop j\right)  z^j S_1  =
-\frac{2}{1-\chi} \left[ (1-\chi) \ln (1+\chi) + (1+\chi) \ln(1-\chi) \right],
\nonumber \\ &&
%\label{sum11000}
\sum\limits_{j=1}^\infty \left( 2j \atop j\right)  \frac{z^j}{j} S_1  =
   \Li{2}{\chi^2} + 2\ln^2 (1+\chi) , 
\nonumber \\&&
\sum\limits_{j=1}^\infty \left( 2j \atop j\right) z^j \bar{S}_1  
= 
\frac{2}{1-\chi} \left[ \chi \ln(1+\chi) - (1+\chi) \ln(1-\chi) \right],
\nonumber \\&&
%\label{sum10100}
\sum\limits_{j=1}^\infty \left( 2j \atop j\right) \frac{z^j}{j} \bar{S}_1  
= 2 \Li{2}{\chi} + \ln^2 (1+\chi) ,
\nonumber \\&&
\sum_{j=1}^\infty \left( 2j \atop j\right) z^j S_2
= -\frac{4 \chi}{1-\chi}  \Li{2}{-\chi}
+ 2 \ln^2 (1+\chi)  ,
\nonumber \\ && 
\sum_{j=1}^\infty \left( 2j \atop j\right) z^j S_1^2
= 
\frac{2}{1\!-\!\chi} \biggl[ 
  4 \chi \Li{2}{\chi}
+ 2 \chi \Li{2}{-\chi}
- (1\!-\! \chi) \ln^2 (1\!+\!\chi) 
+ 2 (1\!+\!\chi) \ln^2 (1\!-\!\chi) 
\biggr] , 
\nonumber \\&& 
\sum_{j=1}^\infty \left( 2j \atop j\right) z^j S_1  \bar{S}_1 
= 
\frac{1}{1-\chi} \biggl[ 
  6 \chi \Li{2}{\chi}
- 2(1+\chi) \ln (1+\chi) \ln(1-\chi) 
\nonumber \\ && \hspace{5cm}
- (1-\chi) \ln^2 (1+\chi) 
+ 4 (1+\chi) \ln^2 (1-\chi) 
\biggr] , 
\nonumber \\ && 
%\label{sum-combination}
\sum_{j=1}^\infty \left( 2j \atop j\right) z^j 
\left(  \bar{S}_2 - \bar{S}_1^2 \right) 
= 
-\frac{2}{1-\chi} \biggl[ 
  2 \chi \Li{2}{\chi}
- 2(1+\chi) \ln(1+\chi) \ln(1-\chi)
\nonumber \\ && \hspace{6cm}
+ \chi \ln^2 (1+\chi) 
+ 2 (1+\chi) \ln^2 (1-\chi) 
\biggr] . 
\label{MBS-list}
\end{eqnarray}

Furthermore, explicit
results for the {\em multiple binomial sums} with $1/j^4$, $S_1/j^2$, 
$\bar{S_1}/j^2$,
$S_2/j$, $S_1^2/j$, $S_1\bar{S}_1/j$, and $(\bar{S}_2-\bar{S}_1^2)/j$
are presented in Eqs.~(A.4)--(A.10) of Ref.~\cite{poleII} (where
the same notation $\chi$ as here was used, with $x$ corresponding to our $z$).  
We confirm all those results. Moreover, using our approach
we have obtained a number of new results
for the {\it multiple binomial sums},
%%%%%%%%%%%%%%%%%%%%%%%%%%%%%%%%%%%%%%%%%%%%%%%%%%%%%%%%%%%%%%%%%%%%%%%%%
\begin{eqnarray}
&& 
\sum_{j=1}^\infty \left( 2j \atop j\right) z^j S_3 
= 
- \tfrac{4}{3} \ln^3 (1\!+\!\chi) 
\!-\! 4 \ln(1\!+\!\chi) \Li{2}{-\chi}
\!-\! \frac{4}{1\!-\!\chi} \left[ 2 \SN{1}{2}{-\chi} + \chi \Li{3}{-\chi} \right] \;, 
\nonumber \\  &&
%%%%%%%%%%%%%%%%%%%%%%%%%%%%%%%%%%%%%%%%%%%%%%%%%%%%%%%%%%%%%%%%%%%%%%%%
\sum_{j=1}^\infty \left( 2j \atop j\right) z^j S_1 S_2  
= 
   \tfrac{4}{3} \ln^3 (1+\chi) 
  + 4 \ln(1+\chi) \Li{2}{-\chi}
  + 4 \ln(1+\chi) \Li{2}{\chi}
\nonumber \\ && \hspace{20mm}
+ \frac{4}{1-\chi} 
\left[
\chi \Li{3}{\chi} + \SN{1}{2}{\chi^2} - 2 \SN{1}{2}{\chi} 
+ (1+\chi) \ln(1-\chi) \Li{2}{-\chi}
\right] \;, 
\nonumber \\ && 
%%%%%%%%%%%%%%%%%%%%%%%%%%%%%%%%%%%%%%%%%%%%%%%%%%%%%%%%%%%%%%%%%%%%%%%%%
\sum_{j=1}^\infty \left( 2j \atop j\right) z^j S_2 \bar{S}_1 
= 
  \tfrac{2}{3} \ln^3 (1+\chi) 
+ 4 \ln (1+\chi) \Li{2}{\chi} 
\nonumber \\ && \hspace{20mm}
+ \frac{2}{1-\chi} 
\left[
  2 \chi \Li{3}{\chi}
- 6 \SN{1}{2}{-\chi}
+ 2 \SN{1}{2}{\chi^2} 
- 4 \SN{1}{2}{\chi}  
- \chi \Li{3}{-\chi}
\right]
\nonumber \\ && \hspace{20mm}
+ \frac{2(1+\chi)}{1-\chi} 
\left[ 2 \ln(1-\chi) - \ln(1+\chi) \right] \Li{2}{-\chi}  \; , 
\nonumber \\ &&
%%%%%%%%%%%%%%%%%%%%%%%%%%%%%%%%%%%%%%%%%%%%%%%%%%%%%%%%%%%%%%%%%%%%%%%%%%%%% 
\sum_{j=1}^\infty \left( 2j \atop j\right) z^j S_1 \left(\bar{S}_2 -  \bar{S}_1^2 \right) = 
 \frac{2}{1-\chi} \left[
  16 \SN{1}{2}{\chi}
- 2 \chi \Li{3}{\chi} 
- 6 \chi \ln(1+\chi) \Li{2}{\chi} 
\right]
\nonumber \\ && \hspace{20mm}
+ \frac{2(1+\chi)}{1-\chi} \left[
  4 \ln^3(1-\chi)
- 4 \ln^2(1-\chi) \ln(1+\chi)
\right. 
\nonumber \\ && \hspace{40mm}
\left.
+ \ln(1-\chi) \ln^2(1+\chi)  
+ 8 \ln(1-\chi) \Li{2}{\chi} 
 \right]
+ \tfrac{2}{3} \ln^3(1+\chi)  \; , 
\nonumber \\ && 
%%%%%%%%%%%%%%%%%%%%%%%%%%%%%%%%%%%%%%%%%%%%%%%%%%%%%%%%%%%%%%%%%%%%%%%%%
\sum_{j=1}^\infty \left( 2j \atop j\right) z^j S_1^2 \bar{S}_1   = 
- \frac{2}{1-\chi} \left[
  2 \SN{1}{2}{\chi^2}
- 6 \SN{1}{2}{-\chi}
+ 16 \SN{1}{2}{\chi}
- \chi \Li{3}{-\chi}
\right.
\nonumber \\ && \hspace{40mm} 
\left.
- 4 \chi \Li{3}{\chi}
+ 2 (1-3\chi) \ln(1+\chi) \Li{2}{\chi}
\right]
\nonumber \\ && \hspace{20mm} 
- \frac{2(1+\chi)}{1-\chi} \left[
4 \ln^3(1-\chi) 
- 2 \ln^2(1-\chi) \ln(1+\chi) 
+ 2 \ln(1-\chi) \Li{2}{-\chi}
\right. 
\nonumber \\ && \hspace{40mm} 
\left.
+ 10 \ln(1-\chi) \Li{2}{\chi} 
- \ln(1+\chi) \Li{2}{-\chi}
\right]
- \tfrac{2}{3} \ln^3(1+\chi)
\; , 
\nonumber \\ && 
%%%%%%%%%%%%%%%%%%%%%%%%%%%%%%%%%%%%%%%%%%%%%%%%%%%%%%%%%%%%%%%%%%%%%%%%%
\sum_{j=1}^\infty \left( 2j \atop j\right) z^j \left( 
3 \bar{S}_1 \bar{S}_2 - 2 \bar{S}_3 - \bar{S}_1^3 \right) 
 = \frac{2}{1\!-\!\chi} 
\left[
12 \SN{1}{2}{\chi} \!-\! 6 \chi \ln(1\!+\!\chi) \Li{2}{\chi} 
\!-\! \chi \ln^3(1\!+\!\chi)  
\right]
\nonumber \\ && \hspace{20mm}
+  \frac{2(1+\chi)}{1-\chi} 
\left[ 
  6 \ln(1-\chi) \Li{2}{\chi}  
- 6 \ln^2(1-\chi) \ln(1+\chi) 
\right.
\nonumber \\ && \hspace{40mm}
\left.
+ 3 \ln(1-\chi) \ln^2(1+\chi) 
+ 4 \ln^3(1-\chi)
\right] \; , 
\nonumber \\ && 
%%%%%%%%%%%%%%%%%%%%%%%%%%%%%%%%%%%%%%%%%%%%%%%%%%%%%%%%%%%%%%%%%%%%%%%%%
\sum_{j=1}^\infty \left( 2j \atop j\right) z^j S_1^3  = 
- \tfrac{4}{3} \ln^3(1+\chi)
- 4 \ln(1+\chi) \Li{2}{-\chi} 
- 12 \ln(1+\chi) \Li{2}{\chi}
\nonumber \\ && \hspace{20mm}
- \frac{4}{1-\chi} \left[
  3 \SN{1}{2}{\chi^2} 
- 4\SN{1}{2}{-\chi}
+ 6 \SN{1}{2}{\chi}
- 2 \chi \Li{3}{-\chi} 
- 3 \chi \Li{3}{\chi}
\right]
\nonumber \\ && \hspace{20mm}
- \frac{4(1+\chi)}{1-\chi} 
\left[ 
  2 \ln^3(1-\chi) 
+ 3 \ln(1-\chi) \Li{2}{-\chi} 
+ 6 \ln(1-\chi) \Li{2}{\chi} 
\right]
\; , 
%%%%%%%%%%%%%%%%%%%%%%%%%%%%%%%%%%%%%%%%%%%%%%%%%%%%%%%%%%%%%%%%%%%%%%%%%
\end{eqnarray}
where $\chi$ is defined in Eq.~(\ref{def_chi}) and, as before, $u=4z$.
These results were recently
used to construct the $\ep$-expansion of two-loop sunset-type diagrams 
of special type (for details, see section~3.1 in~\cite{pole_topI}). 

The results for harmonic sums of the type 
$
\sum_{j=1}^\infty z^j S_{a_1} \ldots  S_{a_k}  \bar{S}_{b_1}  \ldots  \bar{S}_{b_j}  
$
could be deduced from the following relation:
\begin{eqnarray}
\left. _2F_1\left( \begin{array}{c} 1, \; \tfrac{3}{2} \!+\! a \ep \\
                  \tfrac{3}{2} \!+\! b \ep \end{array} \right|z  \right) 
& = & 
\frac{\Gamma(2+2b\ep) \Gamma(1+a\ep) \Gamma(1+(a-b)\ep)}
{\Gamma(1+b\ep) \Gamma(2+2a\ep)}
\frac{4^{(a-b)\ep}}{(1-z)^{1+(a-b)\ep} z^{1/2+b\ep}}
\nonumber \\ && 
- \frac{1+2b\ep}{2 (1+(a-b)\ep)}
\left. _2F_1\left( \begin{array}{c}  1, \; \tfrac{3}{2}\!+\! a \ep \\
                  2 \!+\! (a\!-\!b) \ep \end{array} \right| 1-z  \right) \;,
\end{eqnarray} 
which corresponds to an analytic continuation of the $_2F_1$ function
from the argument $z$ to $(1-z)$.
In particular, we get 
\begin{eqnarray}
\sum_{j=1}^\infty S_1 \bar{S}_1 z^j &=& 
\frac{z}{1-z} 
\Biggl\{ 
  \tfrac{1}{4} \ln^2 (1-z) 
+ \tfrac{1}{2} \ln^2 (1-\sqrt{z}) 
+ \tfrac{1}{2} \ln^2 (1+\sqrt{z}) 
+ \tfrac{1}{2} \Li{2}{z} 
\nonumber \\ && 
+ \frac{1}{2 \sqrt{z}} 
\Biggl[
\Li{2}{\frac{1+\sqrt{z}}{2}} 
- \Li{2}{\frac{1-\sqrt{z}}{2}}  \Biggr]
\nonumber \\ && 
+ \frac{1}{4 \sqrt{z}} \left[\ln^2 (1-\sqrt{z}) 
-  \ln^2 (1+\sqrt{z}) \right] 
- \frac{\ln 2}{2 \sqrt{z}} \ln \left(\frac{1+\sqrt{z}}{1-\sqrt{z}} \right)  
\Biggr] 
\Biggr\}. 
\hspace*{10mm}
\label{representaion:xi}
\end{eqnarray}
Introducing a new variable $\xi$ such that
\begin{equation}
\xi = \frac{1-\sqrt{z}}{1+\sqrt{z}} \; , 
\qquad z=\frac{(1-\xi)^2}{(1+\xi)^2}\; ,
\end{equation}
we can present the above result in a more compact form, 
\begin{equation}
\sum_{j=1}^\infty S_1 \bar{S}_1 z^j = 
\frac{z}{2(1\!-\!z)} \left[ \Li{2}{z} + \ln^2(1\!-\!z) \right]
+ 
\frac{1\!-\!\xi^2}{4 \xi} \left[ \Li{2}{-\xi} +\ln{\xi} \ln{2}
+ \tfrac{1}{2} \zeta_2 \right]
+ \frac{1\!-\!\xi}{8\xi} \ln^2{\xi} \; ,
\label{representaion:xi:2}
\end{equation}
where we have taken into account that 
\[
\Li{2}{\frac{1+\sqrt{z}}{2}} - \Li{2}{\frac{1-\sqrt{z}}{2}}  
= 2 \Li{2}{-\xi} + \ln \xi \ln (1+\xi) + \zeta_2 \;. 
\]
Integrating the representation~(\ref{representaion:xi:2}), it is easy to get 
the following result: 
\begin{eqnarray} 
\sum_{j=1}^\infty S_1 \bar{S}_1 \frac{z^j}{j} &=& 
- \Snp{1,2}{z} - \tfrac{1}{2} \ln (1-z) \Li{2}{z} 
-\tfrac{1}{6}\ln^3(1-z)
- \tfrac{3}{2}\zeta_3 
- \tfrac{1}{4} \ln^2\xi \ln(1-z)
\nonumber \\ &&
- 2 \Li{3}{-\xi}
+ \Li{2}{-\xi} \ln \xi 
- \tfrac{1}{2} \zeta_2 \ln \xi 
+ \tfrac{1}{12} \ln^3 \xi \; .
\label{har_S1s1_1}
\end{eqnarray}

In Appendix E of Ref.~\cite{FKV99} 
the results (up to weight {\bf 4}) for the harmonic sums of the type 
$\sum_{j=1}^\infty  S_{a_1} \ldots  S_{a_k} z^j/j^a$
have been presented.
It is easy to extend those results to the case 
$\sum_{j=1}^\infty \bar{S}_{a_1} \ldots  \bar{S}_{a_k} z^j/j^a$,
using the following property~\cite{generatingfunction}:
\begin{equation}
{\rm if} \quad 
\sum_{j=1}^\infty f(j) \frac{z^j}{j^a} = F(z) 
\qquad {\rm then} \quad
\sum_{j=1}^\infty f(2j) \frac{z^j}{j^a} 
 = 2^{a-1} \left[ F(\sqrt{z}) + F(-\sqrt{z}) \right]\;.
\end{equation}
As a consequence, we get
\begin{eqnarray}
%%%%%%%%%%%%%%%%%%%%%%%%%%%%%%%%%%%%%%%%%%%%%%%%%%%%%%%%%%%%%%
\sum_{j=1}^\infty \bar{S}_a z^j \!\! &=& \!\! 
  \frac{\sqrt{z}}{2(1-\sqrt{z})} \Li{a}{ \sqrt{z}}  
- \frac{\sqrt{z}}{2(1+\sqrt{z})} \Li{a}{-\sqrt{z}}  
\;, 
\nonumber \\ 
%%%%%%%%%%%%%%%%%%%%%%%%%%%%%%%%%%%%%%%%%%%%%%%%%%%%%%%%%%%%%%
\sum_{j=1}^\infty \bar{S}_1 \frac{z^j}{j^a} \!\! &=& \!\!  
2^{a-1} \left[ \SN{a-1}{2}{\sqrt{z}} + \SN{a-1}{2}{-\sqrt{z}} \right] \; .
\end{eqnarray}
As a further illustration,
we also present results for other sums up to weight {\bf 3},
\begin{eqnarray}
%%%%%%%%%%%%%%%%%%%%%%%%%%%%%%%%%%%%%%%%%%%%%%%%%%%%%%%%%%%%%%
\sum_{j=1}^\infty \bar{S}_1^2 z^j  \!\! &=& \!\!  
\frac{\sqrt{z}}{2(1\!-\!\sqrt{z})} 
\left[ 
\Li{2}{ \sqrt{z}}  + \ln^2 (1- \sqrt{z})
\right]
- \frac{\sqrt{z}}{2(1\!+\!\sqrt{z})} 
\left[
\Li{2}{ -\sqrt{z}}  + \ln^2 (1+\sqrt{z})
\right]
\;, 
\nonumber \\ 
%%%%%%%%%%%%%%%%%%%%%%%%%%%%%%%%%%%%%%%%%%%%%%%%%%%%%%%%%%%%%%
\sum_{j=1}^\infty \bar{S}_2 \frac{z^j}{j} \!\! &=& \!\! 
-\ln (1-\sqrt{z} ) \Li{2}{\sqrt{z}} 
-\ln (1+\sqrt{z} ) \Li{2}{-\sqrt{z}} 
- 2 \SN{1}{2}{\sqrt{z}} 
- 2 \SN{1}{2}{-\sqrt{z}} \;, 
\nonumber \\ 
%%%%%%%%%%%%%%%%%%%%%%%%%%%%%%%%%%%%%%%%%%%%%%%%%%%%%%%%%%%%%%
\sum_{j=1}^\infty \bar{S}_1^3 z^j \!\! &=& \!\! 
\frac{\sqrt{z}}{2(1-\sqrt{z})} 
\left[
\Li{3}{\sqrt{z}} - 3 \ln(1-\sqrt{z}) \Li{2}{\sqrt{z}} + \ln^3(1-\sqrt{z})  - 3 \SN{1}{2}{\sqrt{z}}
\right] 
\nonumber \\ && \hspace*{-3mm}
- \frac{\sqrt{z}}{2(1\!+\!\sqrt{z})} 
\left[
\Li{3}{-\sqrt{z}} \!-\! 3 \ln(1\!+\!\sqrt{z}) \Li{2}{-\sqrt{z}} \!+\! \ln^3(1\!+\!\sqrt{z})  
\!-\! 3 \SN{1}{2}{-\sqrt{z}}
\right] , 
\nonumber \\ 
%%%%%%%%%%%%%%%%%%%%%%%%%%%%%%%%%%%%%%%%%%%%%%%%%%%%%%%%%%%%%%
\sum_{j=1}^\infty \bar{S}_1^2 \frac{z^j}{j} \!\! &=& \!\! 
- \ln(1-\sqrt{z}) \Li{2}{\sqrt{z}}   
- \ln(1+\sqrt{z}) \Li{2}{-\sqrt{z}}  
- 2 \SN{1}{2}{\sqrt{z}}
- 2 \SN{1}{2}{-\sqrt{z}} 
\nonumber \\ && 
+ \tfrac{1}{3} \ln^3(1-\sqrt{z})  
+ \tfrac{1}{3} \ln^3(1+\sqrt{z}) \; ,
%%%%%%%%%%%%%%%%%%%%%%%%%%%%%%%%%%%%%%%%%%%%%%%%%%%%%%%%%%%%%%
\nonumber \\ 
\sum_{j=1}^\infty \bar{S}_1 \bar{S}_2 z^j \!\! &=& \!\! 
\frac{\sqrt{z}}{2(1-\sqrt{z})}
\left[ \Li{3}{\sqrt{z}}  - \ln(1-\sqrt{z}) \Li{2}{\sqrt{z}}  
-  \SN{1}{2}{\sqrt{z}}
\right] 
\nonumber \\ && 
-\frac{\sqrt{z}}{2(1+\sqrt{z})}
\left[ \Li{3}{-\sqrt{z}}  - \ln(1+\sqrt{z}) \Li{2}{-\sqrt{z}}  
-  \SN{1}{2}{-\sqrt{z}}
\right] \;. 
%%%%%%%%%%%%%%%%%%%%%%%%%%%%%%%%%%%%%%%%%%%%%%%%%%%%%%%%%%%%%%
\end{eqnarray}
To investigate relations between multiple harmonic sums 
the technique proposed in Ref.~\cite{newII} can be useful. 

As a illustration, we present the hihger order $\ep$-expansion of the 
some of the hypergeometric functions: 

\begin{eqnarray}
&& 
\left. _2F_1\left( \begin{array}{c} 1+a_1 \ep, 1+a_2\ep \\
                 2 + c \ep \end{array} \right| z\right) 
= 
\frac{1+c\ep}{z}
\Biggl(
- \ln (1-z) 
- \ep \Biggl\{
\frac{c-a_1-a_2}{2} \ln^2 (1-z) + c \Li{2}{z}
      \Biggr\}
\nonumber \\ && 
+ \ep^2 \Biggl\{
  \left[ (a_1+a_2)c - c^2 - 2 a_1 a_2  \right] \Snp{1,2}{z} 
+ \left[ (a_1+a_2)c - c^2 - a_1 a_2    \right]  \ln(1-z) \Li{2}{z} 
\nonumber \\ && \hspace{10mm}
+ c^2 \Li{3}{z}
- \frac{(c-a_1-a_2)^2}{6} \ln^3 (1-z) 
      \Biggr\}
\nonumber \\ && 
- \ep^3 \Biggl\{
  c \left[ (a_1+a_2)c - c^2 - 2 a_1 a_2 \right] \Snp{2,2}{z} 
+ c \left[ (a_1+a_2)c - c^2 -   a_1 a_2 \right] \ln(1-z) \Li{3}{z} 
\nonumber \\ && \hspace{10mm}
+ (c-a_1) (c-a_2) (c-a_1-a_2)  \left[  \ln(1-z) \Snp{1,2}{z} + \frac{1}{2}  \ln^2 (1-z) \Li{2}{z} \right]
\nonumber \\ && \hspace{10mm}
+ \frac{(c-a_1-a_2)^3}{24}  \ln^4 (1-z) 
+ c (c - a_1 - a_2)^2 \Snp{1,3}{z} 
+ c^3 \Li{4}{z} 
\Biggr\}
+ {\cal O} (\ep^4)
\Biggr) \; .
\end{eqnarray}

%%%%%%%%%%%%%%%%%%%%%%%%%%%%%%%%%%%%%%%%%%%%%%%%%%%%%%%%%%%%%%%%%%%%%%%%%%%%%%%%%%%%%%%%%
%%%%%%%%%%%%%%%%%%%%%%%%%%%%%%%%%%%%%%%%%%%%%%%%%%%%%%%%%%%%%%%%%%%%%%%%%%%%%%%%%%%%%%%%%
%%%%%%%%%%%%%%%%%%%%%%%%%%%%%%%%%%%%%%%%%%%%%%%%%%%%%%%%%%%%%%%%%%%%%%%%%%%%%%%%%%%%%%%%%
\section{The ${\cal O}(\alpha \alpha_{\rm s})$ corrections 
to the polarization function of  
neutral gauge bosons in arbitrary dimension}
\label{PoleMass}
\setcounter{equation}{0}
%=====================================================================
\begin{figure}[th]
\begin{center}
\centerline{\vbox{\epsfysize=30mm \epsfbox{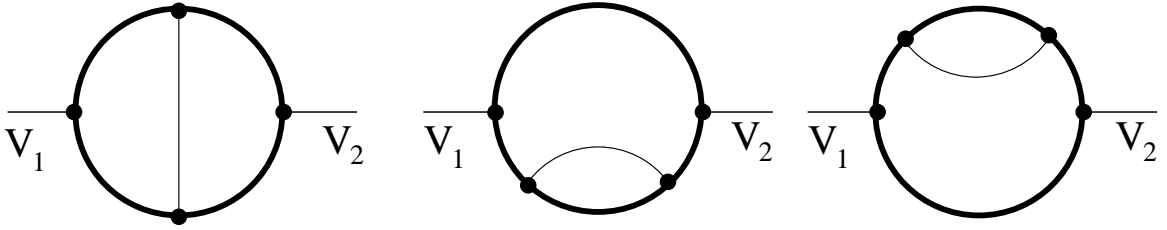}}}
\caption{\label{QCD} 
Two-loop contributions to the off-shell polarization 
function of a neutral gauge boson.
Bold and thin lines correspond to the massive quark propagator
and the massless boson (gluon or photon) propagator, respectively.}
\end{center}
\end{figure}
%
%=====================================================================
Here we present an example of a physically relevant 
calculation\footnote{Another non-trivial example where 
these master integrals appear is given in Ref.~\cite{background}.}
that can be expressed in terms of 
the master integrals $J_{011}$ studied in Section~4.3.
Let us consider the two-loop propagator-type diagrams shown in Fig.~\ref{QCD}.
All of these 
${\cal O}(\alpha \alpha_{\rm s})$ 
contributions to the polarization function of
the gauge bosons 
involve a quark loop with a gluon exchange.
It was analytically calculated in~\cite{qcd},
up to the finite term of the $\ep$-expansion. 
Here we present the bare two-loop results in $n$-dimensional space-time
(see also in Ref.~\cite{poleII}). 
In contrast to the calculations performed in~\cite{qcd}, here we use 
Tarasov's recurrence relations~\cite{T97a} for the reduction of the 
original integrals to the set of master integrals.  
In this Appendix we use the Euclidean notation~\cite{diagramatic},
$P^2\leftrightarrow -p^2$, so that the on-shell limit
would read $P^2\to -m^2$.

Let us decompose the polarization tensor 
into the transverse $\Pi_{\rm T}(P^2)$ and longitudinal 
$\Pi_{\rm L}(P^2)$ parts,
\[
\Pi_{\mu \nu}(P^2) = \left( \delta_{\mu \nu} - \frac{P_\mu P_\nu}{P^2}\right) 
                    \Pi_{\rm T}(P^2) 
                  + \frac{P_\mu P_\nu }{P^2} \Pi_{\rm L}(P^2) \; . 
\]
Then, the two-loop corrections can be written as 
%%%%%%%%%%%%%%%%%%%%%%%%%%%%%%%%%%%%%%%%%%%%%%%%%%%%%%%%%%%%%%%%%%%%%%%%%%%%%%%%%%%%
\begin{eqnarray}
&& \hspace*{-10mm}
\Pi_{\rm T}^{(2)}(P^2) =  \frac{g^2 g_{\rm s}^2}{(4 \pi)^{n/2}} N_c C_F    
\Biggl\{ 
- J_{011}(1,1,1) {\frac {4}{\left (n-1\right )t}}
\Bigl[
-4{ A}n{{ m}}^{2}
+12{ A}{{ m}}^{2}
+{ A}t{n}^{2}
\nonumber \\ && \hspace*{75mm}
-3{ A}tn
+2{ A}t
+St{n}^{2}
-3Stn
+2St
\Bigr]
%%%%%%%%%%%%%%%%%%%%%
\nonumber \\ && 
- J_{011}(1,1,2)
\frac {4}{\left (n-4\right )\left (n-3\right )\left (n-1\right )t}
\Bigl[
112{ A}{{ m}}^{4}n
+{ A}{t}^{2}{n}^{3}
-7{ A}{t}^{2}{n}^{2}
+18{ A}{t}^{2}n
\nonumber \\ && 
+S{t}^{2}{n}^{3}
-7S{t}^{2}{n}^{2}
+18S{t}^{2}n
-16{ A}{{ m}}^{4}{n}^{2}
-192{ A}{{ m}}^{4}
+4{ A}{{ m}}^{2}t{n}^{3}
-32{ A}{{ m}}^{2}t{n}^{2}
-16{ A}{t}^{2}
\nonumber \\ && 
+100{ A}{{ m}}^{2}tn
+4S{{ m}}^{2}t{n}^{3}
-36S{{ m}}^{2}t{n}^{2}
+96S{{ m}}^{2}tn
-112{ A}{{ m}}^{2}t
-16S{t}^{2}
-80S{{ m}}^{2}t
\Bigr]
%%%%%%%%%%%%%%%%%%%%%
\nonumber \\ && 
+ \left[ A_0(m) \right] ^2
{\frac {2 \left (n-2\right )}{\left (4{{ m}}^{2}+t\right )
       {{ m}}^{2}t\left (n-4\right )\left (n-1\right )\left (n-3\right )}}
\Bigl[
-4St{n}^{4}{{ m}}^{2}
+{ A}{t}^{2}{n}^{3}
+4{ A}{{ m}}^{2}t{n}^{3}
\nonumber \\ && 
+36S{{ m}}^{2}t{n}^{3}
+S{t}^{2}{n}^{3}
-32{ A}{{ m}}^{2}t{n}^{2}
-7S{t}^{2}{n}^{2}
-16{ A}{{ m}}^{4}{n}^{2}
-120S{{ m}}^{2}t{n}^{2}
-7{ A}{t}^{2}{n}^{2}
\nonumber \\ && 
+112{ A}{{ m}}^{4}n
+184S{{ m}}^{2}tn
+100{ A}{{ m}}^{2}tn
+18{ A}{t}^{2}n
+18S{t}^{2}n
-192{ A}{{ m}}^{4}
-16{ A}{t}^{2}
\nonumber \\ && 
-112S{{ m}}^{2}t
-112{ A}{{ m}}^{2}t
-16S{t}^{2}
\Bigr]
%%%%%%%%%%%%%%%%%%%%%
\nonumber \\ && 
- A_0(m) B_0(m,m,t) 
{\frac {2 \left (n-2\right )}{{{ m}}^{2}\left (n-3\right )
        \left (4{{ m}}^{2}+t\right )\left (n-4\right )\left (n-1\right )}}
\Bigl[
432{ A}{{ m}}^{4}n
-2{ A}t{n}^{4}{{ m}}^{2}
\nonumber \\ && 
+64{{ m}}^{4}S
+40{{ m}}^{4}Sn
+8{{ m}}^{4}S{n}^{3}
-48{{ m}}^{4}S{n}^{2}
+80{ A}{{ m}}^{4}{n}^{3}
-8{ A}{n}^{4}{{ m}}^{4}
-2St{n}^{4}{{ m}}^{2}
\nonumber \\ && 
+{ A}{t}^{2}{n}^{3}
-7{ A}{t}^{2}{n}^{2}
+18{ A}{t}^{2}n
+S{t}^{2}{n}^{3}
-7S{t}^{2}{n}^{2}
+18S{t}^{2}n
-280{ A}{{ m}}^{4}{n}^{2}
-256{ A}{{ m}}^{4}
\nonumber \\ && 
+24{ A}{{ m}}^{2}t{n}^{3}
-98{ A}{{ m}}^{2}t{n}^{2}
+180{ A}{{ m}}^{2}tn
+24S{{ m}}^{2}t{n}^{3}
-102S{{ m}}^{2}t{n}^{2}
+176S{{ m}}^{2}tn
\nonumber \\ && 
-16{ A}{t}^{2}
-128{ A}{{ m}}^{2}t
-16S{t}^{2}
-96S{{ m}}^{2}t
\Bigr]
%%%%%%%%%%%%%%%%%%%%%%
\nonumber \\ && 
- \left[ B_0(m,m,t) \right]^2
{\frac {2}{\left (4{{ m}}^{2}+t\right )\left (n-4\right )\left (n-1\right )}}
\Bigl[
448{ A}{{ m}}^{4}n
-32{{ m}}^{4}S
+16{ A}{{ m}}^{4}{n}^{3}
\nonumber \\ && 
+{ A}{t}^{2}{n}^{3}
-9{ A}{t}^{2}{n}^{2}
+30{ A}{t}^{2}n
+S{t}^{2}{n}^{3}
-9S{t}^{2}{n}^{2}
+30S{t}^{2}n
-144{ A}{{ m}}^{4}{n}^{2}
-448{ A}{{ m}}^{4}
\nonumber \\ && 
+8{ A}{{ m}}^{2}t{n}^{3}
-72{ A}{{ m}}^{2}t{n}^{2}
+232{ A}{{ m}}^{2}tn
+4S{{ m}}^{2}t{n}^{3}
-44S{{ m}}^{2}t{n}^{2}
+152S{{ m}}^{2}tn
\nonumber \\ && 
-32{ A}{t}^{2}
-240{ A}{{ m}}^{2}t
-32S{t}^{2}
-160S{{ m}}^{2}t
\Bigr] 
\Biggr\} \; ,
\label{PT}
%%%%%%%%%%%%%%%%%%%%%%%%%%%%%%%%%%%%%%%%%%%%%%%%%%%%%%%%%%%%%%%%%%%%%%%%%%%%%%%%%%%%%
%%%%%%%%%%%%%%%%%%%%%%%%%%%%%%%%%%%%%%%%%%%%%%%%%%%%%%%%%%%%%%%%%%%%%%%%%%%%%%%%%%%%%
\\ && \hspace*{-10mm} 
\Pi_{\rm L}^{(2)}(P^2) =  A \frac{g^2 g_{\rm s}^2}{(4 \pi )^{n/2}} N_c C_F    
\Biggl\{ 
- J_{011}(1,1,1) \frac{ 16 m^{2}\left (n-3\right )}{t}
%%%%
\nonumber \\ && 
- J_{011}(1,1,2) \frac {16 m^{2}}{t\left (n-4\right )\left (n-3\right )}
\Bigl[
4   m^{2}{n}^{2}
-28 m^{2}n
+48 m^{2}
+t{n}^{2}-5tn+8t
\Bigr]
%%%%
\nonumber \\ && 
+ \left[ A_0(m) \right] ^2
\frac{8 \left (n-2\right )}{t\left ( 4 m^{2}+t \right )\left (n-4\right )}
\Bigl[
t{n}^{2}-3tn+4{{ m}}^{2}n-16{{ m}}^{2}
\Bigr]
%%%%%%%%%%%%%%%%%%%%%%
\nonumber \\ && 
- \left[ B_0(m,m,t) \right]^2
{\frac {8 {{ m}}^{2}}{\left (4{{ m}}^{2}+t\right )\left (n-4\right )}}
\Bigl[
-4tn+4t+4{{ m}}^{2}{n}^{2}-24{{ m}}^{2}n+t{n}^{2}+40{{m}}^{2}
\Bigr]
%%%%%%%%%%%%%%%%%%%%%
\nonumber \\ && 
- A_0(m) B_0(m,m,t) 
\frac{ 8 \left( n - 2 \right) \left( {n}^{2}-5n+8\right )}{ \left(4 m^{2}+t \right) (n-4) (n-3)}
\Bigl[-2{{ m}}^{2}n+10{{ m}}^{2}+t \Bigr] 
\Biggr\} \;, 
\label{PL}
\end{eqnarray}
%%%%%%%%%%%%%%%%%%%%%%%%%%%%%%%%%%%%%%%%%%%%%%%%%%%%%%%
where the vertices $V_j$ ($j=1,2$) are defined as\footnote{Explicit 
values of coefficients  $v_j$ and $a_j$
for the Standard Model can be extracted from \cite{diagramatic}.}
$$
V_j = {\rm i} g \gamma_\mu (v_j + a_j \gamma_5) \;, 
$$
and we have introduced the following notations:
$$
A = a_1 a_2 \;, 
\quad 
S = v_1 v_2 \;, 
\quad
t=P^2,
$$
$N_c$ is a color factor (equal to 3 for quark and 1 for lepton), 
$C_F$ is the Casimir operator of the fundamental representation
of the Lie algebra (equal to $\tfrac{4}{3}$ for $SU(3)$ and 1 for QED), 
and $m$ is the mass of the loop fermion. Finally,
the occurring integrals are defined as
\begin{eqnarray}
J_{011}(\sigma, \nu_1, \nu_2) & = & 
\frac{\pi^{-n}}{\Gamma^2\left(3-\tfrac{n}{2}\right)}
\int \int \frac{{\rm d}^n K_1 \;\; {\rm d}^n K_2 }
{\left[(K_2-P)^2\right]^{\sigma}
\left[ K_1^2+m^2 \right]^{\nu_1}
\left[ (K_1-K_2)^2 + m^2 \right]^{\nu_2}} \; ,
\nonumber \\ 
B_0(m_a,m_b,t) & = & \frac{\pi^{-n/2}}{\Gamma\left(3-\tfrac{n}{2}\right)}
\int \frac{{\rm d}^n K}
{\left[(K-P)^2+m_a^2\right]
\left[ K^2+m_b^2 \right]} \; ,
\nonumber \\ 
A_0(m) & = & \frac{\pi^{-n/2}}{\Gamma\left(3-\tfrac{n}{2}\right)}
\int \frac{{\rm d}^n K}{K^2+m^2}
\equiv \frac{4 (m^2)^{n-2}}{(n-2)(n-4)} \;.
\end{eqnarray}
We note that the integral $J_{011}$ is defined in the same way
as in Section~4.3, one should only remember to substitute $p^2\to-t$.
 
In particular, for the zero momentum transfer, $P^2=t=0$, we get
\begin{eqnarray}
\Pi_{\rm L}^{(2)}(0) & = & 
- A \frac{g^2 g_{\rm s}^2}{(4 \pi)^{n/2}} N_c  C_F \left[ A_0(m) \right]^2 
\frac{4(n-2)(n^2-5n+7)}{m^2} \;, 
\nonumber \\
\Pi_{\rm T}^{(2)}(0) & = & 
- A \frac{g^2 g_s^2}{(4 \pi)^{n/2}} N_c  C_F \left[ A_0(m) \right]^2 
\frac{4(n-2)(n^3-6n^2+13n-11)}{(n-1)m^2} \;. 
\nonumber 
\end{eqnarray}

To obtain the finite terms of the $\ep$-expansion of the results
given in Eqs.~(\ref{PT}) and (\ref{PL}),
the integral $J_{011}(1,1,2)$ should be expanded up to the $\ep$-part.
Furthermore,
using the approach of Ref.~\cite{T97a}, we obtain the following 
relations between the integrals investigated in 
Section~4.3 and the master integrals $J_{011}(1,1,1)$ and  $J_{011}(1,1,2)$:
\begin{eqnarray}
&& 
\hspace*{-10mm}
J_{011}(1,1,1)\frac{(3n-8)(3n-10)(n-3)^2}{n-4}  = 
\left[ A_0(m)  \right]^2 (n-3)(n-2)^2 
\Biggl[
\frac{t}{4m^4} 
- \frac{(7n-24)}{2m^2 (n-4)}
\Biggr]
\nonumber \\ && 
+J_{011}(1,2,2)
\Biggl[ \frac{8 m^2 (t+m^2)}{n-4}
- 2 t^2 (n-3) + 2 t m^2 (7n-17) + 8 m^4 (2n-5)
\Biggr]
\nonumber \\ && 
+ \left[ J_{011}(1,2,2) + 2 J_{011}(2,1,2) \right] (n-3) (t+4m^2) (t-2 m^2)
\label{J111}
\;, 
\\ 
&& 
\hspace*{-10mm}
J_{011}(1,1,2)\frac{(3n-8)(3n-10)(n-3)^2}{n-4}  =
\left[ A_0(m)  \right]^2 
\frac{(n-3)(2n-7)(3n-8)(n-2)^2}{4m^4 (n-4)}
\nonumber \\ && 
+J_{011}(1,2,2)
\Biggl[
-\frac{ 4 (t+2m^2)}{n-4}
- n^2 \left( \tfrac{15}{2}t + 12m^2 \right) + n \left( \tfrac{79}{2}t+62m^2 \right) - 55 t - 86 m^2
\Biggr]
\nonumber \\ && 
+ \tfrac{1}{2} \left[ J_{011}(1,2,2) + 2 J_{011}(2,1,2) \right] (n-3) (3n-8)(t+4m^2) 
\;. 
\end{eqnarray}
From these relations we see that, in order to obtain the $\ep^j$ terms of 
$J_{011}(1,1,1)$ and  $J_{011}(1,1,2)$, the integral
$J_{011}(1,2,2)$ should be expanded up to $\ep^j$, whereas
the combination $[J_{011}(1,2,2)+2J_{011}(2,1,2)]$ 
up to $\ep^{j-1}$ only.
The expansion of the integrals near the threshold can be performed by using 
the numerical algorithm described in Ref.~\cite{expansion}.

The higher-order terms of the $\ep$-expansion 
of the one-loop integral $B_0(m_1,m_2,t)$ 
can be extracted from Refs.~\cite{D-ep,DK01}. 
Here we present coefficients up to the order $\ep^2$ 
for the particular case $m_1=m_2=m$:
\begin{eqnarray}
\hspace{-5mm}
B_0(m,m,t) & \!\!=\!\! & \frac{(m^2)^{-\ep}}{(1-2\ep) }
\Biggl(
\frac{1}{\ep} + \frac{1+y}{1-y} \Biggl\{ \ln y 
+ \ep \Bigl[  
\tfrac{1}{2} \ln^2 y \!-\! 2 \ln y \ln (1+y) \!-\! 2 \Li{2}{-y} 
\!-\! \zeta_2 
       \Bigr] 
\nonumber \\ && 
+ \ep^2 \Bigl[ 
4 \Snp{1,2}{-y} 
- 2 \Li{3}{-y} 
+ 4 \ln(1+y) \Li{2}{-y} 
- \ln^2 y \ln(1+y) 
\nonumber \\ && 
+ 2 \ln y \ln^2 (1+y) 
+ \tfrac{1}{6} \ln^3 y 
- \zeta_2 \ln y 
+ 2 \zeta_2 \ln(1+y) 
- 2\zeta_3 
\Bigr]  
+ {\cal O}(\ep^3)
\Biggr\}
\Biggr) \;, 
\hspace*{6mm}
\end{eqnarray}
where the variable $y$ is defined in Section~3.

The bare two-loop amplitudes (\ref{PT}) and (\ref{PL}) contain 
subdivergencies which should be canceled by proper counterterms,
$$
\Pi_{\rm CT}(P^2) = 
\delta m^2 \frac{\partial}{\partial m^2 } \Pi^{(1)} (P^2) \;, 
$$
where $\Pi^{(1)}$ is the one-loop amplitude, while $\delta m^2 $ 
is the one-loop mass
counterterm defined in a particular renormalization scheme. 
The derivatives of the bare one-loop amplitudes in $n$ dimensions
read
\begin{eqnarray}
\frac{\partial}{\partial m^2} \Pi_{\rm T}^{(1)}(P^2) 
& \!\!=\!\! & \frac{g^2}{(4 \pi )^{n/2}} N_c 4
\Biggl\{ 
\Biggl[ \frac{S t (3\!-\!n) }{(4m^2\!+\!t)} \!+\! S\!+\!(2\!-\!n)A \Biggr]B_0(m,m,t)
\!+\! \frac{2 S (n\!-\!2) }{(4m^2\!+\!t)} A_0(m)
\Biggr \} \;, 
\nonumber \\
\frac{\partial}{\partial m^2 } \Pi_{\rm L}^{(1)}(P^2) & \!\!=\!\! & 
A \frac{g^2}{(4 \pi )^{n/2}} N_c 4 
\Biggl \{ 
\Biggl[ \frac{t (n-3) }{(4m^2+t)} + 1-n \Biggr]B_0(m,m,t)
+ \frac{2 (2-n) }{(4m^2+t)} A_0(m)
\Biggr \} \;.
\nonumber 
\end{eqnarray}

As an example of application of these formulae, let us consider 
the transversal part of the $\gamma-Z$ propagator $(A=0)$
in the $\overline{MS}$-scheme.
Up to the finite in $\ep$ part, the result for the subtracted quantity 
$\Pi^{\rm sub}(P^2) = \Pi^{(2)}(P^2) + \Pi_{\rm CT}$ is 
\begin{eqnarray}
\label{gammaZ}
\Pi_{{\rm T},\gamma Z}^{\rm sub}(P^2) 
&=& S  \frac{g^2 g_{\rm s}^2}{(4 \pi)^2} N_c C_F    
\Biggl\{ 
- \frac{2t}{\ep} 
+ \left(-\tfrac{55}{3}t + \tfrac{296}{3} m^2 \right)
+ 16 t \zeta_3 \left[ 1 - 4 \frac{y^2}{(1-y)^4} \right]
\nonumber \\ && 
- \tfrac{16}{3} m^2 \frac{(1-4y+y^2)}{y(1-y)^2} 
\left [ \ln(1-y) + 2 \ln(1+y) \right] \ln y 
\left[ (1+y^2) \ln y - 2 (1-y^2) \right]
\nonumber \\ && 
+ \tfrac{8}{3} m^2 \frac{(2+7y-22y^2+6y^3)}{(1-y)^2} \ln^2 y
- 4 m^2 \ln y \frac{(1-6y-46y^2-6y^3+y^4)}{y(1-y^2)}
\nonumber \\ && 
+ 4 t \ln \frac{m^2}{\mu^2} \left[ 1- \frac{12 y}{(1-y)^2} \right] 
- 96 m^2 \frac{y}{1-y^2} \ln y \ln \frac{m^2}{\mu^2}
\nonumber \\ && 
+ \tfrac{32}{3} m^2 \frac{(1-4y+y^2)}{y(1-y)^2} 
\left [ \Li{2}{y} + 2 \Li{2}{-y} \right]  
\left[ 1-y^2 - 2\left(1+y^2\right) \ln y \right] 
\nonumber \\ && 
+ 32 m^2 \frac{(1-4y+y^2)}{y(1-y)^2} (1+y^2) 
\left [\Li{3}{y} + 2 \Li{3}{-y} \right]  
+ {\cal O}(\ep)
\Biggr\} \; , 
\end{eqnarray}
where we have taken into account the one-loop massive counterterm
$$
\delta m^2 =  - 2 \frac{g_{\rm s}^2}{(4\pi)^2} C_F \frac{3}{\ep} m^2 \;. 
$$
The small-momentum expansion of Eq.~(\ref{gammaZ}) is 
\begin{eqnarray}
\hspace{-10mm}
&& 
\left. \Pi_{T,\gamma Z}^{sub}(p^2) \right|_{P^2 \to 0 }= 
S  \frac{g^2 g_{\rm s}^2}{(4 \pi)^2} m^2 N_c C_F  
\Biggl\{ 
\frac{2}{\ep} u
+ \tfrac{13}{3} u + \tfrac{776}{405} u^2 
+ \ln \frac{m^2}{\mu^2 } \left( 4 u + \tfrac{8}{5} u^2 \right)
+ {\cal O}(u^3)
\Biggr\} \;, 
\nonumber 
\end{eqnarray}
where $u=-P^2/m^2$.
Let us remind that in a theory with the spontaneous symmetry breaking, 
like the Standard Model, 
the inclusion of tadpoles~\cite{FJ} is also required 
for the renormalization group 
invariance of the massive parameters~\cite{poleI,poleII,expansion}.
The proper bare two-loop tadpole contribution 
is given in Section~4.3 of Ref.~\cite{poleII}.

%%%%%%%%%%%%%%%%%%%%%%%%%%%%%%%%%%%%%%%%%%%%%%%%%%%%%%%%%%%%%%%%%%%%%%%%%%%%%%%
%%%%%%%%%%%%%%%%%%%%%%%%%%%%%%%%%%%%%%%%%%%%%%%%%%%%%%%%%%%%%%%%%%%%%%%%%%%%%%%

%%%%%%%%%%%%%%%%%%%%%%%%%%%%%%%%%%%%%%%%%%%%%%%%%%%%%%%%%%%%%%%%%%%%%%%%%%%%%%%%%%%%%%%%%
%%%%%%%%%%%%%%%%%%%%%%%%%%%%%%%%%%%%%%%%%%%%%%%%%%%%%%%%%%%%%%%%%%%%%%%%%%%%%%%%%%%%%%%%%

\begin{thebibliography}{99}
\bibitem{dimreg}
G.~'tHooft and M.~Veltman,
Nucl.\ Phys.\ {\bf B44} (1972) 189;\\
%%CITATION = NUPHA,B44,189;%%
C.G.~Bollini and J.J.~Giambiagi,
Nuovo~Cim.\ {\bf 12B} (1972) 20; \\    
%%CITATION = NUCIA,B12,20;%%
J.F.~Ashmore,  
Lett.\ Nuovo Cim.\ {\bf 4} (1972) 289;\\
%%CITATION = NCLTA,4,289;%%
G.M.~Cicuta and E.~Montaldi,
Lett.\ Nuovo Cim.\ {\bf 4} (1972) 329.
%%CITATION = NCLTA,4,329;%%

\bibitem{Lewin}
L.~Lewin, {\it Polylogarithms and associated functions}
(North-Holland, Amsterdam, 1981).

\bibitem{Nielsen}
K.S.~K\"olbig, J.A.~Mignaco and E.~Remiddi, B.I.T. {\bf 10} (1970) 38;\\
%%CITATION = BITXA,10,38;%%
R.~Barbieri, J.A.~Mignaco and E.~Remiddi, Nuovo Cim.\ {\bf A11} (1972) 824;\\
%%CITATION = NUCIA,A11,824;%%
A.~Devoto and D.W.~Duke, Riv.\ Nuovo Cim.\  {\bf 7}, No.6 (1984) 1;\\
%%CITATION = RNCIB,7N6,1;%%
K.S.~K\"olbig, SIAM J.\ Math.\ Anal.\ {\bf 17} (1986) 1232.
%%CITATION = SJMAA,17,1232;%%

\bibitem{RV00}
E.~Remiddi and J.A.M.~Vermaseren,
Int.\ J.\ Mod.\ Phys.\ {\bf A15} (2000) 725.
%%CITATION = IMPAE,A15,725;%%
 
\bibitem{GR2001}
A.B.~Goncharov,
Math.\ Res.\ Lett.\ {\bf 5} (1998) 497; \\
%%CITATION = 00146,5,497;%%
J.M.~Borwein, D.M.~Bradley, D.J.~Broadhurst and P.~Lison\v ek,
Trans.\ Amer.\ Math.\ Soc.\ {\bf 353} (2001) 907; \\
%%CITATION = TAMTA,353,907;%%
T.~Gehrmann and E.~Remiddi,
Nucl.\ Phys.\ {\bf B601} (2001) 248.
%%CITATION = NUPHA,B601,248;%%

\bibitem{nested}
S.~Moch, P.~Uwer and S.~Weinzierl,
J.\ Math.\ Phys.\  {\bf 43} (2002) 3363.
%%CITATION = HEP-PH 0110083;%%

\bibitem{expansionII}
S.~Weinzierl, hep-ph/0402131.
%%CITATION = HEP-PH 0402131;%%


\bibitem{poleII}
F.~Jegerlehner, M.Yu.~Kalmykov and O.~Veretin, 
Nucl.\ Phys.\ {\bf B658} (2003) 49.
%%CITATION = HEP-PH 0212319;%%

\bibitem{dem}
A.V.~Kotikov,
Phys.\ Lett.\ {\bf B254} (1991) 158;
%%CITATION = PHLTA,B254,158;%%
{\bf B259} (1991) 314. 
%%CITATION = PHLTA,B259,314;%%

\bibitem{BD-TMF}
E.E.~Boos and A.I.~Davydychev,
Teor.\ Mat.\ Fiz.\ {\bf 89} (1991) 56.
%%CITATION = TMFZA,89,56;%%

\bibitem{Isaev}
A.P.~Isaev,
%``Multi-loop Feynman integrals and conformal quantum mechanics,''
Nucl.\ Phys.\ {\bf B662} (2003) 461.
%%CITATION = HEP-TH 0303056;%%

\bibitem{FKV98}
J.~Fleischer, A.V.~Kotikov and O.L.~Veretin,
Phys.\ Lett.\ {\bf B417} (1998) 163.
%%CITATION = HEP-PH 9707492;%%

\bibitem{DD} A.I.~Davydychev and R.~Delbourgo, 
J.\ Math.\ Phys.\ {\bf 39} (1998) 4299. 
%%CITATION = JMAPA,39,4299;%%

\bibitem{FKV99}
J.~Fleischer, A.V.~Kotikov and O.L.~Veretin,
Nucl.\ Phys.\ {\bf B547} (1999) 343.
%%CITATION = NUPHA,B547,343;%%

\bibitem{DK-bastei} 
A.I.~Davydychev and M.Yu.~Kalmykov,
Nucl.\ Phys.\ B (Proc.\ Suppl.) {\bf 89} (2000) 283.
%%CITATION = HEP-TH 0005287;%%

\bibitem{DK01}
A.I.~Davydychev and  M.Yu.~Kalmykov,
Nucl.\ Phys.\ {\bf B605} (2001) 266.   
%%CITATION = HEP-TH 0012189;%%

\bibitem{oneloop}
J.~Fleischer, F.~Jegerlehner and O.V.~Tarasov, 
Nucl.\ Phys.\ {\bf B672} (2003) 303.
%%CITATION = HEP-PH 0307113;%%

\bibitem{single}
J.~Fleischer, M.Yu.~Kalmykov and  A.V.~Kotikov,
Phys.\ Lett.\ {\bf B462} (1999) 169; {\bf B467} (1999) 310(E).
%%CITATION = PHLTA,B462,169;%%

\bibitem{harmonic}
D.I.~Kazakov and A.V.~Kotikov, 
Theor.\ Math.\ Phys.\  {\bf 73} (1988) 1264; 
%%CITATION = TMPHA,73,1264;%% 
Nucl.\ Phys.\ {\bf B307} (1988) 721; \\
%%CITATION = NUPHA,B307,721;%% 
D.~Borwein, J.M.~Borwein and R.~Girgensohn,
Proc.\ Edinb.\ Math.\ Soc.\ {\bf 38} (1995) 277;\\
%%CITATION = PEMSA,38,277;%%
J.M.~Borwein and R.~Girgensohn,
Electron.\ J.\ Combin.\ {\bf 3} (1996) R23;\\
%%CITATION = 00185,3,R23;%%
P.~Flajolet and B.~Salvy, Experimental Math.\ {\bf 7} (1998) 15;\\
%%CITATION = 00162,7,15;%%
O.M.~Ogreid and P.~Osland, J.\ Comput.\ Appl.\ Math.\ {\bf 98} (1998) 245;
%%CITATION = JCAMD,98,245;%%
{\bf 136} (2001) 389; \\
%%CITATION = HEP-TH 9904206;%%
V.A.~Smirnov,
Phys.\ Lett.\ {\bf B547} (2002) 239. 
%%CITATION = HEP-PH 0209193;%%

\bibitem{KV00}
M.Yu.~Kalmykov and O.~Veretin, Phys.\ Lett.\ {\bf B483} (2000) 315; \\
%%CITATION = PHLTA,B483,315;%%
J.M.~Borwein and R.~Girgensohn, 
{\em ``Evaluation of Binomial Series''},
CECM preprint CECM-02-188; 
available at {\tt http://www.cecm.sfu.ca/preprints/2002pp.html}.

\bibitem{odd}
J.~Fleischer and M.Yu.~Kalmykov,
%``Single mass scale diagrams: Construction of a basis for the  epsilon-expansion,''
Phys.\ Lett.\ {\bf B470} (1999) 168.
%%CITATION = HEP-PH 9910223;%%

\bibitem{euler}
D.J.~Broadhurst, 
Open University preprint OUT-4102-62 (hep-th/9604128).
%%CITATION = HEP-TH 9604128;%%

\bibitem{B99}
D.J.~Broadhurst, Eur.\ Phys.\ J.\ {\bf C8} (1999) 311.
%%CITATION = EPHJA,C8,311;%%

\bibitem{review}
A.I.~Davydychev and M.Yu.~Kalmykov,
%``Geometrical approach to loop calculations and the epsilon-expansion of  Feynman diagrams,''
Proc. Workshop ``CPP2001'', Tokyo, Japan, November 2001, 
KEK Proceedings 2002-11, p.~169
(hep-th/0203212).
%%CITATION = HEP-TH 0203212;%%

\bibitem{newI}
N.~Batir, 
Appl.\ Math.\ Comput.\ {\bf 147} (2004) 645.

\bibitem{sums}
Z.~Nan-Yue and K.S.~Williams, Pacific J.\ Math.\ {\bf 168} (1995) 271.
%%CITATION = PJMAA,168,271;%%

\bibitem{BBK}
J.M.~Borwein, D.J.~Broadhurst and J.~Kamnitzer,
Experimental Math. {\bf 10} (2001) 25.
%%CITATION = HEP-TH 0004153;%%

\bibitem{DT2}
A.I.~Davydychev and J.B.~Tausk,
Phys.\ Rev.\ {\bf D53} (1996) 7381; \\
%%CITATION = PHRVA,D53,7381;%%

\bibitem{Ls_ex}
A.I.~Davydychev,
%Mainz preprint MZ-TH/99-30
Proc.\ Workshop ``AIHENP-99'', Heraklion, Greece, April 1999 (Parisianou S.A.,
Athens, 2000), p.~219
(hep-th/9908032);\\
%%CITATION = HEP-TH 9908032;%% 
A.V.~Kotikov and L.N.~Lipatov,
Nucl.\ Phys.\ {\bf B582} (2000) 19.
%%CITATION = NUPHA,B582,19;%%

\bibitem{D-ep}
A.I.~Davydychev, Phys.\ Rev.\ {\bf D61} (2000) 087701.
%%CITATION = PHRVA,D61,087701;%%

\bibitem{PBM3} A.P.~Prudnikov,  Yu.A.~Brychkov  and O.I.~Marichev,
{\em Integrals and Series, v.3: More Special Functions},
Gordon and Breach, New York, 1990.

\bibitem{generatingfunction}
H.S.~Wilf,
{\it Generatingfunctionology}, Academic Press, London, 1994,\\
{\tt http://www.math.upenn.edu/$\;\widetilde{}\;$wilf/DownldGF.html}

\bibitem{difference}
O.V.~Tarasov,
%``Connection between Feynman integrals having different values of the  space-time dimension,''
Phys.\ Rev.\ {\bf D54} (1996) 6479;
%%CITATION = HEP-TH 9606018;%%
%``Application and explicit solution of recurrence relations with respect  to space-time dimension,''
Nucl.\ Phys.\ B (Proc.\ Suppl.) {\bf 89} (2000) 237; \\
%%CITATION = HEP-PH 0102271;%%
S.~Moch and J.A.M.~Vermaseren,
%``Deep inelastic structure functions at two loops,''
Nucl.\ Phys.\ {\bf B573} (2000) 853; \\
%%CITATION = HEP-PH 9912355;%%
S.~Moch, J.A.M.~Vermaseren and A.~Vogt,
%``Non-singlet structure functions at three loops: Fermionic contributions,''
Nucl.\ Phys.\ {\bf B646} (2002) 181; \\
%%CITATION = HEP-PH 0209100;%%
S.~Laporta,
%``High-precision calculation of multi-loop Feynman integrals by  difference equations,''
Int.\ J.\ Mod.\ Phys.\ {\bf A15} (2000) 5087; 
%%CITATION = HEP-PH 0102033;%%
%``Calculation of master integrals by difference equations,''
Phys.\ Lett.\ {\bf B504} (2001) 188;
%%CITATION = HEP-PH 0102032;%%
%``High-precision epsilon expansions of massive four-loop vacuum bubbles,''
%Phys.\ Lett.\ 
{\bf B549} (2002) 115.
%%CITATION = HEP-PH 0210336;%%

\bibitem{BFT93}
D.J.~Broadhurst, J.~Fleischer and O.V.~Tarasov,
Z.\ Phys.\ {\bf C60} (1993) 287.
%%CITATION = ZEPYA,C60,287;%%

\bibitem{B90}
D.J.~Broadhurst, Z.\ Phys.\ {\bf C47} (1990) 115.
%%CITATION = ZEPYA,C47,115;%%

\bibitem{D91}
A.I.~Davydychev,
{\em ``Loop calculations in QCD with massive quarks"},
talk at Int.\ Conf.\ ``Relativistic Nuclear Dynamics"
(Vladivostok, Russia, September 1991),\\
{\tt http://wwwthep.physik.uni-mainz.de/$\;\widetilde{}\;$davyd/preprints/vladiv.ps.gz}

\bibitem{sunset:im}
A.~Bashir, R.~Delbourgo and M.~L.~Roberts,
%``Multidimensional phase space and sunset diagrams,''
J.\ Math.\ Phys.\  {\bf 42} (2001) 5553.
%%CITATION = HEP-TH 0101148;%%

\bibitem{T97a}
O.V.~Tarasov, Nucl.\ Phys.\ {\bf B502} (1997) 455.
%%CITATION = NUPHA,B502,455;%%

\bibitem{thresholds}
A.I.~Davydychev and V.A.~Smirnov, Nucl.\ Phys.\ {\bf B554} (1999) 391.
%%CITATION = NUPHA,B554,391;%%

\bibitem{T96}
O.V.~Tarasov, Phys.\ Rev.\ {\bf D54} (1996) 6479.
%%CITATION = HEP-TH 9606018;%%

\bibitem{Hgg:num}
%J.~Fleischer et al.,
J.~Fleischer, V.A.~Smirnov, A.~Frink, J.G.~K\"orner, D.~Kreimer, K.~Schilcher and J.B.~Tausk,
%``Calculation of infrared-divergent Feynman diagrams with zero mass  threshold,''
Eur.\ Phys.\ J. {\bf C2} (1998) 747.
%%CITATION = HEP-PH 9704353;%%

\bibitem{Hgg}
A.~Djouadi, M.~Spira and P.M.~Zerwas,
%``Production of Higgs bosons in proton colliders: QCD corrections,''
Phys.\ Lett.\ {\bf B264} (1991) 440; \\
%%CITATION = PHLTA,B264,440;%%
S.~Dawson and R.~Kauffman,
%``QCD corrections to Higgs boson production: nonleading terms in the heavy quark limit,''
Phys.\ Rev.\ {\bf D49} (1994) 2298; \\
%%CITATION = HEP-PH 9310281;%%
M.~Spira, A.~Djouadi, D.~Graudenz and P.M.~Zerwas,
%``Higgs boson production at the LHC,''
Nucl.\ Phys.\ {\bf B453} (1995) 17.
%%CITATION = HEP-PH 9504378;%%

\bibitem{passarino}
G.~Passarino and S.~Uccirati,
%``Algebraic-numerical evaluation of Feynman diagrams: Two-loop  self-energies,''
Nucl.\ Phys.\ {\bf B629} (2002) 97; \\
%%CITATION = HEP-PH 0112004;%%
A.~Ferroglia, G.~Passarino, S.~Uccirati and M.~Passera,
%``A Frontier In Multi-Scale Multi-Loop Integrals: The Algebraic-Numerical Method,''
Nucl.\ Instrum.\ Meth.\ {\bf A502} (2003) 391; \\
%%CITATION = NUIMA,A502,391;%%
A.~Ferroglia, M.~Passera, G.~Passarino and S.~Uccirati,
%``All-purpose numerical evaluation of one-loop multi-leg Feynman diagrams,''
Nucl.\ Phys.\ {\bf B650} (2003) 162; \\
%%CITATION = HEP-PH 0209219;%%
A.~Ferroglia, G.~Passarino, M.~Passera and S.~Uccirati,
%``Two-Loop Vertices in Quantum Field Theory: Infrared Convergent Scalar Configurations,''
Nucl.\ Phys.\ {\bf B680} (2004) 199.
%%CITATION = HEP-PH 0311186;%%

\bibitem{confirnation_BMR}
R.~Bonciani, P.~Mastrolia and E.~Remiddi,
%``Master Integrals for the 2-loop QCD virtual corrections to the Forward-Backward Asymmetry,''
hep-ph/0311145.
%%CITATION = HEP-PH 0311145;%%


\bibitem{vertex_BMR}
R.~Bonciani, P.~Mastrolia and E.~Remiddi,
%``Vertex diagrams for the QED form factors at the 2-loop level,''
Nucl.\ Phys.\ {\bf B661} (2003) 289.
%%CITATION = HEP-PH 0301170;%%

\bibitem{S_Bha}
V.A.~Smirnov,
Phys.\ Lett.\ {\bf B524} (2002) 129.
%%CITATION = PHLTA,B524,129;%% 

\bibitem{DOS1}
A.I.~Davydychev, P.~Osland and L.~Saks,
%``Quark gluon vertex in arbitrary gauge and dimension,''
Phys.\ Rev.\ D {\bf 63} (2001) 014022.
%[arXiv:hep-ph/0008171].
%%CITATION = PHRVA,D63,014022;%%

\bibitem{vertex_DS}
A.I.~Davydychev and V.A.~Smirnov, 
Nucl.\ Instr.\ Meth.\ {\bf A502} (2003) 621. 
%``Analytical evaluation of certain on-shell two-loop three-point  diagrams,''
% (hep-ph/0210171).
%%CITATION = HEP-PH 0210171;%%

\bibitem{leo96}
L.V.~Avdeev, Comput.\ Phys.\ Commun.\ {\bf 98} (1996) 15.
%%CITATION = CPHCB,98,15;%%

\bibitem{matad}
M.~Steinhauser, Comput.\ Phys.\ Commun.\ {\bf 134} (2001) 335.
%%CITATION = HEP-PH 0009029;%%

\bibitem{B92}
D.J.~Broadhurst, Z.\ Phys.\ {\bf C54} (1992) 599.
%%CITATION = ZEPYA,C54,599;%%

\bibitem{three-different}
A.V.~Kotikov, 
J.\ High Energy Phys.\ {\bf 09} (1998) 001; \\
%%CITATION = JHEPA,9809,001;%%
J.M.~Chung and  B.K.~Chung, 
Phys.\ Rev.\ {\bf D59} (1999) 105014;
%%CITATION = PHRVA,D59,105014;%% 
J.\ Korean Phys.\ Soc.\  {\bf 40} (2002) 435
(hep-ph/0203143).
%%CITATION = HEP-PH 0203143;%%

\bibitem{GKP}
S.~Groote, J.G.~K\"orner and A.A.~Pivovarov,
%``Transcendental numbers and the topology of three-loop bubbles,''
Phys.\ Rev.\ {\bf D60} (1999) 061701.
%%CITATION = HEP-PH 9904304;%%

\bibitem{PSLQ} 
H.R.P.~Ferguson, D.H.~Bailey and S. Arno, 
Math.\ Comput.\ {\bf 68} (1999) 351.
%%CITATION = MCMPA,68,351;%% 

\bibitem{background}
F.~Jegerlehner and O.V.~Tarasov, Nucl.\ Phys.\ {\bf B549} (1999) 481.
%%CITATION = HEP-PH 9809485;%%

\bibitem{qcd}
T.H.~Chang, K.J.~Gaemers and W.L.~van Neerven, Nucl.\ Phys.\ {\bf B202} (1982) 407; \\
%%CITATION = NUPHA,B202,407;%%
A.~Djouadi and C.~Verzegnassi, Phys.\ Lett.\ {\bf B195} (1987) 265; \\
%%CITATION = PHLTA,B195,265;%%
A.~Djouadi, Nuovo Cim.\ {\bf A100} (1988) 357; \\
%%CITATION = NUCIA,A100,357;%%
B.A.~Kniehl, J.H.~K\"uhn and R.G.~Stuart, Phys.\ Lett.\ {\bf B214} (1988) 621;\\
%%CITATION = PHLTA,B214,621;%%
B.A.~Kniehl, Nucl.\ Phys.\ {\bf B347} (1990) 86 ; \\
%%CITATION = NUPHA,B347,86;%%
A.~Djouadi and P.~Gambino, Phys.\ Rev.\ {\bf D49} (1994) 3499;
{\bf D53} (1996)  4111(E).
%%CITATION = HEP-PH 9309298;%%

\bibitem{diagramatic}
M.~Veltman, {\it Diagrammatica}, Cambridge University Press, 1994; \\
D.~Bardin and G.~Passarino, {\it The Standard Model in the Making}, Clarendon, Oxford, 1999.

\bibitem{expansion}
F.~Jegerlehner, M.Yu.~Kalmykov and O.~Veretin, 
Nucl.\ Phys.\ B (Proc.\ Suppl.) {\bf 116} (2003) 382 (hep-ph/0212003);
%%CITATION = HEP-PH 0212003;%%
Nucl.\ Instr.\ Meth.\ {\bf A502} (2003) 618. 

\bibitem{FJ} 
J.~Fleischer and F.~Jegerlehner, 
Phys.\ Rev.\ {\bf D23} (1981) 2001. 
%%CITATION = PHRVA,D23,2001;%%

\bibitem{poleI}
F.~Jegerlehner, M.Yu.~Kalmykov and O.~Veretin, 
Nucl.\ Phys.\ {\bf B641} (2002) 285.
%%CITATION = HEP-PH 0105304;%%

\bibitem{d_3}
K.G.~Chetyrkin and M.~Steinhauser,
Nucl.\ Phys.\ {\bf B573} (2000) 617.
%%CITATION = HEP-PH 9911434;%%

\bibitem{pole_topI}
F.~Jegerlehner and M.Yu.~Kalmykov,
%``The O(alpha alpha(s)) correction to the pole mass of the t-quark 
% within the standard model,''
Nucl.\ Phys.\ {\bf B676} (2004) 365.
%%CITATION = HEP-PH 0308216;%%

\bibitem{newII}
J.~Bl\"umlein,
hep-ph/0311046.
%%CITATION = HEP-PH 0311046;%%

\end{thebibliography}
\end{document}